\documentclass[journal]{IEEEtran}
\IEEEoverridecommandlockouts

\usepackage{xcolor}
\usepackage{amssymb}
\usepackage{amsmath,amsfonts}

\usepackage{enumitem}
\usepackage{array}
\usepackage{multirow}
\usepackage[caption=false,textfont=rm]{subfig}
\usepackage{textcomp}
\usepackage{stfloats}
\usepackage{url}
\usepackage{verbatim}
\usepackage{graphicx}
\usepackage{tikz}

\usetikzlibrary{shapes,arrows,positioning,fit,calc}
\usepackage{booktabs} 
\usepackage{siunitx}
\usepackage{cite} 
\usepackage[colorlinks=true, linkcolor=blue, citecolor=blue, urlcolor=black]{hyperref}
\usepackage{bm}
\usepackage{balance}

\usepackage{thmtools}

\newtheorem{assumption}{Assumption}
    
\newtheorem{remark}{Remark}
\newtheorem{definition}{Definition}

\def\BibTeX{{\rm B\kern-.05em{\sc i\kern-.025em b}\kern-.08emT\kern-.1667em\lower.ex\hbox{E}\kern-.125emX}}

\begin{document}
\bstctlcite{IEEEtranBSTCTL}

\title{Hard/Soft NLoS Detection via Combinatorial Data Augmentation for 6G Positioning}

\author{Sang-Hyeok Kim, Seung Min Yu, Jihong Park,~\IEEEmembership{Senior Member,~IEEE},
and Seung-Woo Ko,~\IEEEmembership{Senior Member,~IEEE}%

\thanks{Sang-Hyeok Kim and Seung-Woo Ko are with Inha University, Incheon 21999, South Korea
(e-mail: sanghyeok.kim@inha.edu, swko@inha.ac.kr).}%
\thanks{Seung Min Yu is with the Korea Railroad Research Institute, Uiwang 16105, South Korea
(e-mail: smyu@krri.re.kr).}%
\thanks{Jihong Park is with Singapore University of Technology and Design (SUTD), Singapore 487372
(e-mail: jihong\_park@sutd.edu.sg).}%
\thanks{An earlier version of this paper was presented in part at IEEE Wireless Communications and Networking Conference (WCNC) 2026~\cite{Kim2026CDAND}.}%
}

\IEEEpubidadjcol

\maketitle

\begin{abstract}
A key enabler for achieving the stringent requirements of 6G positioning is the ability to exploit site-dependent information that governs \textit{line-of-sight} (LoS) and \textit{Non-LoS} (NLoS) propagation. However, acquiring such environmental information in real-time remains challenging in practice. To address this issue, we propose a novel NLoS detection algorithm termed \textit{combinatorial data augmentation-guided NLoS detection} (CDA-ND), which builds upon our prior work. CDA-ND generates numerous \textit{preliminary estimated locations} (PELs) by applying multilateration over many different \textit{gNodeB} (gNB) combinations using a single snapshot of range measurements. When a target gNB is in NLoS, the resulting PELs naturally split into two clusters: one derived using the target gNB’s range measurement, and the other derived without it. The displacement between the two clusters is represented by a single vector, called the \textit{NLoS evidence vector} (NEV), which serves as a key feature for computing an NLoS likelihood score. Based on this score, two modes of NLoS detection are developed. First, each gNB is classified as LoS or NLoS, termed \textit{hard decision} (HD), which can be designed on-the-fly via a simple threshold test on the NLoS likelihood score. Second, each gNB's NLoS confidence is probabilistically quantified, termed \textit{soft decision} (SD), which extends HD with weak site-survey priors, namely empirical NLoS-score samples and the average NLoS probability. We then design the subsequent positioning algorithms tailored to these two modes by excluding gNBs deemed NLoS (for both HD and SD) and re-weighting the remaining gNBs (for SD). The proposed CDA-ND achieves high reliability in indoor factory environments under frequency range~1, attaining NLoS detection accuracies of \SI{96.8}{\%} and \SI{91.5}{\%} when the proportion of NLoS gNBs is approximately \SI{18}{\%} and \SI{56}{\%}, respectively. As a result, a positioning algorithm integrating CDA-ND significantly improves positioning accuracy, achieving reductions of about \SI{22}{\%} and \SI{67}{\%} in mean absolute error under LoS- and NLoS-dominant environments, respectively.
\end{abstract}

\begin{IEEEkeywords}
6G positioning, NLoS detection, combinatorial data augmentation, 3GPP indoor factory, hard/soft decision.
\end{IEEEkeywords}

\section{Introduction}
Positioning is expected to become a core capability of 6G networks, supporting the broader vision of tightly integrating sensing with communications \cite{Chen2022THZ}. Unlike current 5G positioning services, which largely localize a generic \textit{user equipment} (UE) at roughly cell-coverage granularity, 6G positioning is envisioned to be context-specific, adapting to the physical site and its temporal variations (e.g., smart factory and indoor office). Such context-aware designs can exploit environment-dependent characteristics to meet stringent 6G accuracy requirements for positioning and downstream tasks, as exemplified by \textit{non-line-of-sight} (NLoS)-aware design using blockage-related cues (e.g., \cite{Kim2026LMMBM, Henk2025FrugalRIS}). However, obtaining and maintaining such context information in practice is challenging, calling for techniques that infer and leverage it directly from the current measurements without relying on prior knowledge. To address this challenge, this work builds upon our prior work on \textit{combinatorial data augmentation} (CDA) \cite{yu2025combinatorial}, which is effective in extracting environment-dependent context information using only real-time range measurements. We further transform the captured context into a tractable representation that enables \textit{gNodeB} (gNB)-wise NLoS detection in either binary form, termed \textit{hard decision} (HD), or probabilistic form, termed \textit{soft decision} (SD). These decisions are then integrated into novel positioning algorithms designed to meet the accuracy requirement of 6G positioning. 

\vspace{-8pt}

\subsection{Prior Work} \label{sec:prior_work}

In this subsection, we briefly review existing wireless positioning and NLoS detection methodologies, summarizing their representative approaches and inherent limitations.

\subsubsection{Wireless Positioning towards 6G}

Traditional 5G positioning techniques rely heavily on geometry-driven approaches, including time-of-arrival, time-difference-of-arrival, and angle-of-arrival \cite{Guvenc2009Survey}. While these methods perform reliably under \textit{line-of-sight} (LoS) conditions, their efficacy diminishes in NLoS-dominant environments, such as dense urban canyons and complex indoor industrial sites. In these settings, physical obstructions introduce positive biases into range measurements, leading to significant accuracy degradation during the multilateration process \cite{Yang2024positioning}.

6G has introduced several new features to address the challenges of NLoS propagation. For example, \textit{extremely large-scale multiple-input-multiple-output} (XL-MIMO) has been considered a pivotal enabler, as its superior angular resolution allows for the precise separation of the LoS path from NLoS clutter \cite{Li2019MMIMO}. Such resolvability allows multipath components to be leveraged as a new spatial degree of freedom, augmenting localization performance rather than degrading it (see, e.g., \cite{Henk2025JointLoc, Zhou2026MixedNearFarXLMIMO}). Recently, the concept of \textit{movable antenna} (MA) has been introduced, which mechanically adjusts antenna position to mitigate deep fading and optimize channel conditions \cite{Zhu2026MA}. The trajectory of MA forms a synthetic aperture with a limited number of antennas, helping resolve and exploit multipath components similar to XL-MIMO \cite{Xu2025FAS ,Jiang2026MA}. \textit{Reconfigurable intelligent surface} (RIS) is another key technique in 6G, where electromagnetic propagation is intentionally tuned to establish virtual links that bypass blockage \cite{Xia2025RIS} or leverage near-field wavefronts to extract additional spatial information \cite{Kang2025RISLoc}.

Despite their potential, these techniques often rely on idealized assumptions, such as perfect channel information and continuous phase control, which are frequently untenable in practice due to prohibitive signaling overheads and hardware impairments. Moreover, the substantial hardware investment required for these features poses a significant challenge to the 6G mandate for cost-effective network deployment \cite{3GPPTR22.870V1.1.0}.

\subsubsection{AI/ML-Based NLoS Detection \& Positioning} Recent advances in \textit{artificial intelligence} (AI) and \textit{machine learning} (ML) have enabled effective mitigation of the aforementioned NLoS issue. These approaches can be broadly categorized into two methodologies. 

The first approach leverages AI/ML models to identify measurements likely captured under NLoS conditions, and subsequently exploits these labels to refine positioning estimates. Classic NLoS detection has been extensively studied under the premise that measurements adhere to predefined probabilistic models, such as Gaussian processes \cite{Yang2018Gaussianprocess} and Gaussian mixture models \cite{Wang2011GMM}. However, these models often fail to generalize due to frequent discrepancies with real-world data distributions. In contrast, AI/ML-based frameworks are tailored to capture these latent distributions, effectively learning the complex patterns of the propagation environment embedded in the data. 

In \cite{marano2010nlos}, for example, a \textit{support vector machine} (SVM)-based NLoS classifier is trained on \textit{channel impulse responses} (CIRs) measured from ultra-wideband transceivers with LoS/NLoS ground-truth labels. A closely related strategy is adopted for MIMO systems in \cite{huang2020MachineLearningMIMO}, where  multiple classifiers, such as SVM, random forest, and neural networks, are trained on \textit{channel state information} (CSI) to distinguish LoS from NLoS conditions. While such supervised approaches are highly effective, they often rely on substantial labeled datasets, motivating efforts to reduce the labeling burden in  \cite{Tedeschini2023latentSpace}, where LoS-only data samples are used to learn channel embeddings via an autoencoder. NLoS events are then identified  through density-based anomaly detection in the learned latent space. 
Moreover, sequences of consecutive channel-related measurements are often more informative than a single snapshot for NLoS detection, as temporal dynamics can reveal propagation changes and blockage events. To exploit such temporal structure, long short-term memory and transformer-based classifiers are employed in \cite{Choi2018deep} and \cite{Yang2025Fuzzy}, respectively, both of which are well-suited for handling time-varying patterns. Beyond the above binary decisions, the probabilistic description of NLoS propagation is learned in \cite{torsoli2023BI} by training a Real AdaBoost model on labeled CIRs, yielding soft reliability information that can be directly integrated into robust localization.
  
The second approach directly estimates the target position, bypassing the need for explicit NLoS detection. In this paradigm, NLoS-induced distortions are not viewed as errors to be mitigated, but as informative latent features that characterize the specific propagation environment. By training on these complex signal signatures, the model learns a direct mapping from raw measurements to spatial coordinates, effectively exploiting multipath characteristics as a valuable source of localization information rather than a source of interference. 

A representative example is fingerprint-based positioning, where location-dependent fingerprints (e.g., CIR, CSI, or received signal strength) are directly mapped to the target UE's coordinates \cite{Zhu2020FingerprintSurvey}. With recent advances in AI/ML models, such as transformer-based architectures \cite{Xu2024SwinLoc, Wang2026Transformer} and emerging large language model paradigm \cite{bhatia2025indoor}, fingerprinting has been actively studied as these models can capture high-dimensional, nonlinear dependencies in rich channel measurements. However, such end-to-end mappings are often considered black-box predictors that are difficult to interpret and analyze, making it nontrivial to connect the learned model to a well-established localization principle and performance guarantees. A recent study in \cite{Morselli2023SI} proposes learning a mapping from measurements to the likelihood function, rather than the target location, enabling the use of a maximum likelihood estimator, which is theoretically optimal under standard conditions.

Despite their effectiveness, the practical deployment of the aforementioned AI/ML-based NLoS detection and positioning algorithms remains questionable since they typically require dense site-surveys with ground-truth position labels and often demand frequent re-training as the environment changes.

\begin{figure}[!t]  
\centering
\includegraphics[width=0.94\columnwidth]{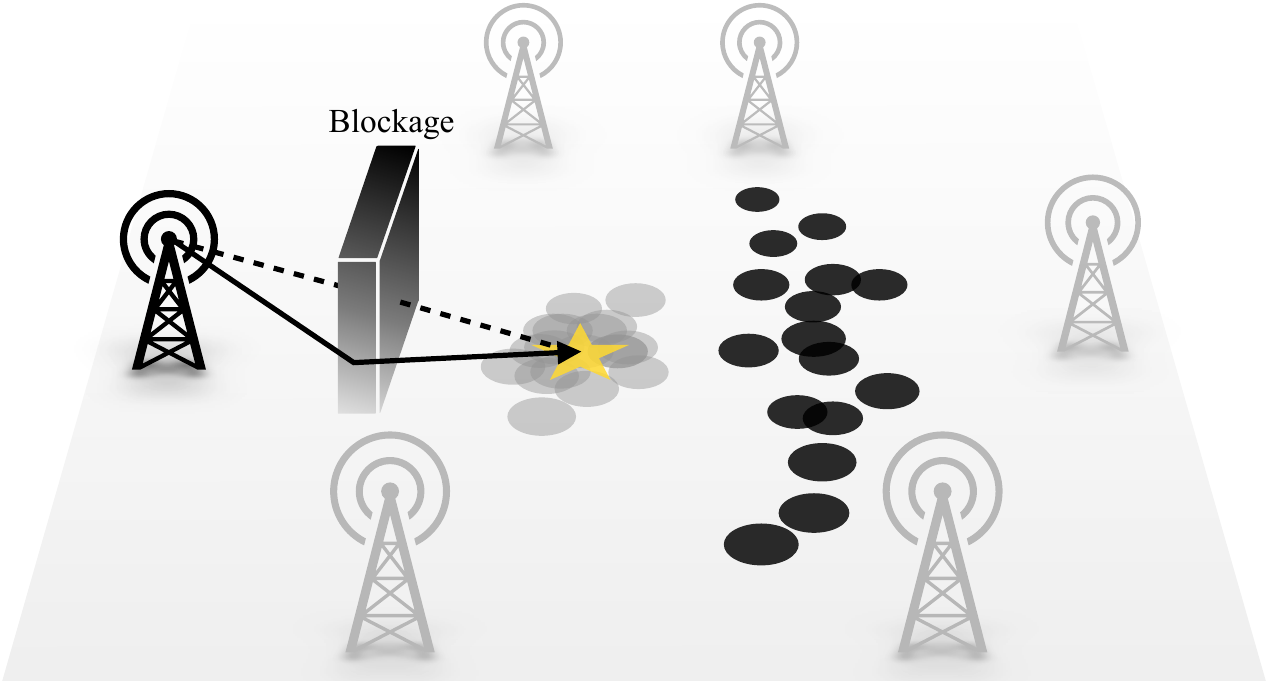}
\vspace{-5pt}

\caption{Graphical illustration of a single snapshot of PELs skewed by a target NLoS gNB. The black circles represent PELs constructed from the target gNB's measurements, while the gray circles represent PELs constructed without it. The yellow star indicates the true UE position.}
\label{fig:hypo}
\vspace{-7pt}
\end{figure}

\vspace{-3pt}

\subsection{Contributions}

As summarized above, the performance gains of many advanced wireless positioning and NLoS detection techniques come at the cost of extra hardware, significant computation and training overhead, and large-scale real-time data collection, all of which remain difficult to sustain in practice. This challenge becomes even more pronounced toward 6G, which is expected to scale across a wide range of deployment sizes \cite{Zhou2024_6G}, including small indoor areas where such expenses are often impractical. Therefore, it is essential to develop a new technique that can operate effectively while minimally paying these costs.

This work addresses the above issue by proposing a technique termed \textit{CDA-guided NLoS detection} (CDA-ND), which fully utilizes the information embedded in a snapshot of range measurements. Specifically, CDA generates a large set of location candidates by applying multilateration to different subsets of the ranges, defined as \textit{preliminary estimated locations} (PELs). The spatial distribution of these PELs carries informative geometric signature that can be leveraged to determine whether a target gNB is in NLoS. As illustrated in Fig. \ref{fig:hypo}, the PELs can be partitioned into two groups: those constructed using the target gNB's range measurement (black circles) and those constructed without it (gray circles). When the target gNB is in NLoS, the first group exhibits a directional shift relative to the second group, typically along the direction opposite to the target gNB. This observation enables the design of CDA-ND within the existing communication architecture, without requiring additional hardware or incurring excessive training and labeling overhead.

The key contributions are summarized as follows.

\begin{figure}[!t]
\centering
\includegraphics[width=0.93\columnwidth]{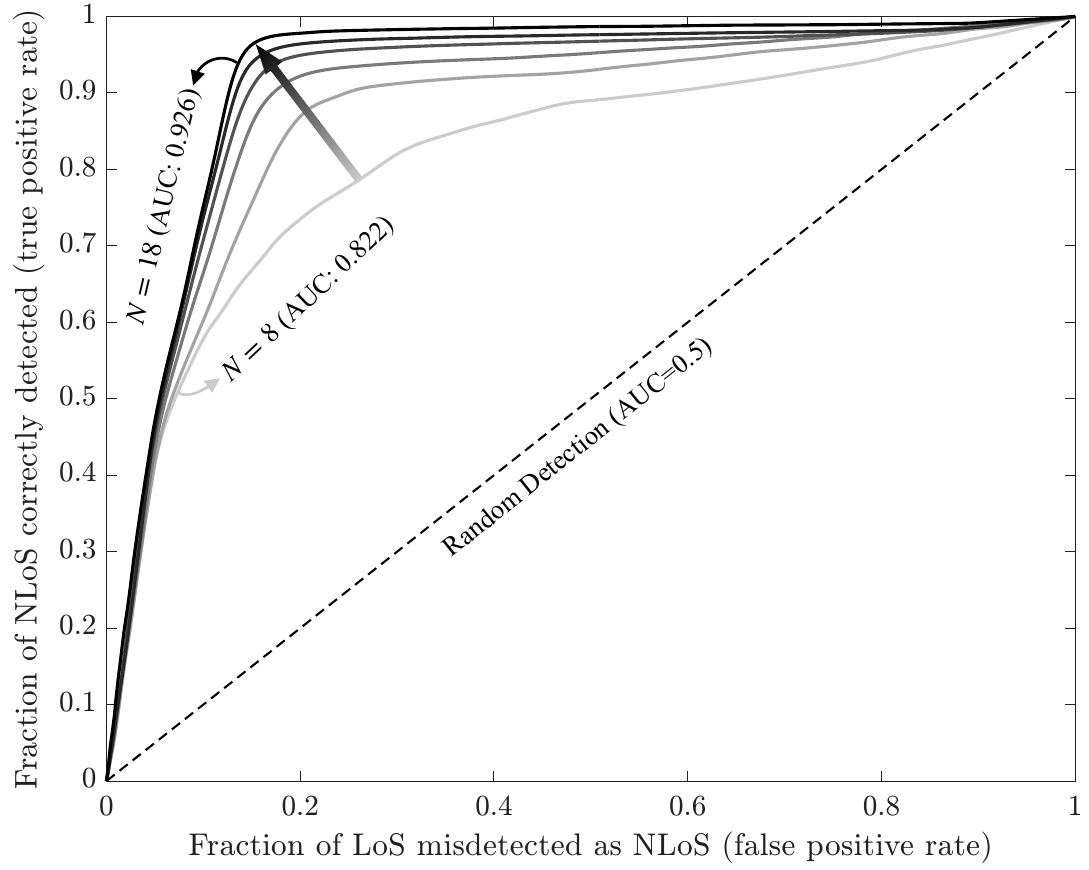}

\vspace{-4pt}
\caption{Receiver operating characteristic curves of the SD output as defined in Sec.~\ref{sec:soft_decision}, under different numbers of gNBs $N\in\{8,10,\dots,18\}$, in the InF-DH scenario under frequency range 1 (settings in Sec.~\ref{sec:sim_setup}). Here, ``positive'' corresponds to NLoS.  Darker curves indicate larger $N$, with higher AUCs demonstrating better separability between LoS and NLoS gNBs.}
\label{fig:AUC}

\vspace{-10pt}
\end{figure}

\begin{itemize}
\item \textbf{CDA-Induced Discriminative Statistics}: The effectiveness of CDA stems from the spatial statistics of the generated PELs, which implicitly encode site-dependent propagation characteristics. Empirically, as the number of participating gNBs increases, the resulting PEL statistics become increasingly stable and discriminative, leading to reliable NLoS detection, as evidenced by the \textit{area under the curve} (AUC) in Fig. \ref{fig:AUC}. Building on this insight, we present our key design components that unleash the full potential of CDA from a positioning perspective.   
\item \textbf{Hard Decision Approach}: To determine whether a target gNB is LoS or NLoS, we introduce the \textit{NLoS evidence vector} (NEV), which summarizes the displacement between the aforementioned PEL groups constructed with and without the target gNB (cf. Fig. \ref{fig:hypo}). Despite its simple form, the NEV provides strong NLoS cues through both its magnitude and direction. We then convert the NEV into an NLoS likelihood score and classify the gNB as NLoS whenever the score exceeds a predetermined threshold. Building on the HD approach, we develop a novel positioning algorithm that integrates CDA with the identified LoS/NLoS states.    
\item \textbf{Soft Decision Approach}: To quantify a target gNB's posterior NLoS probability, we develop an SD function that converts the NLoS score into a posterior probability by incorporating weak site-survey priors, namely, empirical score samples and the average NLoS probability. This probability mapping is then used to refine the NLoS score in the HD framework, whose initial construction may rely on incomplete and biased PEL information. This refinement improves robustness and accuracy, particularly in NLoS-dominant environments. Finally, we enhance the positioning stage by assigning reliability-aware weights to PELs according to the SD values of the gNBs involved in their construction.   
\item \textbf{Verification on 3GPP-Compliant Dataset}: To validate the effectiveness of the proposed CDA-ND, we conduct extensive simulations on two smart factory scenarios and an office scenario defined in the \textit{3rd generation partnership project} (3GPP) \cite{conti2024dataset}, namely \textit{indoor factory-sparse high} (InF-SH), \textit{indoor factory-dense high} (InF-DH), and \textit{indoor open office} (IOO), whose NLoS ratios are \SI{18}{\percent}, \SI{56}{\percent}, and \SI{36}{\percent}, respectively, under both \textit{frequency range~1} (FR1) and \textit{frequency range~2} (FR2) configurations. We show that CDA-ND achieves high NLoS detection reliability; for example, under FR1, it attains accuracies of \SI{94.76}{\%} and \SI{78.04}{\%} with HD, and \SI{96.76}{\%} and \SI{91.50}{\%} with SD, in InF-SH and InF-DH, respectively. Integrating CDA-ND into the positioning stage yields substantial accuracy gains, achieving \textit{mean absolute error} (MAE) reductions of about \SI{47}{\%} with HD and \SI{67}{\%} with SD in the challenging InF-DH scenario. 
\end{itemize}

\IEEEpubidadjcol

\section{System Model}
This section describes the positioning network and measurements. We then formulate two types of NLoS detection problems tackled throughout the work.

\vspace{-2pt}
\subsection{Scenario and Measurements}\label{Sec: Scenario}

Consider a wireless positioning network comprising $N$ gNBs and a single UE. The set of gNB indices is denoted by $\mathcal{N}=\left\{1,\dots,N\right\}$. Each gNB is stationary with known \textit{two-dimensional} (2D) coordinates, denoted by $\bm{z}_n \in \mathbb{R}^{2}$ for $n \in \mathcal{N}$. On the other hand, the UE lies at an unknown 2D position $\bm{p}\in\mathbb{R}^{2}$ within the concerned positioning site, which is to be estimated throughout the work.

We measure \textit{round-trip time} (RTT), denoted by $\tau_n$, to obtain a range measurement for each gNB. It can be obtained by exchanging downlink and uplink reference signals between the UE and gNB $n$ \cite{Jung2025GNN}. This RTT can be translated into an estimated range measurement as 
$d_n = \tau_n \frac{c }{2}$, where \mbox{$c \approx 3\times 10^8 ~ (\mathrm{m/s})$} denotes the speed of light. To capture NLoS effects, we define a binary indicator $\gamma_n\in\{0,1\}$, where $\gamma_n=1$ if gNB $n$ is in NLoS and $\gamma_n=0$ otherwise. The resultant range measurement $d_n$ can be expressed as  
\begin{align}\label{eq:distance_estimate}
d_n=\|\bm{z}_n-\bm{p}\|+\gamma_n b_n+w_n,
\end{align}
where $b_n$ represents the NLoS-induced positive bias and $w_n$ denotes the ranging noise.

The random quantities $\gamma_n$, $b_n$, and $w_n$ in \eqref{eq:distance_estimate} are governed by latent site-dependent parameters, denoted by $\Theta$. The set $\Theta$ captures factors that are difficult to model or obtain in practice, such as obstacle geometry and material properties \cite{kang2026vision}. Throughout this work, $\Theta$ is assumed unknown. 

\begin{figure*}[t!]
    \centering

    \subfloat[]{
        \includegraphics[height=0.27\textwidth]{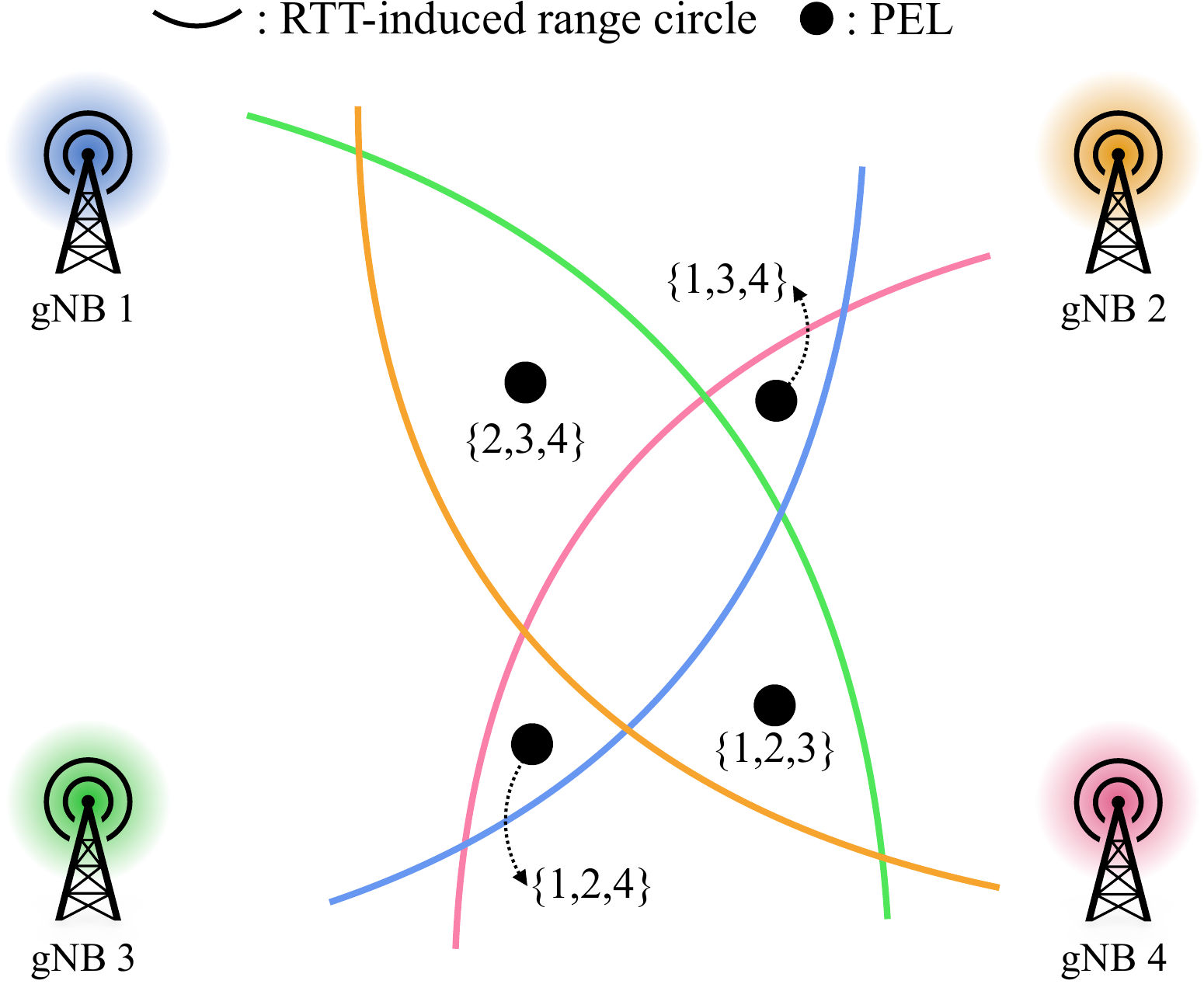}\label{fig:CDA}
    }
    \hfill 
    \subfloat[]{
        \includegraphics[height=0.27\textwidth]{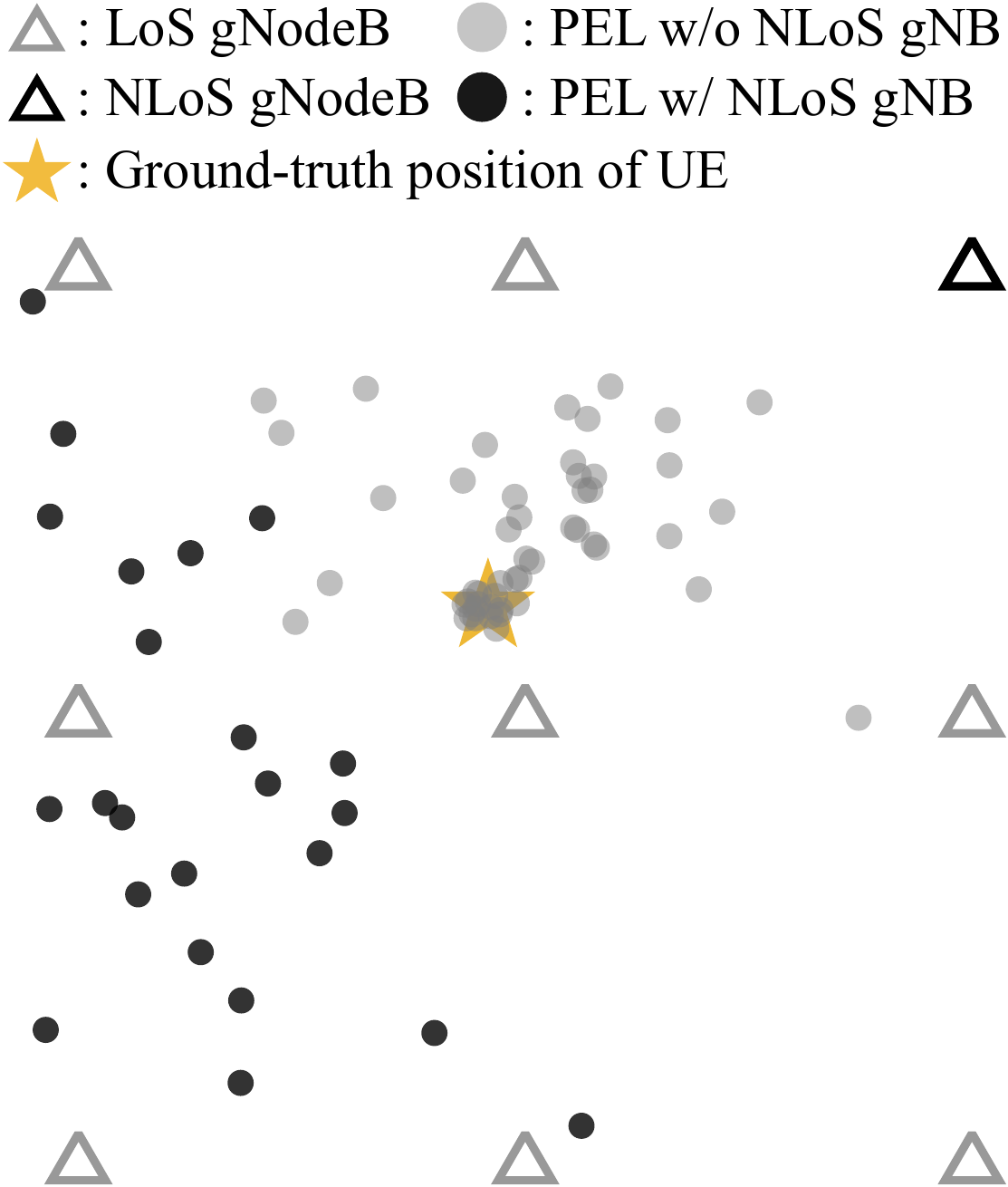}\label{fig:PEL_Dist}
    }
    \hfill 
    \subfloat[]{
        \includegraphics[height=0.27\textwidth]{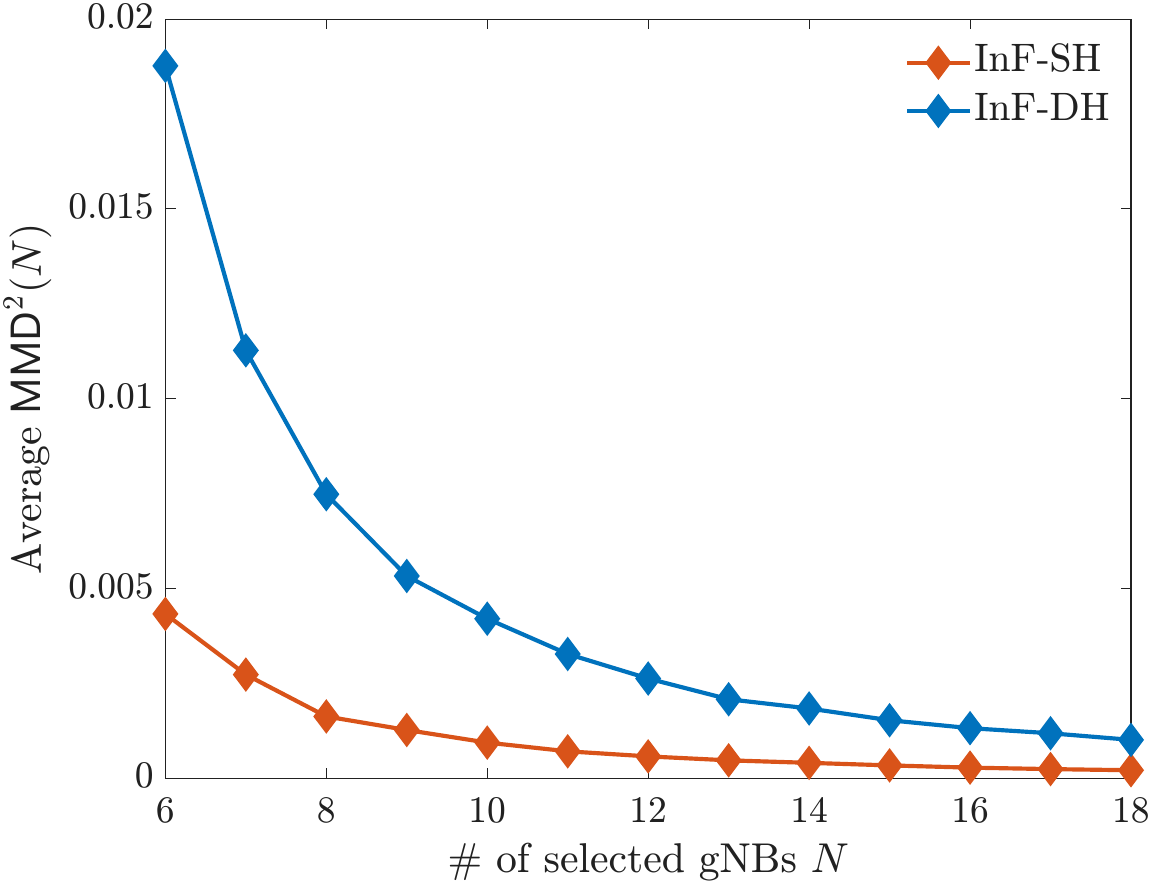}\label{fig:MMD}
    }
    
    \caption{Illustration of CDA and its implications for NLoS detection. (a) CDA generates multiple PELs by applying multilateration to different subsets of gNB range measurements. (b) When a target gNB is NLoS, the PELs formed with that gNB and those formed without it exhibit a systematic spatial shift, yielding a discriminative geometric NLoS signature in the PEL distribution. (c) The PEL distribution empirically converges as the number of participating gNBs $N$ increases, quantified by the average squared MMD between the distribution induced by $N$ and $(N-1)$ gNBs.}
    \vspace{-6pt}

    \label{fig:CDAExample}
\end{figure*}

\subsection{Problem Description}

Given the range measurements $\{d_n\}$ in \eqref{eq:distance_estimate}, a baseline for estimating the UE location $\bm{p}$ is multilateration, which infers $\bm{p}$ from the intersection of RTT-induced range circles. However, the latent NLoS states $\{\gamma_n\}$ introduce unknown positive biases in a subset of range measurements, enlarging the corresponding circles and perturbing their geometric intersections. As a result, multilateration can suffer from large positioning errors. To mitigate the effect, we first aim to identify gNBs that are likely to be in NLoS, and then incorporate this information into the subsequent positioning algorithm. We formulate the following two NLoS detection problems depending on the desired output type.

\subsubsection{Hard Decision} We first determine the NLoS state of each gNB as a binary variable, referred to as a \textit{hard decision} (HD). Ideally, the maximum likelihood detection of NLoS gNBs can be formulated as 
\begin{align} \label{ML_formulation}
    {\{\hat{\gamma}_n\}} = \mathop{\mathrm{arg\,max}}_{\{\gamma_n\} \in \{0, 1\}^N} \mathsf{Pr}\!\left(\{d_n\} \mid \{\gamma_n\}, \Theta \right).
\end{align}
The likelihood in \eqref{ML_formulation} is intractable due to the unknown environmental parameter $\Theta$. To address this, we develop a tractable approximation based only on the snapshot range measurements $\{d_n\}$ and gNB locations $\{\bm{z}_n\}$, which constitutes the main technical contribution in Sec.~\ref{sec:nlos-detection}. 

\subsubsection{Soft Decision}\label{subsubsection: SD definition}
We next assign each gNB an NLoS confidence score (i.e., the posterior NLoS probability), referred to as a \textit{soft decision} (SD), defined by
\begin{align} \label{SI_formulation}
\psi_n=\mathsf{Pr}(\gamma_n=1\mid \{d_n\}, \Theta), \quad \forall n\in\mathcal{N}.
\end{align}
Compared with HD, SD provides richer information that can be exploited in downstream tasks.  
In principle, $\psi_n$ can be obtained by marginalizing the joint posterior  $\mathsf{Pr}\left(\{\gamma_n\} \mid \{d_n\}, \Theta \right)$, which remains intractable for the same reason as above. Building on the proposed HD method, we develop a proxy for $\psi_n$ that leverages site-survey statistics introduced in the sequel, as detailed in Sec. \ref{sec:soft_decision}.

In the following sections, we present the proposed techniques for both HD and SD, and show how each output can be integrated with positioning algorithm tailored to the corresponding output types.

\section{\texorpdfstring{CDA-Guided NLoS Detection:\\ Hard Decision Perspective}
{CDA-Guided NLoS Detection: Hard Decision Perspective}}\label{sec:nlos-detection}

This section introduces CDA-ND, a key enabler for handling the latent parameters for the HD problem of \eqref{ML_formulation}.

\subsection{Preliminary, Hypothesis, and Overview}\label{Sec:Hypothesis}
We begin by briefly reviewing CDA \cite{yu2025combinatorial}, which generates multiple PELs by exploiting different subsets of gNBs, as illustrated in Fig.~\ref{fig:CDA}. Specifically, selecting $M$ gNBs from the total of $N$ yields $\binom{N}{M}$ subsets, among which the subsets of nearly collinear gNBs are excluded since their ranges do not determine a reliable PEL. The remaining $L\le\binom{N}{M}$ subsets are denoted by $\left\{\mathcal{M}^{(\ell)}\right\}_{\ell=1}^L$. For each subset with $M \geq 3$ (we set $M=3$ throughout this paper), the corresponding range measurements are used to compute a PEL as
\begin{align}
    \bm{x}^{(\ell)} = g\left(\left\{d_n\right\}_{n\in \mathcal{M}^{(\ell)}}\right) \in \mathbb{R}^{2},\quad \ell\in\{1,\dots,L\},
\end{align}
where $g(\cdot)$ denotes a conventional multilateration algorithm (e.g., \cite{Mensing2006LM}). The complete set of all PELs, visualized as the distribution of circles in Fig.~\ref{fig:CDA}, is defined as
\begin{align}
\label{eq:PLE_ensemble} 
    \mathcal{X} = \left\{\bm{x}^{(\ell)}\right\}_{\ell=1 }^L,
\end{align}
which encapsulates diverse aspects of the network geometry, such as potential UE locations, spatial spread, and anisotropy that may hint at blockage directions.
For example, as illustrated in Fig. \ref{fig:PEL_Dist}, the PELs exhibit distinct clustering patterns depending on whether the underlying gNB subset is LoS-only or includes NLoS gNBs. PELs computed from all-LoS subsets (gray circles) form a tightly concentrated, nearly isotropic cluster centered near the ground truth. In contrast, subsets containing NLoS gNBs (black circles) generate clusters that are more anisotropic and spatially dispersed, with their centroid biased away from the NLoS gNB (often in the opposite direction). In other words, the spatial distribution of PELs, denoted by $U$, can be interpreted as a proxy for the latent environmental parameter $\Theta$. This allows us to revise the likelihood in \eqref{ML_formulation} as 
\begin{align}\label{Revised_ML_formulation}\tag{P1}
{\{\hat{\gamma}_n\}}=\mathop{\mathrm{arg\,max}}_{\{\gamma_n\} \in \{0, 1\}^N}  \ \mathsf{Pr}\!\left(\{d_n\},\mathcal{X} \mid \{\gamma_n\}, U\right).
\end{align}
Assuming a sufficiently large $N$, we hypothesize that the maximum likelihood estimator based on the revised likelihood in \ref{Revised_ML_formulation} approaches that of the original formulation in \eqref{ML_formulation}. This suggests that  $(\{d_n\}, \mathcal{X})$ can be regarded as an increasingly stable and discriminative set of statistics for NLoS detection.

We empirically investigate the stabilization trend of the PEL spatial distribution as the number of gNBs $N$ increases, considering two indoor factory scenarios with sparse obstacles (InF-SH) or dense obstacles  (InF-DH). Their detailed simulation settings are provided in Sec.~\ref{sec:simulation}. Let $\mathcal{X}(N) = \{\bm{x}_N^{(\ell)}\}_{\ell=1}^{L(N)}$ denote the set of PELs generated from $N$ gNBs, where $L(N)\le\binom{N}{3}$, and view these PELs as samples drawn from the underlying distribution $U(N)$. To show the stabilization of $U(N)$ as $N$ increases, we quantify the distributional discrepancy between $U(N)$ and $U(N-1)$ using the squared \textit{maximum mean discrepancy} (MMD) \cite{gretton2006kernel}, which evaluates the distance between the mean embeddings of two distributions in a reproducing kernel Hilbert space. Its empirical squared form is given by
\begin{align} \label{eq:MMD}
\mathsf{MMD}&^2{(N)} = \frac{1}{L(N)^2} \sum_{i,j=1}^{L(N)} \Upsilon\!\left(\bm{x}_N^{(i)}, \bm{x}_N^{(j)}\right) \nonumber \\
+& \frac{1}{L({N-1})^2} \sum_{i,j=1}^{L({N-1})} \Upsilon\!\left(\bm{x}_{N-1}^{(i)}, \bm{x}_{N-1}^{(j)}\right) \nonumber \\
-& \frac{2}{L(N) L({N-1})} \sum_{i=1}^{L(N)} \sum_{j=1}^{L({N-1})} \Upsilon\!\left(\bm{x}_N^{(i)}, \bm{x}_{N-1}^{(j)}\right),
\end{align}
where $\Upsilon(\bm{a}, \bm{b}) = \exp(-\varsigma \|\bm{a} - \bm{b}\|^2)$ represents the Gaussian radial basis function kernel. To ensure a fair and scale-invariant comparison across $N$, the kernel bandwidth parameter $\varsigma$ is set using the median heuristic computed from pairwise squared Euclidean distances among the PELs in $\mathcal{X}$.

Fig.~\ref{fig:MMD} shows the statistics of $\mathsf{MMD}^2(N)$ in \eqref{eq:MMD} over $100$ random sequences of gNB selections. In each sequence, gNBs are gradually added from $5$ to $18$. The UE's location is independently sampled $1,000$ times. 
Two interesting observations can be made. First, the discrepancy between $U(N-1)$ and $U(N)$ decreases as $N$ increases, indicating that the PEL distribution becomes increasingly stable. Second, stabilization is slower in InF-DH than in InF-SH. The denser-obstacle environment exhibits a richer set of latent propagation factors (captured abstractly by $\Theta$), and thus requires more geometric combinations through more gNBs to sufficiently explain these factors through the induced PEL distribution. Together with the separability trend in Fig.~\ref{fig:AUC}, these two observations provide supporting evidence that CDA, given a sufficient number of gNBs, effectively extracts spatial diversity from multiple gNB combinations, enabling us to capture the environment-dependent geometric features using only snapshot measurements. Appendix~\ref{app:factorization} further evaluates the loss from replacing $\Theta$ by $U$ in \ref{Revised_ML_formulation}, using a synthetic setting where $\Theta$ is known. This loss decreases with $N$ and is larger in harsher NLoS environments, similar to the slower stabilization of InF-DH in Fig.~\ref{fig:MMD}.

In the following, our goal is to derive a tractable form of the maximum likelihood estimator by exploiting the CDA-driven distribution~$U$. First, a key feature extracted from $U$ allows us to factorize the likelihood approximately, which reduces the joint maximum likelihood problem \ref{Revised_ML_formulation} to gNB-wise estimators, as detailed in Sec. \ref{Sec:Factorization}. Next, we construct a score function as a surrogate for the likelihood ratio, which enables the design of a simple threshold-based test for HD, as described in Sec.~\ref{Sec:Threshold-Based Test}. Finally, we propose a new positioning algorithm that integrates HD into the conventional CDA-based positioning framework, as detailed in Sec. \ref{sec:aided_loc}.

\subsection{Factorization via NLoS Evidence Vector} \label{Sec:Factorization}

Problem \ref{Revised_ML_formulation} involves optimizing multiple binary variables $\{\gamma_n\}$ based on their joint probability mass function. However, deriving this distribution is challenging due to the complex correlations among the variables. To address this challenge, we decompose the joint optimization problem into $N$ single-variable optimization problems by introducing the following CDA-driven definition.

\begin{definition}[\textit{NLoS Evidence Vector} (NEV)]\label{Def:NEV} For a given gNB $n$, we partition the set of all PELs $\mathcal{X}$ in \eqref{eq:PLE_ensemble} into two groups: those constructed using the range measurement $d_n$ and the remainder, i.e.,
\begin{align}
    \mathcal{X}_n = \left\{\bm{x}^{(\ell)} \in \mathcal{X} \mid n \in \mathcal{M}^{(\ell)}\right\}, \quad  
    \mathcal{X}_{-n} = \mathcal{X}\setminus \mathcal{X}_n.
\end{align}
We then introduce the NEV as the displacement between the representative points of the two groups:
\begin{align}\label{eq:NEV}
    \bm{r}_n = {\bm{m}}_n - {\bm{m}}_{-n},
\end{align}
where ${\bm{m}}_n = \mathsf{med}(\mathcal{X}_n)$ and ${\bm{m}}_{-n} = \mathsf{med}(\mathcal{X}_{-n})$ denote the coordinate-wise medians of $\mathcal{X}_n$ and $\mathcal{X}_{-n}$, respectively.
\end{definition}

\begin{figure*}[t!]
    \centering 
    \subfloat[]{
        \includegraphics[width=0.31\textwidth]{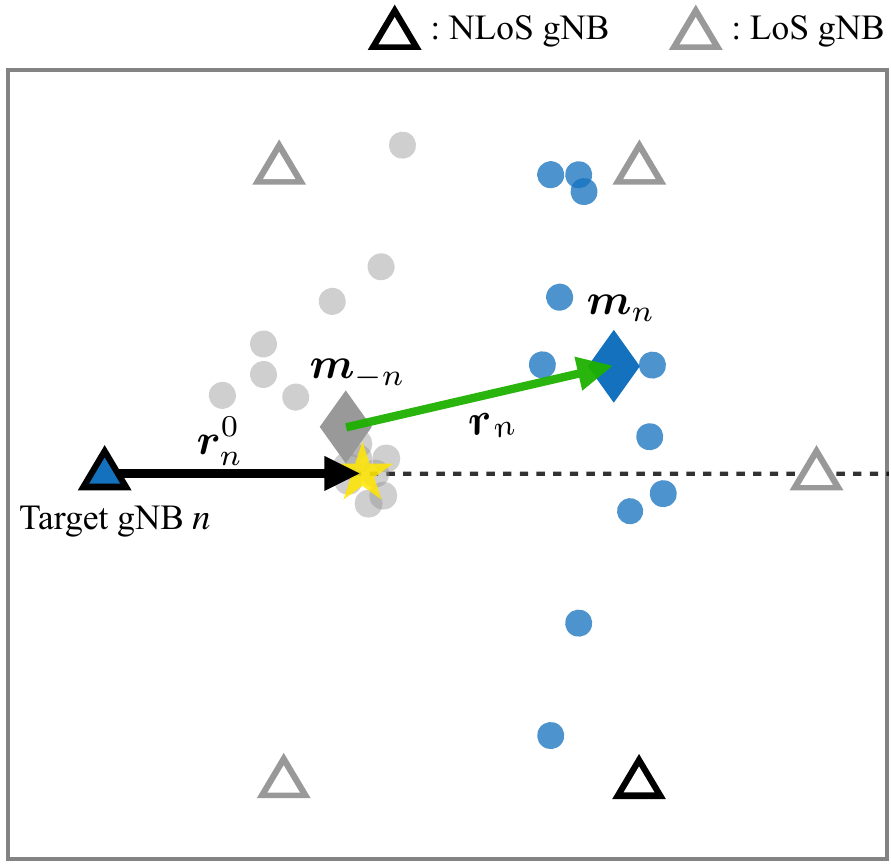}\label{fig:NLoS}
    }
    \subfloat[]{
        \includegraphics[width=0.31\textwidth]{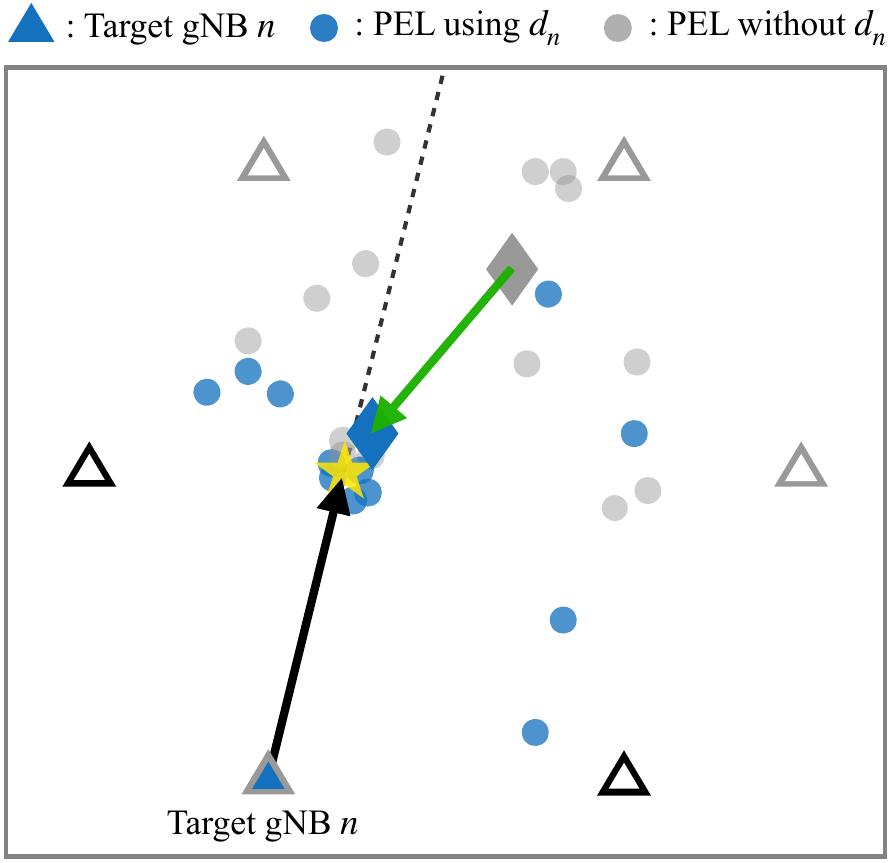}\label{fig:LoS1}
    }
    \subfloat[]{
        \includegraphics[width=0.31\textwidth]{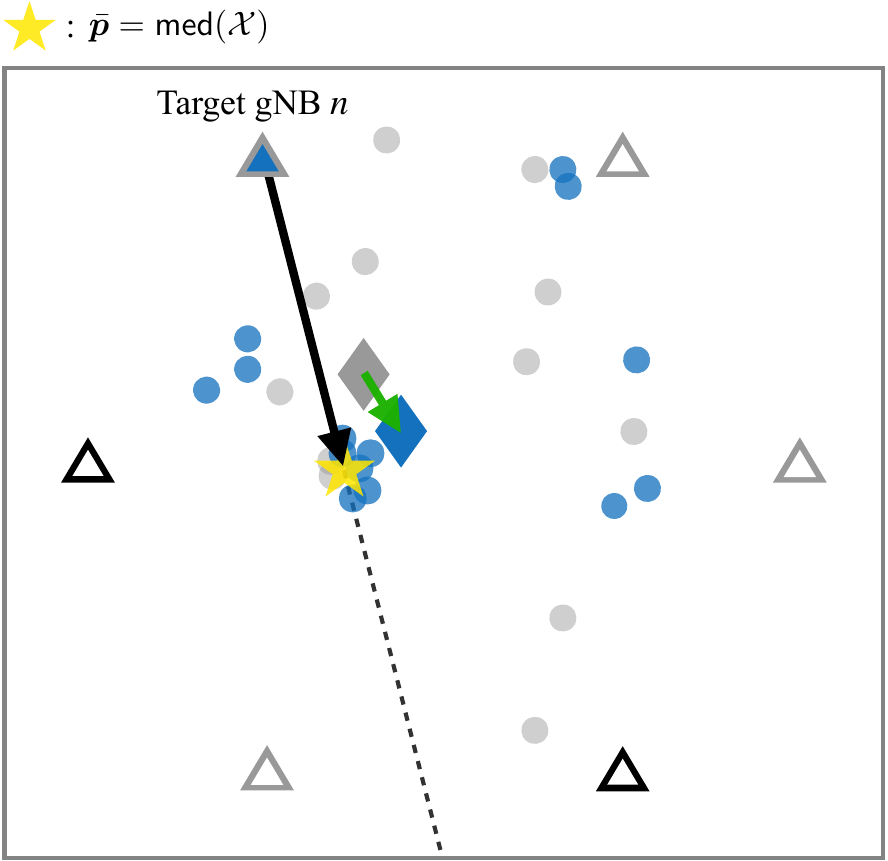}\label{fig:LoS2}
    }

    \caption{Examples of the NEV from \eqref{eq:NEV} in a single snapshot across three target gNB cases: (a) an NLoS case with both sufficient magnitude and directional alignment, (b) a LoS case with sufficient magnitude but directional misalignment, and (c) a LoS case with directional alignment but insufficient magnitude.}
    \label{fig:NEV}

    \vspace{-8pt}
\end{figure*}

Fig. \ref{fig:NLoS} illustrates an example of the NEV $\bm{r}_n$ of \eqref{eq:NEV} for a target gNB $n$ under NLoS. As shown, the NLoS bias shifts the PELs in $\mathcal{X}_n$ in the direction opposite to gNB $n$. Hence, the NEV $\bm{r}_n$ serves as a discriminative feature for identifying whether gNB $n$ is NLoS. Specifically, a gNB is more likely to be NLoS when $\bm{r}_n$ exhibits the following two key characteristics simultaneously: (i) a sufficiently large magnitude, and (ii) a directional alignment with a reference vector defined as 
\begin{align}
\label{eq:idealReference}
   {\bm{r}}_n^{0}= {\bar{\bm{p}}-\bm{z}_n}, 
\end{align}
where $\bar{\bm{p}}=\mathsf{med}(\mathcal{X})$ is a pseudo UE location obtainable from the entire PEL set, and gNB $n$'s location $\bm{z}_n$ is specified in Sec.~\ref{Sec: Scenario}. It is important to note that both features should be observed concurrently. Indeed, neither a large magnitude without alignment nor a well-aligned but weak displacement suffices to indicate NLoS. For better intuition, Fig.~\ref{fig:LoS1} and Fig.~\ref{fig:LoS2} illustrate these two counterexamples: a large but misaligned NEV, and an aligned yet low-magnitude NEV, respectively.

We hypothesize that the pair $\left( \bm{r}_n, \bm{r}_n^0 \right)$ provides sufficient information to infer whether gNB $n$ is NLoS. This allows the joint likelihood in \ref{Revised_ML_formulation} to be approximately factorized as a product of gNB-wise conditional probabilities:
\begin{align} \label{eq:Factorization}
\mathsf{Pr}\left( \{d_n\}, \mathcal{X} \mid \{\gamma_n\}, U \right) \approx \prod\limits_{n\in\mathcal{N}}\mathsf{Pr}\left(d_n, \bm{r}_n\mid \gamma_n, \bm{r}_n^{0}\right).
\end{align} 
Through this factorization, Problem \ref{Revised_ML_formulation} reduces to $N$ single-variable tasks, namely:
\begin{align}\label{Decomposed_Problem}\tag{P2}
\hat\gamma_n=\mathop{\mathrm{arg\,max}}\limits_{{\gamma}\in\{0,1\}}\mathsf{Pr}\left(d_n, \bm{r}_n \mid {\gamma}, \bm{r}_n^0\right),
\end{align}
thereby greatly simplifying the decision-making process.

\begin{remark}[Validity of gNB-Wise Factorization] \label{rem:pseudo_ref}
The factorization from \ref{Revised_ML_formulation} to \ref{Decomposed_Problem} in \eqref{eq:Factorization} requires the gNBs to be decoupled. We assume that the range measurements $\{d_n\}$ in \eqref{eq:distance_estimate} are independent across gNBs given the UE location $\bm{p}$ \cite{Guvenc2009Survey, Morselli2023SI}. Since $\bm{p}$ is unknown in practice, the factorization also assumes that the pseudo location $\bar{\bm{p}}=\mathsf{med}(\mathcal{X})$ in ${\bm{r}}_n^0$ is close to $\bm{p}$, and that $\bm{r}_n$ mainly captures the NLoS bias of gNB~$n$ by separating the PELs constructed with and without this gNB. The accuracy of $\bm{r}_n$ and ${\bm{r}}_n^0$ depends on how heavily the PEL set is biased by NLoS-contaminated gNBs. Both are more accurate in LoS-dominant environments and for larger $N$, where the factorization loss is small, and the loss is larger in the opposite case. Appendix~\ref{app:factorization} quantifies this trend and shows that the factorization loss is mitigated by refining $\bm{r}_n$ and ${\bm{r}}_n^0$ in the SD framework of Sec.~\ref{sec:iterative_update}.
\end{remark}

\subsection{HD: Threshold-Based Test}\label{Sec:Threshold-Based Test}

Assuming that the pseudo location $\bar{\bm{p}}$ in \eqref{eq:idealReference} is an effective representative of the ground-truth location $\bm{p}$ (i.e., $\bar{\bm{p}}=\bm{p}$), the residual between the estimated distance $d_n$ and its geometric estimate $\|\bm{r}_n^0\|$ can be expressed as
\begin{align}
d_n-\left\|\bm{r}_n^0\right\|=\begin{cases}
b_n+w_n &\text{if }\gamma_n=1,\\
w_n &\text{if }\gamma_n=0,
\end{cases}
\end{align}
where the NLoS bias $b_n$ and the measurement noise $w_n$ are specified in \eqref{eq:distance_estimate}.

Accordingly, Problem \ref{Decomposed_Problem} can be reformulated as the following threshold test:
\begin{align}
\frac{f_{b_n+w_n}\left(d_n-\|\bm{r}_n^0\|\right)}{f_{w_n}\left(d_n-\left\|\bm{r}_n^0\right\|\right)}\mathop{\gtrless}_{\hat\gamma_n=0}^{\hat\gamma_n=1}1.
\end{align}
However, the exact forms of the \textit{probability density functions} (PDFs) $f_{b_n+w_n}$ and $f_{w_n}$ are generally unknown, rendering the computation of the likelihood ratio intractable. 

\begin{figure}[!t]
    \centering
    \includegraphics[width=0.93\columnwidth]{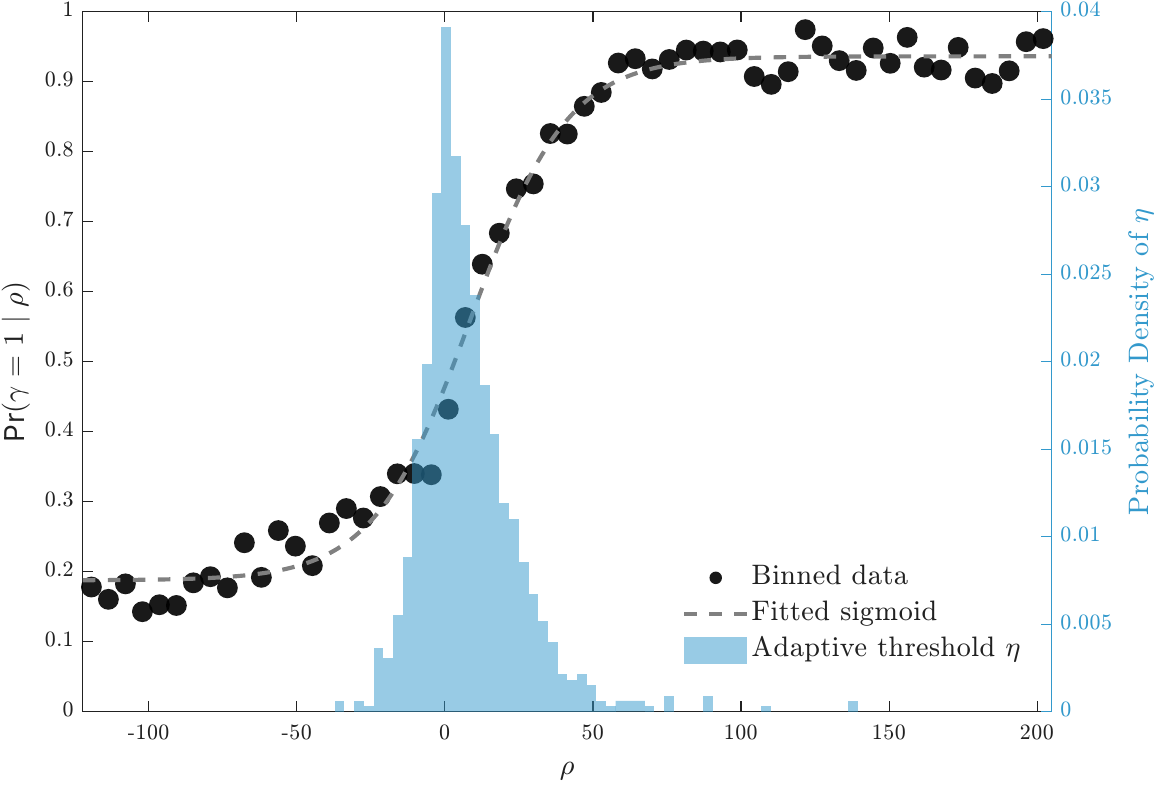}
    \vspace{-4pt}
    \caption{The empirical NLoS probability with respect to score $\rho$ in InF-DH under FR1 (details in Sec.~\ref{sec:sim_setup}), shown to fit a sigmoid curve. The probability density of the adaptive threshold $\eta$ is represented by blue histogram.\vspace{-8pt}}
    \label{fig:sigmoid}
\end{figure}

To address this limitation, we design a surrogate score function that approximates the log-likelihood ratio by leveraging the NEV $\bm{r}_n$ defined in \eqref{eq:NEV}: 
\begin{align}\label{eq:Score}
\rho_n=\left(\frac{\bm{r}_n^\top \bm{r}_n^0}{\|\bm{r}_n^0\|}\right)\sqrt{d_n}.
\end{align}
This score comprises two interpretable components. The first term, $\frac{\bm{r}_n^\top \bm{r}_n^0}{\|\bm{r}_n^0\|}$, represents the scalar projection of the NEV $\bm{r}_n$ onto the reference vector $\bm{r}_n^0$, quantifying their directional alignment as discussed in Sec. \ref{Sec:Factorization}. The second term, $\sqrt{d_n}$, reflects the distance-dependent increase in NLoS probability, which is validated by both well-established channel modeling principles (e.g., 3GPP TR 38.901 \cite{3GPPTR38.901V19.2.0}) and real-world Wi-Fi measurements (e.g., our prior work in \cite{yu2025combinatorial}). By combining these two factors, the score function $\rho_n$ effectively captures the NLoS likelihood: A larger $\rho_n$ indicates a higher probability that gNB $n$ is in NLoS. The monotonic relationship between $\rho_n$ and the NLoS probability is empirically verified, as shown in Fig.~\ref{fig:sigmoid}.

Using the score $\rho_n$, we can determine whether gNB $n$ is NLoS according to the following threshold-based test:
\begin{align} \label{eq:CDA-ND_rule}
    \hat{\gamma}_n = 
    \begin{cases}
        1 & \text{if } \rho_n \ge \eta. \\
        0 & \text{o.w.}
    \end{cases}
    \quad\forall n\in \mathcal{N}
\end{align}
Here, $\eta$ is a decision threshold that balances the trade-off between two performance metrics: 
\begin{itemize}
\item \textbf{Recall}: The ratio of correctly identified NLoS gNBs to the total number of actual NLoS gNBs, reflecting the ability to detect all existing NLoS cases.
\item \textbf{Precision}: The ratio of correctly identified NLoS gNBs to all gNBs classified as NLoS, representing the reliability of NLoS detection results.
\end{itemize}
Increasing $\eta$ decreases recall while improving precision. To achieve a robust balance between the two, we adaptively determine $\eta$ from the distribution of all score values $\{\rho_n\}_{n\in\mathcal{N}}$~as
\begin{align} \label{eq:adaptive_threshold}
    \eta = \mathsf{med}(\{\rho_n\}) + \lambda  \mathsf{med}\left(\{\mathsf{dev}(\rho_n)\}\right),
\end{align}
where $\mathsf{dev}(\rho_n)=|\rho_n-\mathsf{med}(\{\rho_n\})|$ and $\lambda>0$ is a tunable scaling factor. $\lambda$ is set to $\lambda=1.4$ in InF-SH under FR1, $\lambda=1.1$ in InF-SH under FR2, $\lambda=0.5$ in InF-DH under FR1/FR2, and $\lambda=0.6$ in IOO under FR1/FR2, and these values are used throughout the numerical evaluations in Sec.~\ref{sec:simulation}. The first term, $\mathsf{med}(\{\rho_n\})$, provides a baseline threshold, corresponding to approximately half of the gNBs being detected as NLoS and defining the nominal recall level. However, this baseline alone may also include several gNBs whose $\rho_n$ values lie close to the median, which are generally more likely to be LoS rather than NLoS. The second term, $\mathsf{med}\left(\{\mathsf{dev}(\rho_n)\}\right)$, quantifies the overall score dispersion and adjusts the threshold to reduce the false detection of these borderline cases. 
A larger dispersion indicates that the scores $\{\rho_n\}$ are more spread out, meaning that potential NLoS gNBs have clearly higher $\rho_n$ values compared to LoS ones. In such cases, raising the threshold $\eta$ (via a larger dispersion term) helps focus the detection on the more confident, high-score NLoS gNBs, improving precision without a noticeable loss of recall. Conversely, when the dispersion is small, implying weak separability between LoS and NLoS gNBs, a smaller adjustment keeps $\eta$ near the baseline to preserve recall.

\begin{remark} [Sigmoid-like NLoS Probability] \label{rem:monotonic}
Fig.~\ref{fig:sigmoid} illustrates that the empirical curve of the NLoS probability with respect to $\rho_n$ closely follows a sigmoid function (dotted line), exhibiting a sharp transition between its lower and upper saturation levels. Here, the transitional range refers to the interval of $\rho_n$ values where the NLoS probability changes most rapidly, bridging LoS-dominant and NLoS-dominant regions. Interestingly, most threshold values $\eta$, represented by the blue histogram, are concentrated near the boundary between the lower plateau of the sigmoid and this transitional range. 
This indicates that our adaptive rule favors high recall, while maintaining reasonable precision by avoiding the region dominated by clear LoS scores. As a result, the threshold design in \eqref{eq:adaptive_threshold} is well aligned with the empirical score-to-probability relationship.   
\end{remark}

\subsection{Positioning with Hard Decision}
\label{sec:aided_loc}

Let $\mathcal{N}_{\mathrm{HD}}$ denote the set of gNBs that remain after excluding those identified as NLoS using HD, i.e., 
\begin{align} \label{eq:N_ND}
    \mathcal{N}_{\mathrm{HD}}=\{n\in\mathcal{N}\mid\hat\gamma_n=0\}.
\end{align}
Although CDA-ND efficiently filters out most NLoS gNBs, the remaining set $\mathcal{N}_{\mathrm{HD}}$ may still contain undetected NLoS gNBs or LoS gNBs severely corrupted by measurement noise, both of which hinder accurate positioning. 

To overcome this limitation, we propose a new CDA-based algorithm that exploits the gNBs in $\mathcal{N}_{\mathrm{HD}}$ while incorporating two complementary mechanisms, namely, \textit{residual-error} (RE) \& \textit{RTT-sum} (RS) filtering, originally introduced in our prior work \cite{yu2025combinatorial}. These mechanisms jointly enhance the robustness of the location estimate by mitigating the impact of noisy or unreliable measurements within $\mathcal{N}_{\mathrm{HD}}$. The detailed, step-by-step procedure of the proposed algorithm is described below.

\subsubsection{PEL Selection}
In the first stage, among all PELs $\bm{x}^{(\ell)}\in\mathcal{X}$ defined in \eqref{eq:PLE_ensemble}, we extract those that are constructed solely from distance measurements of gNBs in $\mathcal{N}_{\mathrm{HD}}$, given as  
\begin{align}\label{eq:PEL_HD}
    \mathcal{X}_{\mathrm{HD}} = \left\{\bm{x}^{(\ell)} \in \mathcal{X} \mid \mathcal{M}^{(\ell)} \subseteq \mathcal{N}_{\mathrm{HD}}\right\},
\end{align}
where $\mathcal{M}^{(\ell)}$ denotes the subset of gNBs used to generate PEL $\bm{x}^{(\ell)}$, with cardinality $\left|\mathcal{M}^{(\ell)}\right|=M$ as specified in~Sec.~\ref{Sec:Hypothesis}.

\subsubsection{Residual-Error Filtering} In the second stage, we filter the PELs in $\mathcal{X}_{\mathrm{HD}}$ based on their RE value, defined as
\begin{align}
    u^{(\ell)} &= \sum_{n \in \mathcal{M}^{(\ell)}} \left| d_n -\left\|\bm{x}^{(\ell)} - \bm{z}_n\right\| \right|, \label{eq:RE_metric} 
\end{align}
which quantifies the aggregate discrepancies between the measured distances and the corresponding geometric distances from PEL $\bm{x}^{(\ell)}$ to its associated gNBs. A higher RE value indicates that the PEL is less consistent with the measurements and thus less reliable.
To exclude such unreliable PELs, we sort all RE values $\left\{u^{(\ell)}\right\}$ in ascending order and denote the $\ell_{\mathrm{RE}}$-th smallest value as $\bar{u}^{(\ell_{\mathrm{RE}})}$.
We then retain only the PELs whose REs do not exceed this threshold, i.e.,
\begin{align}
    \mathcal{X}_{\mathrm{RE}} = \left\{\bm{x}^{(\ell)} \in \mathcal{X}_{\mathrm{HD}} \mid u^{(\ell)} \le \bar{u}^{(\ell_{\mathrm{RE}})} \right\}.
\end{align}
\subsubsection{RTT-Sum Filtering} In the third stage, we further filter the PELs in $\mathcal{X}_{\textrm{RE}}$ based on their RS values, defined as
\begin{align}
    v^{(\ell)} &= \sum_{n \in \mathcal{M}^{(\ell)}} d_n, \label{eq:RS_metric}
\end{align}
which represents the total sum of distance measurements used to construct PEL $\bm{x}^{(\ell)}$. 
The underlying hypothesis is that a smaller RS value is more likely to be associated with the LoS condition, since NLoS paths typically result in longer RTTs. To retain the PELs most likely to correspond to LoS propagation, we sort all RS values $\left\{v^{(\ell)}\right\}$ in ascending order and denote the $\ell_\mathrm{RS}$-th smallest value as $\bar{v}^{(\ell_\mathrm{RS})}$. The remaining PEL set after RS filtering is given by 
\begin{align}\label{eq: RS set}
\mathcal{X}_{\mathrm{RS}} =  \left\{ \bm{x}^{(\ell)} \in \mathcal{X}_{\mathrm{RE}} \mid v^{(\ell)} \le \bar{v}^{\left(\ell_{\textrm{RS}}\right)}  \right\}.
\end{align}

\subsubsection{Final Location Estimate} In the final stage, the UE’s location is estimated as the median of all PELs that remain after RS filtering, i.e.,
\begin{align}
\label{eq:proposed}
    \hat{\bm{p}}_{\mathrm{HD}} = \mathsf{med}\left(\mathcal{X}_{\mathrm{RS}}\right).
\end{align}
This strategy, which integrates HD with the subsequent RE\&RS filtering, is hereafter termed CDA-ND-RERS (HD).

\begin{remark}[HD vs. RE\&RS Filtering] Although both HD and RE\&RS filtering aim to mitigate the impact of NLoS biases, their specific roles are complementary rather than redundant. HD operates as a gNB-level filtering mechanism, effectively removing dominant NLoS effects, especially in environments where obstacles are spatially clustered (e.g., smart factory). However, because HD excludes all PELs associated with the detected gNB simultaneously, some of the discarded PELs may carry useful geometric information that could improve positioning accuracy. On the other hand, RE\&RS filtering functions as a PEL-level mechanism, efficiently addressing moderate NLoS effects arising from small-scale or scattered blockages. Its effectiveness, however, depends on the initial PEL distribution and may degrade when all PELs are biased toward a particular direction. Consequently, the combination of HD and RE\&RS filtering offers enhanced robustness across diverse propagation environments, as validated in Sec.~\ref{sec:simulation}.      
\label{synergy}

\end{remark}

\section{\texorpdfstring{CDA-Guided NLoS Detection:\\ Soft Decision Perspective}
{CDA-Guided NLoS Detection: Soft Decision Perspective}} \label{sec:soft_decision}

This section extends the preceding HD into an SD approach by estimating the posterior NLoS probability $\psi_n$ in~\eqref{SI_formulation}.

\subsection{Towards Tractable Soft Decision: Key Idea and Overview} \label{sec:tractable_approx}

As described in Sec. \ref{subsubsection: SD definition}, $\psi_n$ is obtained by marginalizing all variables except $\gamma_n$ from the joint posterior distribution $\mathsf{Pr}(\{\gamma_n\} \mid \{d_n\}, \Theta)$, which is in general intractable due to unknown latent parameters in $\Theta$.
On the other hand, leveraging the rationale in Sec.~\ref{sec:nlos-detection}, we derive a tractable approximation of $\psi_n$. 
The key idea is that the scalar score $\rho_n$ in \eqref{eq:Score} serves as a sufficient discriminative feature for the NLoS state of gNB~$n$. Using this score-based representation, we approximate the joint posterior distribution by a gNB-wise factorization: 
\begin{align}
\mathsf{Pr}(\{\gamma_n\} \mid \{d_n\}, \Theta)\approx\prod_{n\in\mathcal{N}} \mathsf{Pr}(\gamma_n \mid \rho_n),
\end{align}
where $\mathsf{Pr}(\gamma_n \mid \rho_n)$ represents the per-gNB posterior conditioned on $\rho_n$. This approximation reduces the original high-dimensional marginalization to a univariate posterior inference problem, leading to
\begin{align} \label{eq:psi_final_approx}
    \psi_n \approx \hat{\psi}_n \triangleq \mathsf{Pr}\left( \gamma_n = 1 \mid \rho_n \right),
\end{align}
which establishes the central link between the SD value and the score $\rho_n$ used throughout this section.  

The remainder of this section consists of the following stages. First, assuming the availability of site-survey data, we develop an estimation method that represents \eqref{eq:psi_final_approx} via a parametric mapping function in Sec.~\ref{sec:estimation}. Second, we refine the score $\rho_n$ by establishing a recursive relationship between $\rho_n$ and $\hat \psi_n$ in Sec.~\ref{sec:iterative_update}, and obtain the SD function for the refined scores in Sec.~\ref{sec:refine_reest}. Finally, we propose an online positioning algorithm in Sec.~\ref{sec:SI_positioning}, which integrates the resulting SD values into the standard CDA framework. The overall pipeline is illustrated in Fig.~\ref{fig:pipeline}.

\begin{figure}[!t]
    \centering
    \resizebox{0.9\columnwidth}{!}{\begin{tikzpicture}[font=\footnotesize, >=latex,
  blk/.style={draw=gray!70, line width=0.9pt, rounded corners=2pt, fill=gray!12, align=center, text width=2.3cm, minimum height=0.8cm, inner sep=2pt},
  frm/.style={draw=black!55, dashed, line width=0.5pt},
  arr/.style={->, draw=gray!75, line width=0.6pt},
  sec/.style={anchor=north west, inner sep=3pt},
  lab/.style={inner sep=1.5pt, fill=white}]
\draw[<->, line width=1.3pt] (0,0.3) -- node[above=1pt] {Offline phase} (3.7,0.3);
\draw[<->, line width=1.3pt] (3.9,0.3) -- node[above=1pt] {Online phase} (8.4,0.3);
\draw[frm] (0,0) rectangle (3.7,-3.0);
\draw[frm] (0,-3.2) rectangle (3.7,-4.85);
\draw[frm] (0,-5.05) rectangle (3.7,-7.05);
\draw[frm] (3.9,0) rectangle (8.4,-7.05);
\node[sec] at (0,0) {Sec.~\ref{sec:estimation}};
\node[sec] at (0,-3.2) {Sec.~\ref{sec:iterative_update}};
\node[sec] at (0,-5.05) {Sec.~\ref{sec:refine_reest}};
\node[sec] at (3.9,0) {Sec.~\ref{sec:SI_positioning}};
\node[blk] (S1) at (1.6,-0.95) {Site-survey in\\Assumption~\ref{Assumption: Weak Statistics}};
\node[blk] (S2) at (1.6,-2.25) {Estimate of\\SD function};
\node[blk] (S3) at (1.6,-4.15) {Score refinement};
\node[blk, minimum height=1.2cm] (S4) at (1.6,-6.2) {Revised\\SD function\\for refined scores};
\draw[arr] (S1) -- node[right=2pt, lab] {$\mathcal{Q}\triangleq\{\rho\},\ \pi$} (S2);
\draw[arr] (S2) -- node[right=2pt, lab] {$\mathcal{Q}$} (S3);
\draw[arr] (S3) -- node[right=2pt, lab] {$\tilde{\mathcal{Q}}=\{\tilde\rho\}$} (S4);
\node (in) at (6.0,-0.42) {$\{\rho_n\}$};
\node[right=-3pt of in, align=left, font=\scriptsize] {computed from\\range snapshot};
\node[blk] (R) at (6.0,-2.25) {Score refinement};
\node[blk] (F) at (6.0,-3.95) {Final SD values\\$\psi^*_n=\tilde h(\tilde\rho_n)$};
\node[blk] (P) at (6.0,-5.55) {Final positioning};
\node (pout) at (6.0,-6.65) {$\hat{\bm{p}}_{\mathrm{SD}}$};
\draw[arr] (in) -- (R);
\draw[arr] (R) -- node[right=2pt, lab] {$\{\tilde\rho_n\}$} (F);
\draw[arr] (F) -- node[right=2pt, lab] {$\{\psi_n^*\}$} (P);
\draw[arr] (P) -- (pout);
\draw[arr] (S2.east) -- (S2.east -| R.west);
\node[lab] (hs) at (3.27,-2.25) {$h^*(\cdot)$};
\draw[arr] ([xshift=8.5pt, yshift=-1.2pt]hs.base west) |- (S3.east);
\draw[arr] (S4.east) -| (4.3,-5.0) |- (F.west);
\node[lab] at (3.27,-6.2) {$\tilde h(\cdot)$};
\end{tikzpicture}}
    
    \caption{The pipeline of the SD framework.}
    \label{fig:pipeline}
    \vspace{-4pt}
\end{figure}

\subsection{Soft-Decision Function}\label{sec:estimation}
A key challenge in evaluating the approximated SD value $\hat \psi_n$ in \eqref{eq:psi_final_approx} is that the stochastic characteristics of the involved variables, namely, the NLoS indicator $\gamma_n$ and the score $\rho_n$, are generally unknown. To make \eqref{eq:psi_final_approx} tractable, we assume that weak prior information can be obtained from an offline site-survey:  
\begin{assumption}[Site-Survey]\label{Assumption: Weak Statistics}
By surveying the positioning environment, we assume the following two site-level statistics are available: 
\begin{itemize}
    \item \textbf{Score Samples}: Range measurements are collected over multiple survey locations and converted into scalar scores $\rho$ using \eqref{eq:Score}. This resulting dataset, denoted by $\mathcal{Q}\triangleq\{\rho\}$, provides an empirical basis for estimating the score PDF $f(\rho)$.
    \item \textbf{Average NLoS Probability}: During the survey, NLoS events are identified (e.g., via vision-based inspection \cite{Zhang2025VisionBlockage}), enabling estimation of the empirical NLoS probability $\pi\triangleq\mathsf{Pr}(\gamma=1)$ by normalizing the number of detected NLoS events by the total number of measurements. 
\end{itemize}
Here, we omit the gNB index in $\rho$ and $\pi$ to emphasize that these two statistics characterize the positioning site rather than a specific gNB.
Beyond $f(\rho)$ and $\pi$, no additional statistical prior information is assumed.  
\end{assumption}

Next, we define an SD function that converts the score to the posterior probability, given as
\begin{align}\label{eq:sd_mapping_def}
    h(\rho) \triangleq \mathsf{Pr}(\gamma=1\mid \rho),
\end{align}
where the SD value in \eqref{eq:psi_final_approx} can be written as $\hat \psi_n = h(\rho_n)$.
In the following, we explain the procedure of how to find $h(\rho)$ using the weak site-survey priors in Assumption \ref{Assumption: Weak Statistics}.

\subsubsection{From Bayes' Rule to a Binary Mixture Model}

Using Bayes' rule, the SD function in \eqref{eq:sd_mapping_def} can be written as
\begin{align}
    h(\rho) = \frac{\pi f(\rho \mid \gamma=1)}{f(\rho)},
\end{align}
where the NLoS-conditional score PDF $f(\rho \mid \gamma=1)$ is the remaining unknown. Note that the observed score PDF $f(\rho)$ is formulated via the total-probability decomposition, given as
\begin{align} \label{eq:binary_mixture}
    f(\rho) = (1-\pi) f(\rho \mid \gamma=0) + \pi f(\rho \mid \gamma=1).
\end{align}
In other words, the observed score PDF is a mixture of the LoS ($\gamma=0$) and NLoS ($\gamma=1$) score distributions with mixture weights $1-\pi$ and $\pi$, respectively.

In the following, we compute $h(\rho)$ through a two-step approach: (i) estimating a flexible posterior mapping using a constrained \textit{Gaussian mixture model} (GMM), as detailed in Sec.~\ref{sec:GMM} and (ii) refining it via a sigmoid fit to better capture tail behavior, as described in Sec.~\ref{sec:sigmoid}.
 
\subsubsection{Gaussian Mixture Modeling}\label{sec:GMM} 
\begin{figure}[!t]
\centering
\includegraphics[width=0.98\columnwidth]{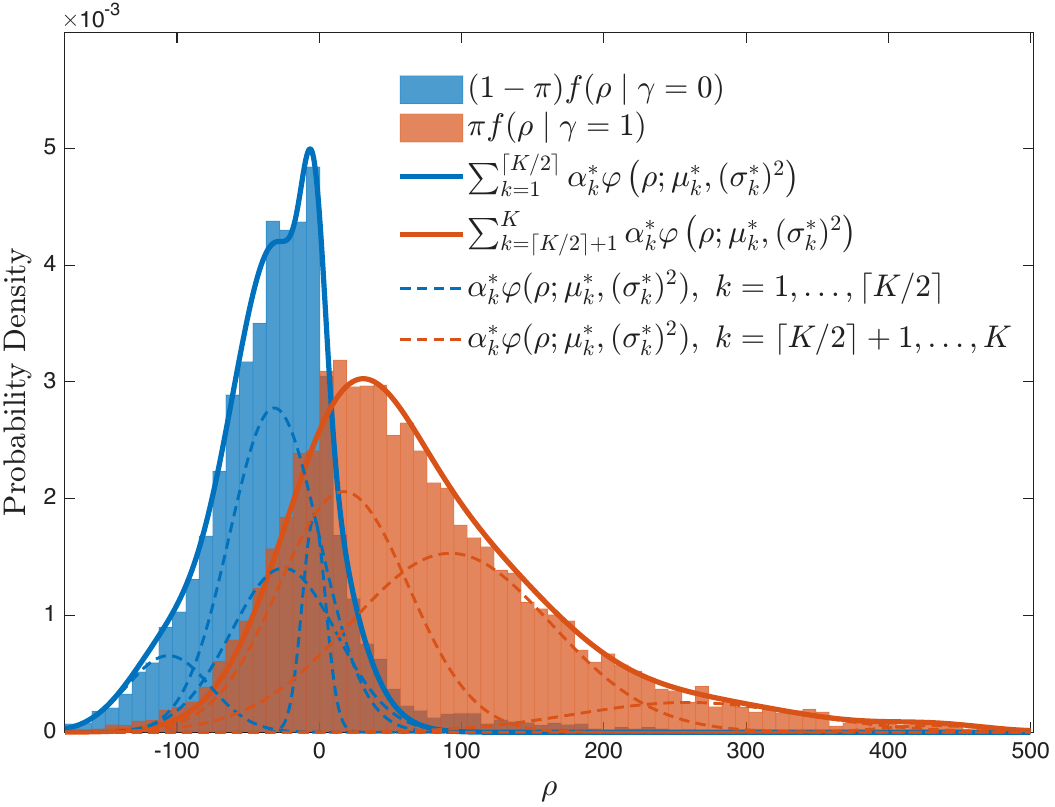}

\vspace{-4pt}
\caption{The empirical PDF of the NLoS score $\rho$ and its GMM-based modeling in InF-DH under FR1 (details in Sec.~\ref{sec:sim_setup}). Histograms represent empirical PDF of LoS and NLoS scores weighted by $1-\pi$ and $\pi$, respectively in \eqref{eq:binary_mixture}. All Gaussian components in \eqref{eq:GMM_compo} are represented by dotted curves, while its LoS and NLoS mixtures are represented by solid curves.}
\label{fig:GMM}

\end{figure}

We approximate the site-level PDF $f(\rho)$ using a $K$-component GMM:
\begin{align} \label{eq:GMM_rep}
   f(\rho)\approx f_{\text{GMM}}(\rho;\Xi) = \sum_{k=1}^K \alpha_k \varphi\left(\rho; \mu_k, \sigma_k^2\right),
\end{align}
where $\varphi\left(\rho; \mu_k, \sigma_k^2\right)$ denotes the Gaussian PDF with mean $\mu_k$ and variance $\sigma^2_k$, and
\begin{align}
\Xi=\left\{\alpha_k,\mu_k,\sigma_k\right\}_{k=1}^K, \quad \alpha_k\geq 0, \quad \sum_{k=1}^K\alpha_k=1.
\end{align}
Motivated by the monotonic relationship between $\rho$ and NLoS probability observed in Remark~\ref{rem:monotonic}, we 
enforce an ordering on the component means, namely, $\mu_1 \le \mu_2 \le \dots \le \mu_K$. We then partition the components into two groups as $\{1,\dots, \lceil K/2 \rceil\}$ and $\{\lceil K/2 \rceil+1,\dots, K\}$, which can be interpreted as predominantly LoS- and NLoS-like, respectively. With this partition, we ensure that the total mixture weight assigned to each group matches the surveyed priors $1-\pi$ and $\pi$. The resultant LoS-conditional and NLoS-conditional score PDFs are given as
\begin{subequations}\label{eq:GMM_compo}
\begin{align}
(1-\pi)f(\rho \mid \gamma=0)&\approx\sum_{k=1}^{\lceil K/2 \rceil} \alpha_k^* \varphi\left(\rho; \mu_k^*, (\sigma_k^*)^2\right),\\
\pi f(\rho \mid \gamma=1)&\approx\sum_{k=\lceil K/2 \rceil+1}^K \alpha_k^* \varphi\left(\rho; \mu_k^*, (\sigma_k^*)^2\right),
\end{align} 
\end{subequations}
where $\Xi^*= \{ \alpha^*_k, \mu^*_k, \sigma^*_k \}_{k=1}^K$ is obtained by the following constrained maximum likelihood estimation:
\begin{align}\label{eq:cem_objective_simplified}\tag{P3}
     \max_{\Xi} & \sum_{\rho\in \mathcal{Q}} \ln \left( {f}_{\text{GMM}}(\rho;\Xi) \right)
    \nonumber\\
    \mathrm{s.t.} \quad & \mu_1 \le \mu_2 \le \dots \le \mu_K, 
    \nonumber\\
    & \sum_{k=1}^{\lceil K/2 \rceil} \alpha_k = 1-\pi, \quad \sum_{k=\lceil K/2 \rceil+1}^{K} \alpha_k = \pi,\nonumber 
\end{align}
which can be efficiently solved using a \textit{constrained expectation-maximization} (CEM) algorithm whose details are well summarized in Appendix~\ref{app:CEM}. The effectiveness of the algorithm is well visualized in Fig.~\ref{fig:GMM}, where the LoS- and NLoS-conditional PDF derived from $f_{\text{GMM}}(\rho; \Xi^*)$ well aligns with the ground-truth distributions using two weak prior statistics in Assumption \ref{Assumption: Weak Statistics}.

With $\Xi^*$, the SD function is approximated by the GMM-induced posterior function, given as
\begin{align} \label{eq:gmm_posterior}
    h(\rho)\approx\Psi_{\mathrm{GMM}}(\rho; {\Xi}^*) \triangleq \frac{\sum_{k=\lceil K/2 \rceil+1}^{K} {\alpha}^*_k \varphi\left(\rho; {\mu}^*_k, ({\sigma}_k^{*})^2\right)}{{f}_{\text{GMM}}(\rho; {\Xi}^*)}.
\end{align} 
\subsubsection{Sigmoid Fitting}\label{sec:sigmoid}
While $\Psi_{\mathrm{GMM}}(\rho; {\Xi}^*)$ is effective near each component's mode, it may be less accurate in the tails due to score-sample sparsity in these regions. This can be overcome by fitting $\Psi_{\mathrm{GMM}}(\rho; {\Xi}^*)$ to a generalized sigmoid model: 
\begin{align} \label{eq:generalized_sigmoid}
    \Psi(\rho; \Phi) = \frac{\phi_1}{1 + \exp({-\phi_2(\rho - \phi_3)})} + \phi_4,
\end{align}
where $\Phi = \{\phi_1, \phi_2, \phi_3, \phi_4\}$. The optimal parameters ${\Phi}^*$ are obtained by least-squares fitting over a set of sampled score points in $\mathcal{Q}$:
\begin{subequations}\label{eq:weak_optim_all}
\begin{align} \label{eq:weak_optim}
    \Phi^* = \mathop{\mathrm{arg\,min}}_{{\Phi}} & \sum_{\rho\in\mathcal{Q}} \left| \Psi_{\mathrm{GMM}}(\rho;\Xi^*) - \Psi(\rho; {\Phi}) \right|^2 \\
    \mathrm{s.t.} &\quad  {\phi}_1 + {\phi}_4 \le 1, \quad {\phi}_1, {\phi}_2, {\phi}_4 \ge 0. 
\end{align}
\end{subequations}

Here, the constraint $\phi_2 \ge 0$ ensures that the mapping is monotonically increasing, while the remaining constraints guarantee a valid probability range of $[0, 1]$. 

Finally, the SD function in \eqref{eq:sd_mapping_def} is derived as 
\begin{align}\label{eq: SD value}
h^*(\rho)=\Psi(\rho; \Phi^*),
\end{align}
where $h^*$ indicates the best SD function obtained through the above procedure.

\subsection{Iterative Score Refinement} \label{sec:iterative_update}

As recalled in Sec.~\ref{Sec:Threshold-Based Test}, the score $\rho_n$ depends on the representative points $\bm{m}_n, \bm{m}_{-n},$ and $\bar{\bm{p}}$, which are computed by treating all PELs as equally reliable. This unweighted HD process may not match the underlying propagation environment. In particular, when a substantial fraction of PELs are generated using NLoS-contaminated gNBs, the representative points become biased, causing $\rho_n$ to lose its discriminative feature. Such a bias can lead to incorrect NLoS detection and degrade the subsequent positioning performance.

On the other hand, the SD function in \eqref{eq: SD value} enables us to quantify the reliability of each PEL. Specifically, assuming conditional independence across gNBs, we can define the reliability of PEL $\ell$ as the probability that all gNBs involved in constructing that PEL are in LoS, namely, 
\begin{align} \label{eq:soft_weight}
    \omega^{(\ell)} = \prod_{n \in \mathcal{M}^{(\ell)}} (1- h^*(\rho_n)).
\end{align}
This reliability down-weights PELs constructed from gNBs with high NLoS probabilities.

Using $\{\omega^{(\ell)}\}$, we re-estimate the representative points $\bm{m}_n, \bm{m}_{-n},$ and $\bar{\bm{p}}$ as \textit{weighted medians} by minimizing weighted $\mathsf{L}_1$-distances:

\begin{align}
    \tilde{\boldsymbol{m}}_n &= \mathop{\arg\min}_{{\boldsymbol{m}}} \sum_{\{\ell \mid \boldsymbol{x}^{(\ell)} \in \mathcal{X}_n \cap\mathcal{X}_{\mathrm{HD}}\}} \omega^{(\ell)} \left\| \boldsymbol{x}^{(\ell)} - {\boldsymbol{m}} \right\|_1, \label{eq:restored_rep1}\\
    \tilde{\boldsymbol{m}}_{-n} &= \mathop{\arg\min}_{{\boldsymbol{m}}} \sum_{\{\ell \mid \boldsymbol{x}^{(\ell)} \in \mathcal{X}_{-n}\cap\mathcal{X}_{\mathrm{HD}}\}} \omega^{(\ell)} \left\| \boldsymbol{x}^{(\ell)} - {\boldsymbol{m}} \right\|_1,\label{eq:restored_rep2} \\
    \tilde{\boldsymbol{p}} &= \mathop{\arg\min}_{{\boldsymbol{m}}} \sum_{\{\ell \mid \boldsymbol{x}^{(\ell)} \in \mathcal{X}_{\mathrm{HD}}\}} \omega^{(\ell)} \left\| \boldsymbol{x}^{(\ell)} - {\boldsymbol{m}} \right\|_1,\label{eq:restored_rep3}
\end{align}  
where we restrict the PELs used for re-estimation to the subset $\mathcal{X}_{\mathrm{HD}}$ specified in \eqref{eq:PEL_HD}. This restriction improves robustness by excluding low-confidence PELs that are filtered out in the HD stage.
The resultant score, denoted by $\tilde{\rho}_n$, is refined as
\begin{align} \label{eq:refined_score_func}
    \tilde{\rho}_n = \left(\frac{\tilde{\bm{r}}_n^\top \tilde{\bm{r}}_n^0}{\|\tilde{\bm{r}}_n^0\|}\right) \sqrt{d_n},
\end{align}
where 
\begin{align}
    \tilde{\bm{r}}_n = \tilde{\bm{m}}_n - \tilde{\bm{m}}_{-n},\quad 
    \tilde{\bm{r}}_n^0 = \tilde{\bm{p}} - \bm{z}_n.
\end{align}

It is worth noting that $\tilde{\rho}_n$ is recursive: the refined scores can determine the PEL reliabilities through \eqref{eq:soft_weight}, while the reliabilities in turn reshape the representative points in \eqref{eq:restored_rep1}–\eqref{eq:restored_rep3}, which subsequently further refine the scores. This motivates an iterative refinement procedure summarized below.

\subsubsection{Initialization}\label{sec:refine_init} Let ${\tilde{\rho}_n}[t]$ denote gNB $n$'s score at iteration $t$, initialized with the score in \eqref{eq:Score}, i.e.,  ${\tilde{\rho}_n}[0]={{\rho}_n}$.

\subsubsection{Weight and Score Updates} 
 Given $\{\tilde\rho_n[t]\}$, we recompute the PEL weights and representative points via \eqref{eq:soft_weight}--\eqref{eq:restored_rep3}, and then obtain the next iteration score $\{\tilde{\rho}_n[t+1]\}$ using \eqref{eq:refined_score_func}.  

\subsubsection{Convergence Check}\label{sec:refine_conv}
We iterate until
\begin{align}
{\sum_{n\in\mathcal{N}_\mathrm{HD}}\left(h^*(\tilde{\rho}_n[t+1])-h^*\left(\tilde{\rho}_n[t]\right)\right)^2} \le \epsilon,
\end{align}
where $\epsilon > 0$ is a predefined threshold. The process terminates when the above condition is satisfied  or when the maximum iteration count $T_{\max}$ is reached. 

Upon termination at iteration $t^*$, we obtain the refined scores $\{\tilde\rho_n[t^*]\}_{n\in \mathcal{N}_{\mathrm{HD}}}$. Hereafter, we omit the iteration index and write $\tilde\rho_n[t^*]$ as $\tilde\rho_n$.

\subsection{Revised Soft-Decision Function for Refined Scores}\label{sec:refine_reest}
{
The refinement in Sec.~\ref{sec:iterative_update} improves the separability of the scores, and the reliability of $h^*(\tilde\rho_n)$ as the NLoS probability can be further enhanced, since $h^*$ in \eqref{eq: SD value} is fitted to the score samples $\mathcal{Q}$ before refinement. We therefore revise the SD function for the refined scores. Let $\tilde{\mathcal{Q}}=\{\tilde\rho_q\}$ denote the refined scores obtained by repeating the procedure in Sec.~\ref{sec:iterative_update} at each survey location. Since HD excludes the samples detected as NLoS from the refinement, the average NLoS probability over $\tilde{\mathcal{Q}}$ differs from $\pi$ and is estimated as
\begin{align}\label{eq:recal_pik}
    \tilde\pi = \frac{1}{|\tilde{\mathcal{Q}}|}\sum_{q=1}^{|\tilde{\mathcal{Q}}|} h^*(\rho_q),
\end{align}
where $\rho_q\in\mathcal{Q}$ is the sample corresponding to $\tilde\rho_q$ before refinement. Applying the GMM estimation in \ref{eq:cem_objective_simplified} and the sigmoid fitting in \eqref{eq:weak_optim_all} to $\tilde{\mathcal{Q}}$, with $\tilde\pi$ in place of $\pi$, yields $\tilde\Phi$, and the SD function for the refined scores is given as
\begin{align}\label{eq:reest_SD}
    \tilde h(\tilde\rho)=\Psi(\tilde\rho;\tilde\Phi),
\end{align}
where $\Psi$ is the sigmoid model in \eqref{eq:generalized_sigmoid}.
}

\subsection{Positioning with Soft Decision} \label{sec:SI_positioning}

{
In the online phase, the scores $\{\rho_n\}$ in \eqref{eq:Score} are computed from a range snapshot of the target UE and refined by the procedure in Sec.~\ref{sec:iterative_update}. The final SD values, i.e., the estimates of $\hat\psi_n$, are given as
\begin{align} \label{eq:final_SD_Val}
     \psi^*_n = \begin{cases}
\tilde h(\tilde\rho_n) &\text{if }n \in\mathcal{N}_\mathrm{HD},\\
h^*(\rho_n) &\text{if }n \in\mathcal{N}\setminus\mathcal{N}_\mathrm{HD},
\end{cases}
\end{align}
where the gNBs retained by HD use the refined score $\tilde\rho_n$ with the SD function $\tilde h$ in \eqref{eq:reest_SD}, while those detected as NLoS use the original score $\rho_n$ with $h^*$.
}

We exploit the above SD values $\{{\psi}_n^*\}$ to compute the UE location more accurately by taking the weighted median of $\mathcal{X}_{\mathrm{RS}}$, given as
\begin{align} \label{eq:SD_positioning}
    \hat{\boldsymbol{p}}_{\mathrm{SD}} = \mathop{\mathrm{arg\,min}}_{{\boldsymbol{p}} \in \mathbb{R}^{2}} \sum_{\{\ell \mid \boldsymbol{x}^{(\ell)} \in \mathcal{X}_{\mathrm{RS}}\}} \omega^{*(\ell)} \left\| \boldsymbol{x}^{(\ell)} - {\boldsymbol{p}} \right\|_1,
\end{align}
where
\begin{align} \label{eq:reliability}
\omega^{*(\ell)} = \prod_{n \in \mathcal{M}^{(\ell)}} \left(1- \psi^*_n\right),
\end{align}
and the RE\&RS filtered PEL subset $\mathcal{X}_{\mathrm{RS}}$ is specified in \eqref{eq: RS set}. This strategy is hereafter termed CDA-ND-RERS (SD).

{CDA-ND-RERS (SD) in \eqref{eq:SD_positioning} belongs to the class of subset-weighting approaches~\cite{chen1999RWGH, Guvenc2009Survey}. Its distinguishing feature over other subset-weighting schemes is that it constructs the PEL weight $\omega^{*(\ell)}$ from gNB-wise NLoS evidence in the following steps. First, the score $\rho_n$ in \eqref{eq:Score} isolates the effect of gNB $n$ on the PEL distribution by comparing the PELs constructed with and without the gNB (Sec.~\ref{sec:nlos-detection}). Second, the SD function obtained from the weak site-survey priors maps the score to the posterior NLoS probability (Sec.~\ref{sec:estimation}). Third, the score is refined by establishing a recursive relationship with the NLoS probability through the reliability-weighted medians of the PELs (Sec.~\ref{sec:iterative_update}). This refinement reduces the bias of the PEL statistics and improves the separability of the scores, particularly in NLoS-dominant environments. Finally, the SD function is revised for the refined scores to enhance the reliability of the NLoS probability (Sec.~\ref{sec:refine_reest}). The contribution of each step is examined in the ablation study (see Appendix~\ref{app:ablation}).}

\begin{figure}[!t]
    \centering
    \subfloat[FR1.]{\includegraphics{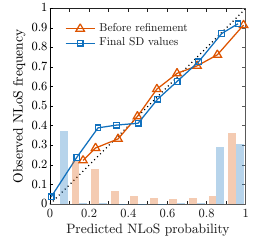}\label{fig:rel_dh_FR1}}
    \hfill
    \subfloat[FR2.]{\includegraphics{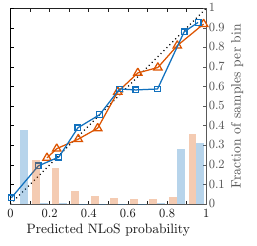}\label{fig:rel_dh_FR2}}
    \caption{Reliability diagrams of the NLoS probability before refinement, $h^*(\rho_n)$, and of the final SD values, $\psi_n^*$, $n\in\mathcal{N}$, in InF-DH (details in Sec.~\ref{sec:sim_setup}). Bars (right axis): fraction of samples. Dotted line: ideal case.}
    \label{fig:rel_dh}
    
    \vspace{-6pt}
\end{figure}

\begin{remark}[Reliability of the NLoS Probability] \label{rem:reliability}
Fig.~\ref{fig:rel_dh} shows the reliability diagrams of the NLoS probability $h^*(\rho_n)$ from the initial SD function in Sec.~\ref{sec:estimation} and of the final SD values $\psi^*_n$. Both lie close to the ideal line, where the observed NLoS frequency equals the NLoS probability, and are thus reliable. On the other hand, the final SD values have better separability owing to the refinement: most $\psi^*_n$ values fall outside the uncertain range ($0.2$--$0.8$), where many $h^*(\rho_n)$ values lie. Other reliability metrics and other scenarios are evaluated in Sec.~\ref{sec:sim_calib}.
\end{remark}

\subsection{Complexity Analysis}\label{sec:complexity}
{
CDA-ND comprises four steps: PEL generation, score generation and HD, score refinement, and positioning. Given the number of gNBs $N$ and the number of PELs $L\le\binom{N}{3}$, the computational complexity of each step is analyzed as follows.
\begin{itemize}
    \item \textbf{PEL Generation (Sec.~\ref{Sec:Hypothesis})}: As in \cite{yu2025combinatorial}, deriving one PEL from $M=3$ ranges requires a bounded number of operations, and generating all PELs requires $O(L)=O(N^3)$.
    \item \textbf{Score Generation and HD (Secs.~\ref{Sec:Factorization} and \ref{Sec:Threshold-Based Test})}: The score $\rho_n$ in \eqref{eq:Score} requires the medians of $\mathcal{X}_n$ and $\mathcal{X}_{-n}$. Once the PELs are sorted along each coordinate by an efficient sorting algorithm, e.g., merge sort, with complexity $O(L\log L)$, the medians for all $n$ are obtained by scanning the sorted PELs with complexity $O(NL)$. Since the threshold in \eqref{eq:adaptive_threshold} and the test in \eqref{eq:CDA-ND_rule} require $O(N)$, the complexity becomes $O(NL)=O(N^4)$.
    \item \textbf{Score Refinement (Sec.~\ref{sec:iterative_update})}: Each iteration computes the weights in \eqref{eq:soft_weight} with complexity $O(L)$ and the weighted medians in \eqref{eq:restored_rep1}--\eqref{eq:restored_rep3} for all $n$ by scanning the sorted PELs with complexity $O(NL)$. Since the number of iterations is bounded by the constant $T_{\max}$, the complexity becomes $O(N^4)$.
    \item \textbf{HD and SD Positioning (Secs.~\ref{sec:aided_loc} and \ref{sec:SI_positioning})}: As in \cite{yu2025combinatorial}, RE\&RS filtering sorts the PELs twice with complexity $O(L\log L)=O(N^3\log N)$, and the median in \eqref{eq:proposed} for HD or the weighted median in \eqref{eq:SD_positioning} for SD of the remaining PELs requires $O(L)$.
\end{itemize}
As a result, the overall complexity becomes $O(N^4)$ for both HD and SD. Compared to CDA-RERS \cite{yu2025combinatorial} with complexity $O(N^3\log N)$, the complexity of CDA-ND is higher by a factor of $O(N/\log N)$, which is reasonable in comparison to its performance enhancement in Sec.~\ref{sec:simulation}. The SD functions $h^*$ and $\tilde h$ are obtained offline and do not increase the online complexity. For instance, in InF-DH FR1 with $N=18$, the average runtime per snapshot on a single core of an Apple M4 Pro CPU (MATLAB R2025a) is \SI{2.4}{ms} for HD and \SI{3.8}{ms} for SD. The runtime and the performance for different $N$ are examined in Appendix~\ref{app:gnb}.
}

\begin{table}[t]
\centering
\setlength{\tabcolsep}{4.0pt}
\fontsize{7.5pt}{8.5pt}\selectfont
\caption{Range-measurement error statistics (mean/Std) and NLoS ratio $\pi$ for each scenario--carrier configuration.}

\vspace{-4pt}

\label{tab:range_stats_grid}
\begin{tabular}{lcccccc}
\toprule
\multirow{2}{*}{Scenario} & \multicolumn{3}{c}{FR1} & \multicolumn{3}{c}{FR2} \\
\cmidrule(lr){2-4}\cmidrule(lr){5-7}
& LoS [\si{m}] & NLoS [\si{m}] & $\pi$ & LoS [\si{m}] & NLoS [\si{m}] & $\pi$ \\
\midrule
InF-SH & \num{1.5}/\num{5.9} & \num{26.1}/\num{20.1} & \num{0.176} & \num{4.4}/\num{48.4} & \num{64.7}/\num{318.1} & \num{0.183} \\
IOO & \num{0.1}/\num{1.7} & \num{9.1}/\num{9.0} & \num{0.357} & \num{0.2}/\num{1.3} & \num{68.9}/\num{306.5} & \num{0.364} \\
InF-DH & \num{4.0}/\num{14.0} & \num{25.1}/\num{19.2} & \num{0.564} & \num{3.2}/\num{11.6} & \num{26.8}/\num{22.7} & \num{0.560} \\
\bottomrule
\end{tabular}

\vspace{-6pt}

\end{table}

\section{Numerical Results}
\label{sec:simulation}
\subsection{Simulation Setup}\label{sec:sim_setup}
We validate CDA-ND using the publicly available 3GPP-compliant positioning dataset in \cite{conti2024dataset}. We consider two indoor factory scenarios: InF-SH and InF-DH, representing LoS-dominant and NLoS-dominant environments whose NLoS ratios $\pi$ are about $0.18$ and $0.56$, respectively. 
Following the standard deployments, we use $N=\num{18}$ gNBs in both scenarios. The factory area is \SI{300}{m}$\times$\SI{150}{m} for InF-SH and \SI{120}{m}$\times$\SI{60}{m} for InF-DH. We also consider the IOO scenario with $N=\num{12}$ gNBs in a \SI{120}{m}$\times$\SI{50}{m} area, whose NLoS ratio $\pi$ is about $0.36$. These values of $\pi$ are computed over the entire dataset of each configuration and are listed in Table~\ref{tab:range_stats_grid}. 

For each UE realization, we use the dataset-provided gNB coordinates $\{\bm z_n\}$, the binary LoS/NLoS labels $\{\gamma_n\}$, and RTT-based range measurements $\{d_n\}$, where $d_n$ is obtained by averaging the downlink and uplink range estimates from reference signals. Two FR configurations with \SI{100}{MHz} bandwidth are considered: FR1 with \SI{3.5}{GHz} and FR2 with \SI{28}{GHz} carriers in InF-SH and InF-DH, and FR1 with \SI{4}{GHz} and FR2 with \SI{30}{GHz} carriers in IOO.  All other simulation settings, including the 3GPP standard-based wireless channel generation, the reference signal configurations, and scenario-specific parameters, follow the dataset generation procedure in \cite{conti2024dataset}. The resulting range-measurement statistics of each configuration are summarized in Table \ref{tab:range_stats_grid}. CDA-ND is also validated with real Wi-Fi RTT measurements \cite{han2024mobility} in Appendix~\ref{app:wifi}.

For each scenario--frequency configuration, the dataset provides \num{100} spatially consistent scenario instantiations (with spatially consistent channels and NLoS maps) and \num{10} UE drops per instantiation, resulting in \num{1000} UE realizations per configuration. We partition these \num{1000} UE realizations into $10$ subsets: one subset is used to evaluate the NLoS detection and positioning performance, while the remaining nine subsets are used to construct the SD functions in Secs.~\ref{sec:estimation} and \ref{sec:refine_reest}. We repeat this process $10$ times by changing the evaluation subset, thereby performing $10$-fold cross-validation. In each fold, $\pi$ is computed from the nine training subsets only and differs from that of the entire dataset by at most $\num{0.004}$.

\subsection{NLoS Detection Performance}\label{sec:sim_identification}

\begin{table*}[!t]
\centering
\setlength{\tabcolsep}{4pt}
\fontsize{7.2pt}{8.5pt}\selectfont
\caption{NLoS detection performance of CDA-ND and the benchmarks (accuracy, recall, and precision in \%). RTT dataset and LoS/NLoS labeling: required (\checkmark) or not (--). The proposed methods are highlighted in bold.}

\vspace{-4pt}

\label{tab:detection_perf}\label{tab:app_detection}
\begin{tabular}{llcccccccccc}
\toprule
\multirow{2}{*}{Scenario} & \multirow{2}{*}{Method} & \multirow{2}{*}{\shortstack{RTT\\dataset}} & \multirow{2}{*}{\shortstack{LoS/NLoS\\labeling}} & \multicolumn{4}{c}{FR1} & \multicolumn{4}{c}{FR2} \\
\cmidrule(lr){5-8}\cmidrule(lr){9-12}
& & & & Accuracy & Recall & Precision & AUC & Accuracy & Recall & Precision & AUC \\
\midrule
\multirow{6}{*}{InF-SH} & LMedS \cite{casas2006LMedS} & -- & -- & \num{93.24} & \num{97.31} & \num{73.11} & \num{0.976} & \num{87.72} & \num{94.80} & \num{60.47} & \num{0.934} \\
 & EM \cite{yin2014EM} & -- & -- & \num{95.57} & \num{95.19} & \num{82.36} & \num{0.959} & \num{88.02} & \num{89.36} & \num{61.94} & \num{0.868} \\
 & \textbf{Hard Decision} & -- & -- & \num{94.76} & \num{90.29} & \num{81.79} & \num{0.967} & \num{88.28} & \num{82.83} & \num{63.84} & \num{0.918} \\
\cmidrule(l){2-12}
 & RF classifier & \checkmark & \checkmark & \num{97.86} & \num{96.43} & \num{91.81} & \num{0.990} & \num{90.94} & \num{83.44} & \num{71.68} & \num{0.960} \\
 & MLP classifier & \checkmark & \checkmark & \num{97.70} & \num{94.31} & \num{92.73} & \num{0.991} & \num{91.20} & \num{85.11} & \num{71.91} & \num{0.962} \\
 & \textbf{Soft Decision} & \checkmark & $\pi$ only & \num{96.76} & \num{97.63} & \num{85.85} & \num{0.979} & \num{89.58} & \num{83.90} & \num{67.24} & \num{0.939} \\
\midrule
\multirow{6}{*}{IOO} & LMedS \cite{casas2006LMedS} & -- & -- & \num{83.56} & \num{61.84} & \num{88.76} & \num{0.860} & \num{78.67} & \num{51.26} & \num{83.89} & \num{0.810} \\
 & EM \cite{yin2014EM} & -- & -- & \num{83.92} & \num{67.97} & \num{83.99} & \num{0.812} & \num{75.10} & \num{52.88} & \num{71.30} & \num{0.672} \\
 & \textbf{Hard Decision} & -- & -- & \num{84.52} & \num{64.13} & \num{89.64} & \num{0.845} & \num{78.85} & \num{52.04} & \num{83.69} & \num{0.819} \\
\cmidrule(l){2-12}
 & RF classifier & \checkmark & \checkmark & \num{90.67} & \num{78.76} & \num{94.18} & \num{0.937} & \num{86.38} & \num{71.29} & \num{89.10} & \num{0.897} \\
 & MLP classifier & \checkmark & \checkmark & \num{90.79} & \num{79.56} & \num{93.74} & \num{0.940} & \num{85.47} & \num{68.32} & \num{89.26} & \num{0.885} \\
 & \textbf{Soft Decision} & \checkmark & $\pi$ only & \num{90.76} & \num{80.91} & \num{92.29} & \num{0.927} & \num{85.97} & \num{69.99} & \num{89.15} & \num{0.881} \\
\midrule
\multirow{6}{*}{InF-DH} & LMedS \cite{casas2006LMedS} & -- & -- & \num{59.63} & \num{34.61} & \num{84.72} & \num{0.725} & \num{62.35} & \num{37.84} & \num{88.10} & \num{0.753} \\
 & EM \cite{yin2014EM} & -- & -- & \num{77.27} & \num{72.03} & \num{85.35} & \num{0.780} & \num{78.31} & \num{71.24} & \num{87.68} & \num{0.794} \\
 & \textbf{Hard Decision} & -- & -- & \num{78.04} & \num{73.24} & \num{85.72} & \num{0.842} & \num{78.54} & \num{73.48} & \num{86.14} & \num{0.836} \\
\cmidrule(l){2-12}
 & RF classifier & \checkmark & \checkmark & \num{88.62} & \num{88.62} & \num{90.95} & \num{0.954} & \num{89.36} & \num{90.44} & \num{90.54} & \num{0.953} \\
 & MLP classifier & \checkmark & \checkmark & \num{88.59} & \num{88.11} & \num{91.34} & \num{0.955} & \num{89.53} & \num{90.50} & \num{90.77} & \num{0.954} \\
 & \textbf{Soft Decision} & \checkmark & $\pi$ only & \num{91.50} & \num{96.59} & \num{89.22} & \num{0.926} & \num{92.28} & \num{96.95} & \num{90.02} & \num{0.933} \\
\bottomrule
\end{tabular}

\vspace{-4pt}
\end{table*}

We evaluate gNB-wise LoS/NLoS detection against the dataset-provided ground-truth labels.
{
HD and SD are compared with the following benchmarks:
\begin{itemize}
    \item \textit{Least median of squares} (LMedS) \cite{casas2006LMedS}: We adopt the original localization algorithm, which estimates the UE position from the PEL with the minimum median of the squared residuals of all $N$ ranges. We classify gNBs with residuals much larger than this median as NLoS. 
    \item \textit{Expectation-maximization} (EM) \cite{yin2014EM}: The range errors are modeled by a mixture of two Gaussian distributions for LoS and NLoS, whose parameters are estimated by the iterative EM algorithm. The gNB is considered NLoS if its posterior probability, computed by inputting range measurements into the Gaussian distributions, is above the threshold. 
    \item \textit{Random forest} (RF) and \textit{multilayer perceptron} (MLP): These binary classifiers are trained on the features extracted from the range snapshot with LoS/NLoS labels. The input features and the model structures are given in Appendix~\ref{app:bench_det}.
\end{itemize}
}

We use the following metrics to summarize detection performance in Table~\ref{tab:detection_perf}.
\begin{itemize}
\item \textit{Recall}: The number of gNBs correctly detected as NLoS, normalized by the total number of actual NLoS gNBs.
\item \textit{Precision}: The number of gNBs correctly detected as NLoS normalized by the total number of gNBs detected as NLoS.
\item \textit{Accuracy}: The fraction of gNBs whose LoS/NLoS states are correctly classified, i.e., the number of correctly classified gNBs normalized by the total number of gNBs.
\item \textit{AUC}: We report the AUCs of the score $\rho_n$ in \eqref{eq:Score} for HD and the final SD value $\psi^*_n$ in \eqref{eq:final_SD_Val} for SD, and those of the normalized residual for LMedS and the NLoS probability for the other benchmarks. 
\end{itemize}
The NLoS threshold for the accuracy, recall, and precision in Table~\ref{tab:detection_perf} and for the confusion matrices in Fig.~\ref{fig:conf_SH_FR1}--\ref{fig:conf_DH_FR2} follows \eqref{eq:adaptive_threshold} for HD and the outlier test of \cite{casas2006LMedS} for LMedS. For SD, EM, RF, and MLP, which output the NLoS probability, it is set to $0.5$, i.e., the maximum a posteriori decision. While the accuracy, recall, and precision depend on this threshold, the AUC measures the separability between LoS and NLoS gNBs over all thresholds.

In general, both HD and SD modes provide reliable performance across all scenario-frequency configurations, while several interesting observations are made.

\definecolor{gray}{RGB}{200, 200, 200}
\definecolor{lightgray}{RGB}{240, 240, 240}
\begin{figure*}[t!]
  \centering
  
  \subfloat[InF-SH FR1.]{
  \begin{tikzpicture}[font=\fontsize{7.3pt}{8.3pt}\selectfont]
    \def\cs{1.3}
    \fill[lightgray] (0,0) rectangle (1.3*\cs,\cs);
    \fill[lightgray] (1.3*\cs,\cs) rectangle (2.6*\cs,2*\cs);
    \fill[gray] (0,\cs) rectangle (1.3*\cs,2*\cs);
    \fill[gray] (1.3*\cs,0) rectangle (2.6*\cs,\cs);
    \draw[black] (0,0) rectangle (2.6*\cs,2.0*\cs);
    \draw[black] (1.3*\cs,0) -- (1.3*\cs,2*\cs);
    \draw[black] (0,\cs) -- (2.6*\cs,\cs);
    \node[align=left] at (0.65*\cs, 1.5*\cs) {HD: \SI{90.29}{\percent} \\ SD: \SI{97.63}{\percent}};
    \node[align=left] at (1.95*\cs, 1.5*\cs) {HD: \SI{9.71}{\percent} \\ SD: \SI{2.37}{\percent}};
    \node[align=left] at (0.65*\cs, 0.5*\cs) {HD: \SI{4.29}{\percent} \\ SD: \SI{3.43}{\percent}};
    \node[align=left] at (1.95*\cs, 0.5*\cs) {HD: \SI{95.71}{\percent} \\ SD: \SI{96.57}{\percent}};
    \node at (0.65*\cs, -0.15) {NLoS};
    \node at (1.95*\cs, -0.15) {LoS};
    \node at (1.3*\cs,  -0.45) {Predicted};
    \node[rotate=90] at (-0.15, 1.5*\cs) {NLoS};
    \node[rotate=90] at (-0.15, 0.5*\cs) {LoS};
    \node[rotate=90] at (-0.45, \cs)     {Actual};
  \end{tikzpicture}
  \label{fig:conf_SH_FR1}
  }
  \hspace{0.2cm}
  \subfloat[InF-SH FR2.]{
  \begin{tikzpicture}[font=\fontsize{7.3pt}{8.3pt}\selectfont]
    \def\cs{1.3}
    \fill[lightgray] (0,0) rectangle (1.3*\cs,\cs);
    \fill[lightgray] (1.3*\cs,\cs) rectangle (2.6*\cs,2*\cs);
    \fill[gray] (0,\cs) rectangle (1.3*\cs,2*\cs);
    \fill[gray] (1.3*\cs,0) rectangle (2.6*\cs,\cs);
    \draw[black] (0,0) rectangle (2.6*\cs,2.0*\cs);
    \draw[black] (1.3*\cs,0) -- (1.3*\cs,2*\cs);
    \draw[black] (0,\cs) -- (2.6*\cs,\cs);
    \node[align=left] at (0.65*\cs, 1.5*\cs) {HD: \SI{82.83}{\percent} \\ SD: \SI{83.90}{\percent}};
    \node[align=left] at (1.95*\cs, 1.5*\cs) {HD: \SI{17.17}{\percent} \\ SD: \SI{16.10}{\percent}};
    \node[align=left] at (0.65*\cs, 0.5*\cs) {HD: \SI{10.50}{\percent} \\ SD: \SI{9.14}{\percent}};
    \node[align=left] at (1.95*\cs, 0.5*\cs) {HD: \SI{89.50}{\percent} \\ SD: \SI{90.86}{\percent}};
    \node at (0.65*\cs, -0.15) {NLoS};
    \node at (1.95*\cs, -0.15) {LoS};
    \node at (1.3*\cs,  -0.45) {Predicted};
    \node[rotate=90] at (-0.15, 1.5*\cs) {NLoS};
    \node[rotate=90] at (-0.15, 0.5*\cs) {LoS};
    \node[rotate=90] at (-0.45, \cs)     {Actual};
  \end{tikzpicture}
  \label{fig:conf_SH_FR2}
  }
  \hspace{0.2cm}
  \subfloat[InF-DH FR1.]{
  \begin{tikzpicture}[font=\fontsize{7.3pt}{8.3pt}\selectfont]
    \def\cs{1.3}
    \fill[lightgray] (0,0) rectangle (1.3*\cs,\cs);
    \fill[lightgray] (1.3*\cs,\cs) rectangle (2.6*\cs,2*\cs);
    \fill[gray] (0,\cs) rectangle (1.3*\cs,2*\cs);
    \fill[gray] (1.3*\cs,0) rectangle (2.6*\cs,\cs);
    \draw[black] (0,0) rectangle (2.6*\cs,2.0*\cs);
    \draw[black] (1.3*\cs,0) -- (1.3*\cs,2*\cs);
    \draw[black] (0,\cs) -- (2.6*\cs,\cs);
    \node[align=left] at (0.65*\cs, 1.5*\cs) {HD: \SI{73.24}{\percent} \\ SD: \SI{96.59}{\percent}};
    \node[align=left] at (1.95*\cs, 1.5*\cs) {HD: \SI{26.76}{\percent} \\ SD: \SI{3.41}{\percent}};
    \node[align=left] at (0.65*\cs, 0.5*\cs) {HD: \SI{15.76}{\percent} \\ SD: \SI{15.07}{\percent}};
    \node[align=left] at (1.95*\cs, 0.5*\cs) {HD: \SI{84.24}{\percent} \\ SD: \SI{84.93}{\percent}};
    \node at (0.65*\cs, -0.15) {NLoS};
    \node at (1.95*\cs, -0.15) {LoS};
    \node at (1.3*\cs,  -0.45) {Predicted};
    \node[rotate=90] at (-0.15, 1.5*\cs) {NLoS};
    \node[rotate=90] at (-0.15, 0.5*\cs) {LoS};
    \node[rotate=90] at (-0.45, \cs)     {Actual};
  \end{tikzpicture}
  \label{fig:conf_DH_FR1}
  }
  \hspace{0.2cm}
  \subfloat[InF-DH FR2.]{
  \begin{tikzpicture}[font=\fontsize{7.3pt}{8.3pt}\selectfont]
    \def\cs{1.3}
    \fill[lightgray] (0,0) rectangle (1.3*\cs,\cs);
    \fill[lightgray] (1.3*\cs,\cs) rectangle (2.6*\cs,2*\cs);
    \fill[gray] (0,\cs) rectangle (1.3*\cs,2*\cs);
    \fill[gray] (1.3*\cs,0) rectangle (2.6*\cs,\cs);
    \draw[black] (0,0) rectangle (2.6*\cs,2.0*\cs);
    \draw[black] (1.3*\cs,0) -- (1.3*\cs,2*\cs);
    \draw[black] (0,\cs) -- (2.6*\cs,\cs);
    \node[align=left] at (0.65*\cs, 1.5*\cs) {HD: \SI{73.48}{\percent} \\ SD: \SI{96.95}{\percent}};
    \node[align=left] at (1.95*\cs, 1.5*\cs) {HD: \SI{26.52}{\percent} \\ SD: \SI{3.05}{\percent}};
    \node[align=left] at (0.65*\cs, 0.5*\cs) {HD: \SI{15.03}{\percent} \\ SD: \SI{13.66}{\percent}};
    \node[align=left] at (1.95*\cs, 0.5*\cs) {HD: \SI{84.97}{\percent} \\ SD: \SI{86.34}{\percent}};
    \node at (0.65*\cs, -0.15) {NLoS};
    \node at (1.95*\cs, -0.15) {LoS};
    \node at (1.3*\cs,  -0.45) {Predicted};
    \node[rotate=90] at (-0.15, 1.5*\cs) {NLoS};
    \node[rotate=90] at (-0.15, 0.5*\cs) {LoS};
    \node[rotate=90] at (-0.45, \cs)     {Actual};
  \end{tikzpicture}
  \label{fig:conf_DH_FR2}
  }

  \caption{Confusion matrix of binary NLoS detection results where HD and SD denote HD-based and SD-based results, respectively.}
  \label{fig:conf_DH}

\vspace{-6pt}

\end{figure*}

\subsubsection{HD vs. SD}
HD relies on unweighted representative points to form $\bm r_n$ and $\bm r_n^0$ (Sec.~\ref{Sec:Factorization}), hence its score $\rho_n$ in \eqref{eq:Score} can lose discriminability when many PELs are affected by NLoS-contaminated gNBs (Remark~\ref{rem:pseudo_ref}).
SD mitigates this issue by incorporating weak site-survey priors (Assumption~\ref{Assumption: Weak Statistics}) and applying the reliability-weighted refinement in Sec.~\ref{sec:iterative_update}, which consistently improves separability.
For example, Table~\ref{tab:detection_perf} shows that the AUC increases from $\num{0.842}$ to $\num{0.926}$ in InF-DH FR1 when moving from HD to SD, together with the corresponding accuracy improvements from \SI{78.04}{\percent} to \SI{91.50}{\percent} (InF-DH FR1). 
This improvement holds in all configurations, e.g., in IOO, SD increases the AUC from $\num{0.845}$ to $\num{0.927}$ in FR1 and from $\num{0.819}$ to $\num{0.881}$ in FR2.

This gain in separability from the refinement is also reflected in Fig.~\ref{fig:conf_SH_FR1}--\ref{fig:conf_DH_FR2}, where SD reduces both the fraction of actual NLoS gNBs detected as LoS (top-right cell) and that of actual LoS gNBs detected as NLoS (bottom-left cell) in all configurations. For instance, in InF-DH FR2, the former decreases from \SI{26.52}{\percent} to \SI{3.05}{\percent} and the latter from \SI{15.03}{\percent} to \SI{13.66}{\percent}.

\subsubsection{FR1 vs. FR2}
The difference between FR1 and FR2 is largely driven by ranging quality differences in Table~\ref{tab:range_stats_grid}.
In particular, in InF-SH the LoS ranging standard deviation increases from \SI{5.9}{m} in FR1 to \SI{48.4}{m} in FR2, which increases the likelihood that noise-driven outliers produce NLoS-like score patterns in \eqref{eq:Score}.
Consistently, Table~\ref{tab:detection_perf} shows that the HD AUC decreases from $\num{0.967}$ to $\num{0.918}$ and the accuracy decreases from \SI{94.76}{\percent} to \SI{88.28}{\percent} when moving from FR1 to FR2 in InF-SH. 
This trend is also reflected in Fig.~\ref{fig:conf_SH_FR1}--\ref{fig:conf_SH_FR2}, where the fraction of actual LoS gNBs excluded as NLoS (bottom-left cell) increases from \SI{4.29}{\percent} to \SI{10.50}{\percent} in HD.

In InF-DH, NLoS bias effects are dominant in both FR1 and FR2 (Table~\ref{tab:range_stats_grid}), and SD reduces NLoS misses via score refinement-based PEL re-weighting.
Accordingly, Table~\ref{tab:detection_perf} shows that SD maintains high AUC values around $\num{0.93}$ for both FR1 and FR2, and Fig.~\ref{fig:conf_DH_FR1}--\ref{fig:conf_DH_FR2} shows that the fractions of actual NLoS gNBs detected as LoS remain around a few percent under SD.

\subsubsection{InF-SH vs. IOO vs. InF-DH}
In InF-SH, the NLoS ratio $\pi$ is relatively low. Consequently, the score $\rho_n$ in \eqref{eq:Score} already captures the underlying geometry reasonably well even without SD refinement. 
For instance, Table~\ref{tab:detection_perf} reports high AUC values under HD, namely $\num{0.967}$ in FR1 and $\num{0.918}$ in FR2, together with accuracies of \SI{94.76}{\percent} and \SI{88.28}{\percent}, respectively. 
In this LoS-dominant regime, SD provides an additional but moderate gain, e.g., the AUC increases to $\num{0.979}$ and $\num{0.939}$ in FR1 and FR2, respectively (Table~\ref{tab:detection_perf}), which mainly corrects borderline cases through reliability-aware refinement.

On the other hand, InF-DH has a higher NLoS ratio $\pi$ than InF-SH, so a substantial fraction of PELs are generated from subsets containing NLoS-contaminated gNBs, making the representative points and the resulting score $\rho_n$ more vulnerable to bias (Remark~\ref{rem:pseudo_ref}). 
This is reflected in Table~\ref{tab:detection_perf}, where the HD AUC drops to $\num{0.842}$ in FR1 and $\num{0.836}$ in FR2, with accuracies of \SI{78.04}{\percent} and \SI{78.54}{\percent}. 
In this NLoS-dominant regime, SD becomes particularly effective: Table~\ref{tab:detection_perf} shows that the AUC increases to $\num{0.926}$ in FR1 and $\num{0.933}$ in FR2, and Fig.~\ref{fig:conf_DH_FR1}--\ref{fig:conf_DH_FR2} confirms that the fractions of actual NLoS gNBs detected as LoS are reduced from about \SI{27}{\percent} in HD to about \SI{3}{\percent} in SD.

\begin{table}[!t]
\centering
\setlength{\tabcolsep}{3pt}
\fontsize{7.2pt}{8.5pt}\selectfont
\caption{Characteristics of the range measurements. The range error is $e_n = d_n - \|\bm z_n - \bm p\|$, and $\sigma_{\mathrm r}$ is the normalized median absolute deviation in \cite{rousseeuw1993MAD}, computed only from the range errors of the LoS measurements ($\gamma_n = 0$). Ranges with $e_n > 3\sigma_{\mathrm r}$ are NLoS-like \cite{pearson2002hampel}, and the others are LoS-like.}

\vspace{-4pt}

\label{tab:app_dataset}
\begin{tabular}{lcccccc}
\toprule
\multirow{2}{*}{Metric} & \multicolumn{2}{c}{InF-SH} & \multicolumn{2}{c}{IOO} & \multicolumn{2}{c}{InF-DH} \\
\cmidrule(lr){2-3}\cmidrule(lr){4-5}\cmidrule(lr){6-7}
 & FR1 & FR2 & FR1 & FR2 & FR1 & FR2 \\
\midrule
$\sigma_{\mathrm r}$ [m] & \num{0.95} & \num{1.07} & \num{0.92} & \num{0.94} & \num{1.08} & \num{1.09} \\
\midrule
$\mathsf{Pr}(e_n < 3\sigma_{\mathrm r} \mid \gamma_n = 1)$ [\%] & \num{2.1} & \num{1.5} & \num{29.0} & \num{42.1} & \num{2.2} & \num{2.0} \\
$\mathsf{Pr}(\gamma_n = 1 \mid e_n > 3\sigma_{\mathrm r})$ [\%] & \num{85.3} & \num{63.9} & \num{97.5} & \num{93.3} & \num{90.0} & \num{91.0} \\
\bottomrule
\end{tabular}

\end{table}

Although IOO has a lower $\pi$ than InF-DH (Table~\ref{tab:range_stats_grid}), its AUCs are comparable to or lower than those in InF-DH, and its recalls are the lowest among the three scenarios in both FR1 and FR2, e.g., \SI{64.13}{\percent} and \SI{52.04}{\percent} for HD (Table~\ref{tab:detection_perf}). Two factors contribute to this degradation. IOO has fewer gNBs than the InF scenarios ($N=12$ against $18$). Moreover, as shown in Table~\ref{tab:app_dataset}, \SI{29.0}{\percent} and \SI{42.1}{\percent} of its NLoS ranges in FR1 and FR2 are LoS-like, far more than in InF-SH and InF-DH. Since these NLoS ranges have small errors, they are hardly distinguishable from the LoS ranges by the PEL geometry. On the other hand, the precisions in IOO remain high, e.g., \SI{89.64}{\percent} and \SI{92.29}{\percent} for HD and SD in FR1.

\subsubsection{CDA-ND vs. Benchmarks}\label{sec:sim_bench}
{
As shown in Table~\ref{tab:detection_perf}, among the methods without training, HD attains accuracies and AUCs comparable to those of LMedS in the LoS-dominant InF-SH and IOO.
In the NLoS-dominant InF-DH ($\pi=0.56$), however, the AUCs of LMedS decrease to $\num{0.725}$ and $\num{0.753}$ in FR1 and FR2, below those of HD, $\num{0.842}$ and $\num{0.836}$, as LMedS is reported to degrade when more than half of the gNBs are NLoS~\cite{rousseeuw1987robust}.
HD is more robust than LMedS to a high NLoS ratio.
EM attains accuracies similar to those of HD except in IOO FR2, but lower AUCs in all configurations, e.g., $\num{0.672}$ against $\num{0.819}$ in IOO FR2.

The RF and MLP classifiers in Table~\ref{tab:detection_perf} are trained with gNB-wise LoS/NLoS labels, whereas SD uses only the weak site-survey priors in Assumption~\ref{Assumption: Weak Statistics}.
Nevertheless, Table~\ref{tab:detection_perf} shows that SD attains accuracies at most \num{1.62} percentage points below those of the RF and MLP classifiers in InF-SH and IOO and \num{2.75} to \num{2.92} percentage points higher in InF-DH.
}

\begin{table}[!t]
\centering
\setlength{\tabcolsep}{5pt}
\fontsize{7.2pt}{7.8pt}\selectfont
\caption{Brier score and ECE of the NLoS probability.}

\label{tab:app_prob}
\begin{tabular}{llcccc}
\toprule
\multirow{2}{*}{Scenario} & \multirow{2}{*}{Method} & \multicolumn{2}{c}{FR1} & \multicolumn{2}{c}{FR2} \\
\cmidrule(lr){3-4}\cmidrule(lr){5-6}
& & Brier & ECE & Brier & ECE \\
\midrule
\multirow{4}{*}{InF-SH} & RF classifier & \num{0.019} & \num{0.005} & \num{0.060} & \num{0.008} \\
 & MLP classifier & \num{0.019} & \num{0.007} & \num{0.059} & \num{0.005} \\
\cmidrule(l){2-6}
 & Before refinement & \num{0.038} & \num{0.025} & \num{0.098} & \num{0.090} \\
 & \textbf{Soft Decision} & \num{0.030} & \num{0.016} & \num{0.078} & \num{0.050} \\
\midrule
\multirow{4}{*}{IOO} & RF classifier & \num{0.075} & \num{0.011} & \num{0.106} & \num{0.013} \\
 & MLP classifier & \num{0.073} & \num{0.008} & \num{0.114} & \num{0.023} \\
\cmidrule(l){2-6}
 & Before refinement & \num{0.118} & \num{0.025} & \num{0.147} & \num{0.054} \\
 & \textbf{Soft Decision} & \num{0.077} & \num{0.024} & \num{0.116} & \num{0.064} \\
\midrule
\multirow{4}{*}{InF-DH} & RF classifier & \num{0.082} & \num{0.006} & \num{0.080} & \num{0.007} \\
 & MLP classifier & \num{0.082} & \num{0.007} & \num{0.079} & \num{0.009} \\
\cmidrule(l){2-6}
 & Before refinement & \num{0.158} & \num{0.056} & \num{0.157} & \num{0.051} \\
 & \textbf{Soft Decision} & \num{0.075} & \num{0.027} & \num{0.069} & \num{0.023} \\
\bottomrule
\end{tabular}

\vspace{-9pt}
\end{table}

\subsubsection{Reliability of the NLoS Probability}
\label{sec:sim_calib}

Table~\ref{tab:app_prob} lists the Brier score \cite{brier1950verification} and the \textit{expected calibration error} (ECE) with ten bins \cite{guo2017calibration}, which measure the reliability of the NLoS probability.
It compares the NLoS probabilities before refinement, $h^*(\rho_n)$, and the final SD values $\psi^*_n$ in \eqref{eq:final_SD_Val}, $n\in\mathcal{N}$, with the benchmarks.
The score refinement and the SD function revision in Secs.~\ref{sec:iterative_update} and \ref{sec:refine_reest} reduce the Brier score in all configurations, and SD achieves ECEs of at most $\num{0.064}$.
The RF and MLP classifiers achieve lower ECEs, while SD attains comparable Brier scores, within $\num{0.02}$ of theirs.

\vspace{-5pt}

\subsection{Positioning Performance}\label{sec:sim_positioning}
\begin{figure*}[t!]
    \centering
    \subfloat[HD-based positioning in InF-SH scenario.]{%
        \includegraphics[height=134bp]{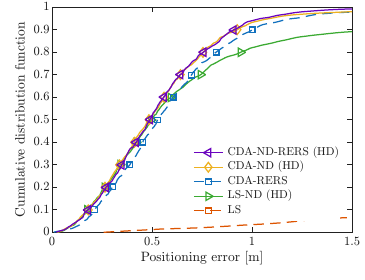}\label{fig:SHFR1_HD}%
    }
    \hfill
    \subfloat[HD-based positioning in IOO scenario.]{%
        \includegraphics[height=134bp]{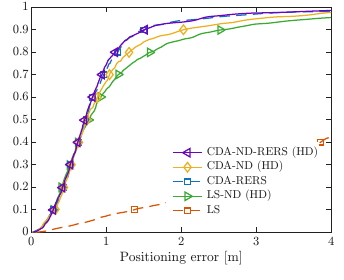}\label{fig:IOOFR1_HD}%
    }
    \hfill
    \subfloat[HD-based positioning in InF-DH scenario.]{%
        \includegraphics[height=134bp]{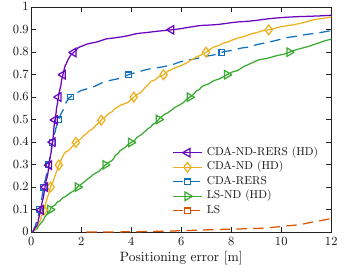}\label{fig:DHFR1_HD}%
    }
    \\
    \subfloat[SD-based positioning in InF-SH scenario.]{%
        \includegraphics[height=134bp]{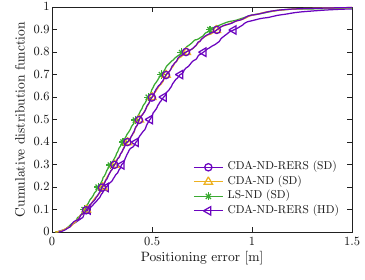}\label{fig:SHFR1_SD}%
    }
    \hfill
    \subfloat[SD-based positioning in IOO scenario.]{%
        \includegraphics[height=134bp]{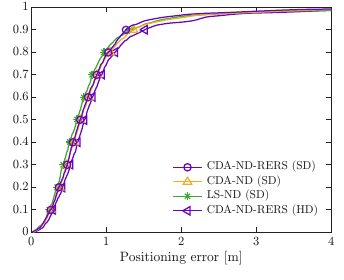}\label{fig:IOOFR1_SD}%
    }
    \hfill
    \subfloat[SD-based positioning in InF-DH scenario.]{%
        \includegraphics[height=134bp]{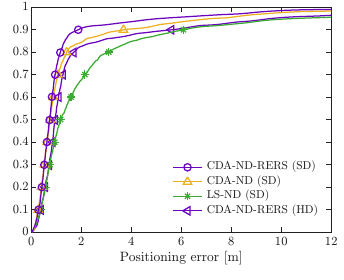}\label{fig:DHFR1_SD}%
    }
    \caption{Positioning performance under FR1. The performance is reported in terms of CDF of the positioning error.}
    \label{fig:FR1_CDF}

    \vspace{-4pt}
\end{figure*}

\begin{figure*}[t!]
    \centering
    \subfloat[HD-based positioning in InF-SH scenario.]{%
        \includegraphics[height=134bp]{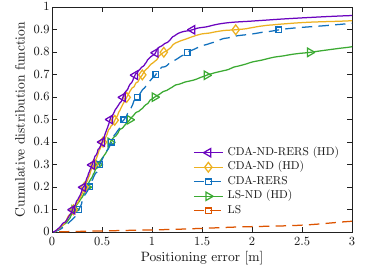}\label{fig:SHFR2_HD}%
    }
    \hfill
    \subfloat[HD-based positioning in IOO scenario.]{%
        \includegraphics[height=134bp]{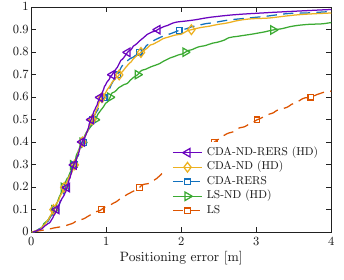}\label{fig:IOOFR2_HD}%
    }
    \hfill
    \subfloat[HD-based positioning in InF-DH scenario.]{%
        \includegraphics[height=134bp]{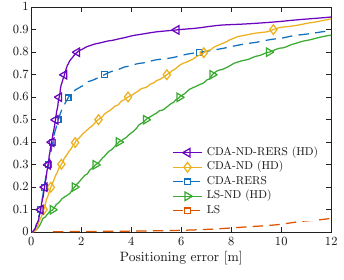}\label{fig:DHFR2_HD}%
    }
    \\
    \subfloat[SD-based positioning in InF-SH scenario.]{%
        \includegraphics[height=134bp]{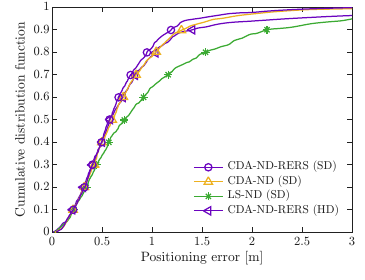}\label{fig:SHFR2_SD}%
    }
    \hfill
    \subfloat[SD-based positioning in IOO scenario.]{%
        \includegraphics[height=134bp]{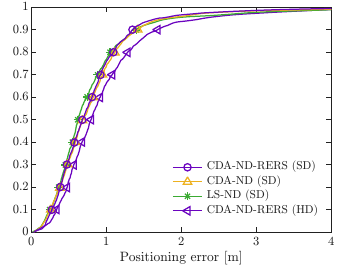}\label{fig:IOOFR2_SD}%
    }
    \hfill
    \subfloat[SD-based positioning in InF-DH scenario.]{%
        \includegraphics[height=134bp]{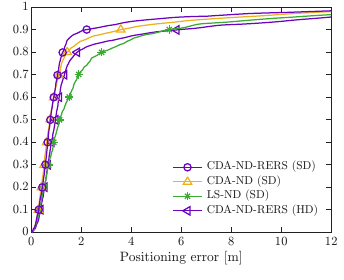}\label{fig:DHFR2_SD}%
    }
    \caption{Positioning performance under FR2. The performance is reported in terms of CDF of the positioning error.}
    \label{fig:FR2_CDF}

    \vspace{-4pt}

\end{figure*}

Positioning performance is evaluated in terms of the 2D error. Fig.~\ref{fig:FR1_CDF} and Fig.~\ref{fig:FR2_CDF} show the \textit{cumulative distribution functions} (CDFs) of positioning error for FR1 and FR2, respectively, while Table~\ref{tab:positioning_perf} reports the mean, standard deviation, and 95th percentile errors. Appendix~\ref{app:positioning} extends Table~\ref{tab:positioning_perf} with the percentiles of the positioning error from the 10th to the 99th. 

The compared methods are grouped as follows.

\noindent\textit{Benchmarks without training:}
\begin{itemize}
    \item \textit{Least-squares (LS):} Standard multilateration using all gNB measurements in $\mathcal{N}$ \cite{Mensing2006LM}.
    \item \textit{Residual weighting} (RWGH) \cite{chen1999RWGH}: Weighted average of the LS position estimates from all subsets of at least three gNBs, with weights inversely proportional to their normalized residuals.
    \item \textit{LMedS} \cite{casas2006LMedS}: The same as the LMedS benchmark for NLoS detection in Sec.~\ref{sec:sim_identification}.
    \item \textit{EM} \cite{yin2014EM}: The same as the EM benchmark for NLoS detection in Sec.~\ref{sec:sim_identification}.
    \item \textit{Squared-range-based maximum correntropy criterion} (SR-MCC) \cite{xiong2021MCC}: LS on the squared ranges, robustified by the maximum correntropy criterion that down-weights the ranges with large errors.
    \item \textit{CDA-RERS} \cite{yu2025combinatorial}: RE\&RS filtering in \eqref{eq:RE_metric}--\eqref{eq:RS_metric} applied to the full PEL set $\mathcal{X}$, followed by $\mathsf{med}(\cdot)$ of the retained PELs.
\end{itemize}

\noindent\textit{Benchmarks trained with ground-truth (GT) positions:}
\begin{itemize}
    \item \textit{Soft information} (SI) \cite{Morselli2023SI}: The joint distribution of the measured and true ranges is learned by a GMM from the GT positions, which gives the likelihood of each range measurement, and the UE position is estimated by maximum likelihood.
    \item \textit{RF and MLP}: The range error of each gNB is predicted from the features in Appendix~\ref{app:bench_pos}, and the corrected ranges are used for LS weighted by the predicted error magnitude.
\end{itemize}

\noindent\textit{CDA-ND-based methods:} The proposed CDA-ND-RERS for both HD and SD, and their variants.
\begin{itemize}
    \item \textit{LS-ND (HD):} LS applied to the LoS-identified gNB subset $\mathcal{N}_{\mathrm{HD}}$ in \eqref{eq:N_ND}.
    \item \textit{CDA-ND (HD):} Unweighted median over $\mathcal{X}_{\mathrm{HD}}$ in \eqref{eq:PEL_HD}.
    \item \textit{LS-ND (SD):} Weighted LS applied to $\mathcal{N}_{\mathrm{HD}}$ with per-gNB weight $(1-\psi_n^*)$, where $\psi_n^*$ is given in \eqref{eq:final_SD_Val}.
    \item \textit{CDA-ND (SD)}: Reliability-aware weighted median over $\mathcal{X}_{\mathrm{HD}}$ using $\omega^{*(\ell)}$ in \eqref{eq:reliability}.
    \item \textit{CDA-ND-RERS (HD):} Form $\mathcal{X}_{\mathrm{HD}}$ in \eqref{eq:PEL_HD}, apply RE\&RS filtering to obtain $\mathcal{X}_{\mathrm{RS}}$ in \eqref{eq: RS set}, and compute $\mathsf{med}(\mathcal{X}_{\mathrm{RS}})$, yielding $\hat{\bm{p}}_{\mathrm{HD}}$ in \eqref{eq:proposed}.
    \item \textit{CDA-ND-RERS (SD):} Same as CDA-ND-RERS (HD), but replacing $\mathsf{med}(\mathcal{X}_{\mathrm{RS}})$ with the reliability-aware weighted median over $\mathcal{X}_{\mathrm{RS}}$, yielding $\hat{\bm{p}}_{\mathrm{SD}}$ in \eqref{eq:SD_positioning}.
\end{itemize}

\begin{table}[t]
\centering
\setlength{\tabcolsep}{3pt}
\fontsize{7.2pt}{7.8pt}\selectfont
\caption{Hyperparameter settings for RE\&RS filtering.}

\label{tab:hyperparams_rers}
\begin{tabular}{llcccccc}
\toprule
\multirow{2}{*}{Method} & \multirow{2}{*}{Parameter} &
\multicolumn{2}{c}{InF-SH} &
\multicolumn{2}{c}{IOO} &
\multicolumn{2}{c}{InF-DH} \\
\cmidrule(lr){3-4}\cmidrule(lr){5-6}\cmidrule(lr){7-8}
& & FR1 & FR2 & FR1 & FR2 & FR1 & FR2 \\
\midrule
\multirow{2}{*}{CDA-RERS}
& ${\ell_{\mathrm{RE}}}/{L}$ & \num{0.63} & \num{0.53} & \num{0.34} & \num{0.50} & \num{0.15} & \num{0.15} \\
& ${\ell_{\mathrm{RS}}}/{L}$ & \num{0.36} & \num{0.26} & \num{0.12} & \num{0.37} & \num{0.08} & \num{0.08} \\
\midrule
\multirow{2}{*}{CDA-ND-RERS (HD)}
& ${\ell_{\mathrm{RE}}}/{L}$ & \num{0.88} & \num{0.85} & \num{0.85} & \num{0.70} & \num{0.14} & \num{0.34} \\
& ${\ell_{\mathrm{RS}}}/{L}$ & \num{0.83} & \num{0.75} & \num{0.26} & \num{0.35} & \num{0.08} & \num{0.09} \\
\midrule
\multirow{2}{*}{CDA-ND-RERS (SD)}
& ${\ell_{\mathrm{RE}}}/{L}$ & \num{1.00} & \num{0.68} & \num{0.79} & \num{0.86} & \num{0.32} & \num{0.32} \\
& ${\ell_{\mathrm{RS}}}/{L}$ & \num{0.98} & \num{0.64} & \num{0.29} & \num{0.51} & \num{0.20} & \num{0.12} \\
\bottomrule
\end{tabular}

\vspace{-4pt}

\end{table}

The RE\&RS retained ratios (reported as $\ell_{\mathrm{RE}}/L$ and $\ell_{\mathrm{RS}}/L$) are given in Table~\ref{tab:hyperparams_rers}.
For the SD functions, we use the GMM in \eqref{eq:GMM_rep} with $K=8$; the iterative refinement in Sec.~\ref{sec:iterative_update} uses $\epsilon=10^{-3}$ and $T_{\max}=25$.

\begin{table*}[t]
\centering
\setlength{\tabcolsep}{3.1pt}
\fontsize{7.2pt}{7.9pt}\selectfont
\caption{Positioning error statistics (\si{m}) for each scenario--frequency configuration. The proposed methods are highlighted in bold.}

\vspace{-3pt}

\label{tab:positioning_perf}
\begin{tabular}{lcccccccccccccccccccc}
\toprule
\multicolumn{1}{l}{\multirow{3}{*}{Method}} &
\multirow{3}{*}{\shortstack{RTT\\dataset}} & \multirow{3}{*}{Labeling} &
\multicolumn{6}{c}{InF-SH} &
\multicolumn{6}{c}{IOO} &
\multicolumn{6}{c}{InF-DH} \\
\cmidrule(lr){4-9}\cmidrule(lr){10-15}\cmidrule(lr){16-21}
\multicolumn{1}{c}{} & & &
\multicolumn{3}{c}{FR1} & \multicolumn{3}{c}{FR2} &
\multicolumn{3}{c}{FR1} & \multicolumn{3}{c}{FR2} &
\multicolumn{3}{c}{FR1} & \multicolumn{3}{c}{FR2} \\
\cmidrule(lr){4-6}\cmidrule(lr){7-9}\cmidrule(lr){10-12}\cmidrule(lr){13-15}\cmidrule(lr){16-18}\cmidrule(lr){19-21}
\multicolumn{1}{c}{} & & &
Mean & Std & 95th &
Mean & Std & 95th &
Mean & Std & 95th &
Mean & Std & 95th &
Mean & Std & 95th &
Mean & Std & 95th \\
\midrule
LS& -- & -- &
\num{8.07} & \num{5.53} & \num{18.0} &
\num{20.4} & \num{40.4} & \num{69.5} &
\num{5.19} & \num{3.57} & \num{11.7} &
\num{28.2} & \num{59.5} & \num{170} &
\num{23.1} & \num{9.24} & \num{38.8} &
\num{22.7} & \num{7.98} & \num{36.1} \\
RWGH \cite{chen1999RWGH}& -- & -- &
\num{1.59} & \num{3.95} & \num{4.25} &
\num{2.61} & \num{6.64} & \num{10.4} &
\num{2.50} & \num{4.22} & \num{8.71} &
\num{2.13} & \num{3.39} & \num{6.13} &
\num{12.2} & \num{13.2} & \num{36.6} &
\num{11.7} & \num{13.3} & \num{36.4} \\
LMedS \cite{casas2006LMedS}& -- & -- &
\num{0.80} & \num{1.01} & \num{1.49} &
\num{0.98} & \num{1.38} & \num{1.99} &
\num{1.50} & \num{3.78} & \num{3.34} &
\num{1.31} & \num{1.33} & \num{3.70} &
\num{12.5} & \num{13.4} & \num{40.3} &
\num{11.2} & \num{12.4} & \num{34.5} \\
EM \cite{yin2014EM}& -- & -- &
\num{0.53} & \num{0.64} & \num{1.07} &
\num{1.57} & \num{4.02} & \num{8.74} &
\num{1.39} & \num{2.97} & \num{4.29} &
\num{1.74} & \num{2.71} & \num{5.35} &
\num{7.47} & \num{10.4} & \num{26.8} &
\num{7.11} & \num{10.3} & \num{25.6} \\
SR-MCC \cite{xiong2021MCC}& -- & -- &
\num{1.26} & \num{3.11} & \num{2.25} &
\num{2.71} & \num{4.99} & \num{14.1} &
\num{2.90} & \num{4.96} & \num{11.9} &
\num{4.29} & \num{33.8} & \num{5.93} &
\num{16.9} & \num{12.1} & \num{40.1} &
\num{16.1} & \num{10.6} & \num{34.8} \\
CDA-RERS \cite{yu2025combinatorial}& -- & -- &
\num{0.60} & \num{0.45} & \num{1.21} &
\num{1.19} & \num{1.83} & \num{3.76} &
\num{0.95} & \num{1.27} & \num{2.32} &
\num{1.06} & \num{0.89} & \num{2.83} &
\num{3.97} & \num{5.51} & \num{16.1} &
\num{3.73} & \num{5.42} & \num{16.3} \\
\midrule
SI \cite{Morselli2023SI}& \checkmark & GT pos. &
\num{0.38} & \num{0.22} & \num{0.76} &
\num{0.58} & \num{0.49} & \num{1.39} &
\num{0.68} & \num{0.71} & \num{1.47} &
\num{0.72} & \num{0.54} & \num{1.59} &
\num{1.08} & \num{2.66} & \num{2.23} &
\num{1.33} & \num{3.69} & \num{2.92} \\
MLP& \checkmark & GT pos. &
\num{0.62} & \num{0.50} & \num{1.18} &
\num{0.80} & \num{0.91} & \num{1.67} &
\num{0.94} & \num{1.15} & \num{2.27} &
\num{1.16} & \num{1.15} & \num{3.06} &
\num{3.35} & \num{3.94} & \num{10.7} &
\num{2.80} & \num{4.31} & \num{9.60} \\
RF& \checkmark & GT pos. &
\num{0.48} & \num{0.35} & \num{0.98} &
\num{0.63} & \num{0.61} & \num{1.36} &
\num{0.93} & \num{1.45} & \num{2.18} &
\num{0.98} & \num{1.59} & \num{2.26} &
\num{3.19} & \num{3.75} & \num{10.7} &
\num{2.66} & \num{4.09} & \num{8.60} \\
\midrule
LS-ND (HD)& -- & -- &
\num{0.85} & \num{1.53} & \num{2.90} &
\num{2.06} & \num{3.84} & \num{8.25} &
\num{1.22} & \num{1.59} & \num{3.82} &
\num{1.84} & \num{5.44} & \num{4.57} &
\num{7.09} & \num{7.90} & \num{21.4} &
\num{6.28} & \num{6.47} & \num{16.8} \\
CDA-ND (HD)& -- & -- &
\num{0.55} & \num{0.40} & \num{1.09} &
\num{1.05} & \num{1.72} & \num{3.75} &
\num{1.05} & \num{1.19} & \num{3.04} &
\num{1.10} & \num{1.08} & \num{3.06} &
\num{4.09} & \num{4.03} & \num{11.7} &
\num{4.13} & \num{4.48} & \num{12.1} \\
\textbf{CDA-ND-RERS (HD)}& -- & -- &
\num{0.53} & \num{0.33} & \num{1.05} &
\num{0.89} & \num{1.32} & \num{2.38} &
\num{0.90} & \num{0.91} & \num{2.25} &
\num{0.96} & \num{0.71} & \num{2.31} &
\num{2.12} & \num{3.64} & \num{9.60} &
\num{2.19} & \num{3.62} & \num{11.4} \\
\midrule
LS-ND (SD)& \checkmark & $\pi$ only &
\num{0.48} & \num{0.72} & \num{0.95} &
\num{1.11} & \num{2.34} & \num{3.01} &
\num{0.80} & \num{1.13} & \num{1.86} &
\num{0.83} & \num{0.94} & \num{1.86} &
\num{2.93} & \num{6.30} & \num{10.7} &
\num{2.45} & \num{4.42} & \num{9.68} \\
CDA-ND (SD)& \checkmark & $\pi$ only &
\num{0.47} & \num{0.26} & \num{0.96} &
\num{0.72} & \num{0.52} & \num{1.69} &
\num{0.83} & \num{0.96} & \num{1.96} &
\num{0.86} & \num{1.09} & \num{1.80} &
\num{1.64} & \num{2.95} & \num{7.18} &
\num{1.67} & \num{3.50} & \num{7.41} \\
\textbf{CDA-ND-RERS (SD)}& \checkmark & $\pi$ only &
\num{0.47} & \num{0.25} & \num{0.96} &
\num{0.67} & \num{0.45} & \num{1.46} &
\num{0.79} & \num{0.79} & \num{1.66} &
\num{0.82} & \num{0.94} & \num{1.67} &
\num{1.30} & \num{2.37} & \num{5.15} &
\num{1.42} & \num{2.62} & \num{5.26} \\
\bottomrule
\end{tabular}

\vspace{-4pt}

\end{table*}

\subsubsection{Analysis of CDA-ND-Based Methods}\label{sec:sim_pos_variants}
Fig.~\ref{fig:FR1_CDF} reports the FR1 results. 
In the LoS-dominant InF-SH case, Figs.~\ref{fig:SHFR1_HD} and~\ref{fig:SHFR1_SD} show that CDA-based estimators concentrate in the sub-meter regime, whereas LS-ND (HD) exhibits a pronounced right shift with a long tail. 
Table~\ref{tab:positioning_perf} quantifies this contrast: LS has a MAE of \SI{8.07}{m}, while CDA-RERS reduces it to \SI{0.60}{m} and the proposed CDA-ND-RERS for both HD and SD reaches \SI{0.53}{m} and \SI{0.47}{m}, respectively. 
This small HD--SD gap (\SI{0.53}{m} vs.\ \SI{0.47}{m}) matches the NLoS-detection reliability in Sec.~\ref{sec:sim_identification}: in InF-SH FR1, HD already attains accuracy \SI{94.76}{\percent} and AUC \num{0.967}, and SD increases them to \SI{96.76}{\percent} and \num{0.979} in Table~\ref{tab:detection_perf}. 
This behavior is also consistent with the PEL stabilization trend in Fig.~\ref{fig:MMD}, which is faster in InF-SH, reducing the need for additional SD-based refinement beyond gNB-level exclusion of HD.

In the NLoS-dominant InF-DH scenario under FR1, Fig.~\ref{fig:DHFR1_HD} shows that LS remains unreliable even after the HD-based gNB-level exclusion, and that combining ND with PEL-level RE\&RS filtering yields a substantial improvement. 
Specifically, Table~\ref{tab:positioning_perf} shows that LS has a MAE of \SI{23.1}{m} and LS-ND (HD) still yields \SI{7.09}{m}, whereas CDA-ND-RERS (HD) reduces it to \SI{2.12}{m}. 
Fig.~\ref{fig:DHFR1_SD} further indicates a pronounced tail improvement under SD-based processing; correspondingly, the 95th percentile decreases from \SI{9.60}{m} to \SI{5.15}{m} and the MAE decreases from \SI{2.12}{m} to \SI{1.30}{m} when moving from CDA-ND-RERS (HD) to CDA-ND-RERS (SD) in Table~\ref{tab:positioning_perf}. 
This gain is consistent with the detection results in InF-DH (Table~\ref{tab:detection_perf}).

Fig.~\ref{fig:FR2_CDF} reports the FR2 results. 
In InF-SH FR2, Fig.~\ref{fig:SHFR2_HD} shows a larger performance degradation than in FR1 (Fig.~\ref{fig:SHFR1_HD}). 
Table~\ref{tab:positioning_perf} summarizes the gap for the proposed method: CDA-ND-RERS (HD) increases from \SI{0.53}{m} to \SI{0.89}{m}, whereas it increases from \SI{0.47}{m} to \SI{0.67}{m} in CDA-ND-RERS (SD). 
This is consistent with NLoS detection performance (HD AUC \num{0.967} $\rightarrow$ \num{0.918} in Table~\ref{tab:detection_perf}) and ranging statistics (LoS standard deviation \SI{5.9}{m} $\rightarrow$ \SI{48.4}{m} in Table~\ref{tab:range_stats_grid}).

In InF-DH, the FR1/FR2 gap is modest. 
This is also reflected in Fig.~\ref{fig:FR1_CDF}--Fig.~\ref{fig:FR2_CDF}, where the InF-DH SD-based CDFs in Fig.~\ref{fig:DHFR1_SD} and Fig.~\ref{fig:DHFR2_SD} exhibit similar shapes. 
Numerically, CDA-ND-RERS (SD) achieves MAEs of \SI{1.30}{m} in FR1 and \SI{1.42}{m} in FR2, with 95th percentiles of \SI{5.15}{m} and \SI{5.26}{m}, respectively (Table~\ref{tab:positioning_perf}).

In IOO, CDA-ND-RERS for both HD and SD achieves MAEs of \SI{0.90}{m} and \SI{0.79}{m}, respectively, in FR1. Although the LoS-like NLoS ranges in Table~\ref{tab:app_dataset} are hard to detect, they degrade the positioning only slightly owing to their small errors. In IOO FR2, where the NLoS range errors have a heavy tail (Table~\ref{tab:range_stats_grid}), LS yields an MAE of \SI{28.2}{m}, whereas CDA-ND-RERS for both HD and SD yields \SI{0.96}{m} and \SI{0.82}{m}, respectively. Moreover, the gain of SD over HD tends to increase with $\pi$, e.g., from \SI{0.06}{m} in InF-SH to \SI{0.11}{m} in IOO and \SI{0.82}{m} in InF-DH under FR1.

Finally, Fig.~\ref{fig:FR1_CDF}--Fig.~\ref{fig:FR2_CDF} show that gNB-level exclusion and PEL-level RE\&RS filtering are complementary, and that SD further improves robustness by applying the PEL-weighted median in \eqref{eq:SD_positioning}.

\subsubsection{CDA-ND vs. Benchmarks}\label{sec:sim_pos_bench}
As shown in Table~\ref{tab:positioning_perf}, among the methods without training, CDA-ND-RERS (HD) achieves MAEs lower than or equal to those of the others in all configurations. In the NLoS-dominant InF-DH, RWGH, LMedS, EM, and SR-MCC yield MAEs from \SI{7.47}{m} to \SI{16.9}{m} in FR1, whereas CDA-ND-RERS (HD) yields \SI{2.12}{m}. CDA-ND remains robust in this environment by combining the gNB-level exclusion and the PEL-level RE\&RS filtering.

Among the methods trained with GT positions, SI achieves the lowest MAEs in all configurations. Although CDA-ND-RERS (SD) requires no GT positions, its MAEs are comparable to those of SI, higher by only \SI{0.09}{m} to \SI{0.22}{m} in all configurations, and lower than those of RF and MLP in most configurations. This suggests that the NLoS likelihood score and the iterative score refinement of CDA-ND largely narrow the gap to the methods trained with GT positions, using only the $\pi$ and score samples without labels.

\begin{figure}[!t]
\centering
\includegraphics[width=\columnwidth]{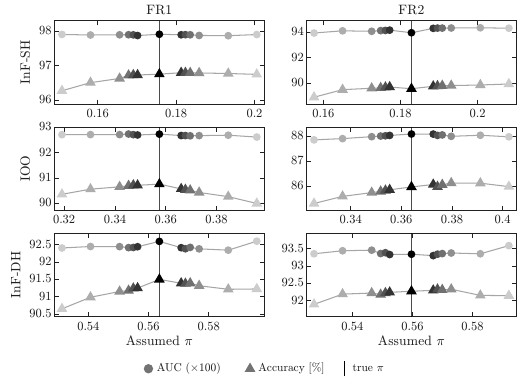}

\vspace{-8pt}

\caption{AUC and accuracy of SD under the assumed prior $\pi \pm \delta(V)$, $V \in \{50, 100, 300, 500, 700, 900\}$. Darker markers correspond to larger $V$.}
\label{fig:sens_prior}

\vspace{-6pt}

\end{figure}

\subsection{Sensitivity to the NLoS Prior}\label{sec:sim_sens}
{
We construct the SD functions $h^*$ and $\tilde h$ with the assumed prior $\pi \pm \delta(V)$ in place of $\pi$, where $\delta(V) = {Q^{-1}\big((1-0.95)/2\big)}\sqrt{\pi(1-\pi)/(NV)}$ is the half-width of the approximate \SI{95}{\percent} confidence interval of $\pi$ estimated from the $NV$ measurements at $V$ survey locations, $Q^{-1}(\cdot)$ is the inverse Q-function, and $V \in \{50, 100, 300, 500, 700, 900\}$.

As shown in Fig.~\ref{fig:sens_prior}, the AUC, which measures the separability of the SD values independently of the decision threshold, is nearly unchanged for all $V$. Although the assumed $\pi$ shifts the decision threshold of SD and trades recall for precision, the accuracy varies little, because few SD values lie near the decision threshold owing to the high separability. The accuracy of SD decreases by at most $\num{0.85}$ percentage points, in InF-DH FR1 at $\pi - \delta(V)$ with $V = 50$, and remains above that of HD in all configurations.
}

\section{Concluding Remarks}

This work has developed CDA-ND, a novel NLoS detection algorithm that exploits discriminative geometric signatures manifested in CDA-generated PELs, which are constructed solely from real-time gNB-UE range measurements. 
These signatures are compactly captured by a single vector, termed NEV, which enables a tractable HD detector that operates online. When weak site-survey priors are available, the framework naturally extends to the SD mode that quantifies posterior NLoS probabilities. Both HD and SD outputs are seamlessly integrated into CDA-based positioning algorithms, whose effectiveness is extensively validated through simulation studies on a 3GPP-compliant dataset.

The current work can be extended in several directions. First, although this work focuses on positioning, the CDA-ND features can benefit a broader range of 6G tasks, such as LoS/NLoS-aware beamforming and link adaptation. 
Second, gNBs identified as NLoS implicitly provide hints on dominant reflections. Leveraging these hints to infer coarse reflection-point locations could complement vision-based sensing and help mitigate drift. 
Last, integrating CDA-ND with emerging 6G technologies, such as MA and RIS, will be another interesting direction.

\bibliography{bibtex/IEEEabrv, references}

@STRING{IEEE_J_VT         = "{IEEE} Trans. Veh. Technol."}

@STRING{IEEE_J_STSP       = "{IEEE} J. Sel. Topics Signal Process."}

@STRING{IEEE_J_COML       = "{IEEE} Commun. Lett."}

@STRING{IEEE_J_JSAC       = "{IEEE} J. Sel. Areas Commun."}

@STRING{IEEE_J_COM        = "{IEEE} Trans. Commun."}

@STRING{IEEE_J_WCOM       = "{IEEE} Trans. Wireless Commun."}

@STRING{IEEE_J_WCOML      = "{IEEE} Wireless Commun. Lett."}

@STRING{IEEE_J_IOT        = "{IEEE} Internet Things J."}

@STRING{IEEE_J_MC         = "{IEEE} Trans. Mobile Comput."}

@STRING{IEEE_J_IM         = "{IEEE} Trans. Instrum. Meas."}

@STRING{IEEE_J_SENSOR     = "{IEEE} Sensors J."}

@STRING{IEEE_O_CSTO       = "{IEEE} Commun. Surveys Tuts."}

@IEEEtranBSTCTL{IEEEtranBSTCTL,
  CTLuse_forced_etal       = "yes",
  CTLmax_names_forced_etal = "6"
}

@inproceedings{Kim2026CDAND,
  author    = {Sang-Hyeok Kim and Seung Min Yu and Jihong Park and Seung-Woo Ko},
  title     = {On-the-fly {NLoS} detection for wireless positioning: {C}ombinatorial data augmentation approach},
  booktitle = {Proc. IEEE Wireless Commun. Netw. Conf.},
  year      = {2026},
  pages     = {1--6}
}

@ARTICLE{Chen2022THZ,
  author={Chen, Hui and Sarieddeen, Hadi and Ballal, Tarig and Wymeersch, Henk and Alouini, Mohamed-Slim and Al-Naffouri, Tareq Y.},
  journal=IEEE_O_CSTO,
  title={A tutorial on terahertz-band localization for {6G} communication systems},
  year={2022},
  volume={24},
  number={3},
  pages={1780-1815},
  doi={10.1109/COMST.2022.3178209}
}

@ARTICLE{Kim2026LMMBM,
  author={Kim, Seungnyun and Saha, Subham and Jeong, Seokhyun and Shim, Byonghyo and Win, Moe Z.},
  journal=IEEE_J_JSAC,
  title={Large multimodal model-based environment-aware beam management},
  year={2026},
  volume={44},
  number={},
  pages={991-1007},
  doi={10.1109/JSAC.2025.3611419}
}

@ARTICLE{Henk2025FrugalRIS,
  author={Ettefagh, Yasaman and Keskin, Musa Furkan and Keykhosravi, Kamran and Seco-Granados, Gonzalo and Wymeersch, Henk},
  journal=IEEE_J_VT,
  title={Frugal {RIS}-aided {3D} localization with {CFO} under {LoS} and {NLoS} conditions},
  year={2025},
  volume={},
  number={},
  pages={1-14},
  doi={10.1109/TVT.2025.3638972}
}

@ARTICLE{Guvenc2009Survey,
  author={Guvenc, Ismail and Chong, Chia-Chin},
  journal=IEEE_O_CSTO,
  title={A survey on {TOA} based wireless localization and {NLOS} mitigation techniques},
  year={2009},
  volume={11},
  number={3},
  pages={107-124},
  doi={10.1109/SURV.2009.090308}
}

@ARTICLE{Yang2024positioning,
  author={Yang, Yang and Chen, Mingzhe and Blankenship, Yufei and Lee, Jemin and Ghassemlooy, Zabih and Cheng, Julian and Mao, Shiwen},
  journal=IEEE_J_JSAC,
  title={Positioning using wireless networks: {A}pplications, recent progress, and future challenges},
  year={2024},
  volume={42},
  number={9},
  pages={2149-2178},
  doi={10.1109/JSAC.2024.3423629}
}

@ARTICLE{Li2019MMIMO,
  author={Li, Xuhong and Leitinger, Erik and Oskarsson, Magnus and Åström, Kalle and Tufvesson, Fredrik},
  journal=IEEE_J_WCOM,
  title={Massive {MIMO}-based localization and mapping exploiting phase information of multipath components},
  year={2019},
  volume={18},
  number={9},
  pages={4254-4267},
  doi={10.1109/TWC.2019.2922264}
}

@ARTICLE{Henk2025JointLoc,
  author={Fascista, Alessio and Deutschmann, Benjamin J. B. and Keskin, Musa Furkan and Wilding, Thomas and Coluccia, Angelo and Witrisal, Klaus and Leitinger, Erik and Seco-Granados, Gonzalo and Wymeersch, Henk},
  journal=IEEE_J_STSP,
  title={Joint localization, synchronization and mapping via phase-coherent distributed arrays},
  year={2025},
  volume={19},
  number={2},
  pages={412-429},
  doi={10.1109/JSTSP.2025.3533111}
}

@article{Zhou2026MixedNearFarXLMIMO,
  author        = {Zhou, Cong and You, Changsheng and Zhou, Chao and Cheng, Hongqiang and Shi, Shuo},
  title         = {Mixed near-field and far-field localization in extremely large-scale {MIMO} systems},
  journal        = {arXiv:2503.04681v3},
  year          = {2026}
}

@ARTICLE{Zhu2026MA,
  author={Zhu, Lipeng and Ma, Wenyan and Mei, Weidong and Zeng, Yong and Wu, Qingqing and Ning, Boyu and Xiao, Zhenyu and Shao, Xiaodan and Zhang, Jun and Zhang, Rui},
  journal=IEEE_O_CSTO,
  title={A tutorial on movable antennas for wireless networks},
  year={2026},
  volume={28},
  number={},
  pages={3002-3054},
  doi={10.1109/COMST.2025.3546373}
}

@ARTICLE{Xu2025FAS,
  author={Xu, Hao and Wong, Kai-Kit and New, Wee Kiat and Ghadi, Farshad Rostami and Zhou, Gui and Murch, Ross and Chae, Chan-Byoung and Zhu, Yongxu and Jin, Shi},
  journal=IEEE_J_COM,
  title={Capacity maximization for {FAS}-assisted multiple access channels},
  year={2025},
  volume={73},
  number={7},
  pages={4713-4731},
  doi={10.1109/TCOMM.2024.3516499}
}

@ARTICLE{Jiang2026MA,
  author={Jiang, Qijun and Shao, Xiaodan and Zhang, Rui},
  journal=IEEE_J_WCOM,
  title={Statistical channel-based low-complexity rotation and position optimization for {6D} movable antennas enabled wireless communication},
  year={2026},
  volume={25},
  number={},
  pages={8874-8889},
  doi={10.1109/TWC.2025.3647782}
}

@ARTICLE{Xia2025RIS,
  author={Xia, Wenchao and Zhao, Ben and Tang, Wankai and Zhu, Yongxu and Wong, Kai-Kit and Lambotharan, Sangarapillai and Shin, Hyundong},
  journal=IEEE_J_COM,
  title={{RIS}-empowered integrated location sensing and communication with superimposed pilots},
  year={2025},
  volume={73},
  number={10},
  pages={9644-9657},
  doi={10.1109/TCOMM.2025.3557994}
}

@ARTICLE{Kang2025RISLoc,
  author={Kang, Jeongwan and Ko, Seung-Woo and Kim, Sunwoo},
  journal=IEEE_J_WCOM,
  title={Near-field localization with {RIS} via two-dimensional signal path classification},
  year={2025},
  volume={24},
  number={4},
  pages={3417-3432},
  doi={10.1109/TWC.2025.3531055}
}

@techreport{3GPPTR22.870V1.1.0,
  
  author       = {{Technical Specification Group}},
  type         = {document},
  number       = {TR 22.870 V1.1.0, {R}elease 20},
  year         = {2026},
  month        = {Jan.},
  title = {Study on {6G} Use Cases and Service Requirements; Stage 1},
  institution={\textit{3rd Generation Partnership Project (3GPP)}}
}

@techreport{3GPPTR38.901V19.2.0,
  
  author       = {{Technical Specification Group Radio Access Network}},
  type         = {document},
  number       = {TR 38.901 V19.2.0, {R}elease 19},
  year         = {2025},
  month        = {Dec.},
  title = {Study on channel model for frequencies from 0.5 to 100 {GHz}},
  institution={\textit{3rd Generation Partnership Project (3GPP)}}
}

@ARTICLE{Jung2025GNN,
  author={Jung, Hongseok and Ko, Seung-Woo and Kim, Sunwoo},
  journal=IEEE_J_WCOML,
  title={Toward cooperative localization with implicit connectivity: {G}raph neural network approach},
  year={2025},
  volume={14},
  number={10},
  pages={3184-3188},
  doi={10.1109/LWC.2025.3588339}
}

@article{kang2026vision,
  title={{V}ision-language-model-guided differentiable ray tracing for fast and accurate multi-material {RF} parameter estimation},
  author={Kang, Zerui and Lim, Yishen and Gu, Zhouyou and Ko, Seung-Woo and Quek, Tony QS and Park, Jihong},
  journal={arXiv:2601.18242},
  year={2026}
}

@ARTICLE{Zhang2025VisionBlockage,
  author={Zhang, Tengyu and Wang, Yucong and Ouyang, Ming and Xing, Ling and Ma, Shaodan and Gao, Feifei},
  journal=IEEE_J_COM,
  title={An {RFSoC} prototype for third-party camera aided {mmWave} communications},
  year={2025},
  volume={73},
  number={4},
  pages={2769-2785},
  doi={10.1109/TCOMM.2024.3475275}
}

@article{marano2010nlos,
  title={{NLOS} identification and mitigation for localization based on {UWB} experimental data},
  author={Marano, Stefano and Gifford, Wesley M and Wymeersch, Henk and Win, Moe Z},
  journal=IEEE_J_JSAC,
  volume={28},
  pages={1026--1035},
  year={2010}
}

@ARTICLE{huang2020MachineLearningMIMO,
  author={Huang, Chen and Molisch, Andreas F. and He, Ruisi and Wang, Rui and Tang, Pan and Ai, Bo and Zhong, Zhangdui},
  journal=IEEE_J_WCOM,
  title={Machine learning-enabled {LOS}/{NLOS} identification for {MIMO} systems in dynamic environments},
  year={2020},
  volume={19},
  number={6},
  pages={3643-3657},
  doi={10.1109/TWC.2020.2967726}
}

@ARTICLE{Tedeschini2023latentSpace,
  author={Tedeschini, Bernardo Camajori and Nicoli, Monica and Win, Moe Z.},
  journal=IEEE_J_JSAC,
  title={On the latent space of {mmWave} {MIMO} channels for {NLOS} identification in {5G}-advanced systems},
  year={2023},
  volume={41},
  number={6},
  pages={1655-1669},
  doi={10.1109/JSAC.2023.3273769}
}

@ARTICLE{Choi2018deep,
  author={Choi, Jeong-Sik and Lee, Woong-Hee and Lee, Jae-Hyun and Lee, Jong-Ho and Kim, Seong-Cheol},
  journal=IEEE_J_VT,
  title={Deep learning based {NLOS} identification with commodity {WLAN} devices},
  year={2018},
  volume={67},
  number={4},
  pages={3295-3303},
  doi={10.1109/TVT.2017.2780121}
}

@ARTICLE{Yang2025Fuzzy,
  author={Yang, Hongchao and Wang, Yunjia and Seow, Chee Kiat and Sun, Meng and Coene, Sander and Huang, Lu and Joseph, Wout and Plets, David},
  journal=IEEE_J_IM,
  title={Fuzzy transformer machine learning for {UWB} {NLOS} identification and ranging mitigation},
  year={2025},
  volume={74},
  pages={1-17},
  doi={10.1109/TIM.2025.3548180}
}

@ARTICLE{torsoli2023BI,
  author={Torsoli, Gianluca and Win, Moe Z. and Conti, Andrea},
  journal=IEEE_J_JSAC,
  title={Blockage intelligence in complex environments for beyond {5G} localization},
  year={2023},
  volume={41},
  number={6},
  pages={1688-1701},
  doi={10.1109/JSAC.2023.3275612}
}

@ARTICLE{Zhu2020FingerprintSurvey,
  author={Zhu, Xiaoqiang and Qu, Wenyu and Qiu, Tie and Zhao, Laiping and Atiquzzaman, Mohammed and Wu, Dapeng Oliver},
  journal=IEEE_O_CSTO,
  title={Indoor intelligent fingerprint-based localization: {P}rinciples, approaches and challenges},
  year={2020},
  volume={22},
  number={4},
  pages={2634-2657},
  doi={10.1109/COMST.2020.3014304}
}

@ARTICLE{Xu2024SwinLoc,
  author={Xu, Xiaodong and Zhu, Fangzhou and Han, Shujun and Yu, Zhongyao and Zhao, Hangyu and Wang, Bizhu and Zhang, Ping},
  journal=IEEE_j_VT,
  title={Swin-Loc: {T}ransformer-based {CSI} fingerprinting indoor localization with {MIMO} {ISAC} system},
  year={2024},
  volume={73},
  number={8},
  pages={11664-11679},
  doi={10.1109/TVT.2024.3381433}
}

@ARTICLE{Wang2026Transformer,
  author={Wang, Xiping and Guan, Ke and He, Danping and Ai, Bo and Liu, Ruiqi and Yu, Keping and Zhong, Zhangdui and Hrovat, Andrej and Cui, Zhuangzhuang and Pollin, Sofie},
  journal=IEEE_J_WCOM,
  title={Resilient {3D} indoor localization using a masked {T}ransformer encoder with multi-band {CSI} fingerprints},
  year={2026},
  volume={25},
  number={},
  pages={4391-4404},
  doi={10.1109/TWC.2025.3610624}
}

@article{bhatia2025indoor,
  title={Indoor localization using compact, telemetry-agnostic, transfer-learning enabled decoder-only {T}ransformer},
  author={Bhatia, Nayan Sanjay and Kocheta, Pranay and Elliott, Russell and Kuttivelil, Harikrishna S and Obraczka, Katia},
  journal={arXiv:2510.11926},
  year={2025}
}

@ARTICLE{Morselli2023SI,
  author={Morselli, Flavio and Modarres Razavi, Sara and Win, Moe Z. and Conti, Andrea},
  journal=IEEE_J_WCOM,
  title={Soft information-based localization for {5G} networks and beyond},
  year={2023},
  volume={22},
  number={12},
  pages={9923-9938},
  doi={10.1109/TWC.2023.3275122}
}

@ARTICLE{Yang2018Gaussianprocess,
  author={Yang, Xiaofeng},
  journal=IEEE_J_SENSOR,
  title={{NLOS} mitigation for {UWB} localization based on sparse pseudo-input {G}aussian process},
  year={2018},
  volume={18},
  number={10},
  pages={4311-4316},
  doi={10.1109/JSEN.2018.2818158}
}

@ARTICLE{Wang2011GMM,
  author={Wang, Qinghua and Balasingham, Ilangko and Zhang, Miaomiao and Huang, Xin},
  journal=IEEE_J_COML,
  title={Improving {RSS}-based ranging in {LOS}-{NLOS} scenario using {GMMs}},
  year={2011},
  volume={15},
  number={10},
  pages={1065-1067},
  doi={10.1109/LCOMM.2011.080811.111087}
}

@article{yu2025combinatorial,
  title={Combinatorial data augmentation: {A} key enabler to bridge geometry- and data-driven {WiFi} positioning},
  author={Yu, Seung Min and Han, Kyuwon and Park, Jihong and Kim, Seong-Lyun and Ko, Seung-Woo},
  journal=IEEE_J_MC,
  volume={24},
  pages={306--320},
  year={2025}
}

@article{han2024mobility,
  title={Mobility-induced graph learning for {WiFi} positioning},
  author={Han, Kyuwon and Yu, Seung Min and Kim, Seong-Lyun and Ko, Seung-Woo},
  journal=IEEE_J_JSAC,
  volume={42},
  number={9},
  pages={2487--2502},
  year={2024}
}

@inproceedings{gretton2006kernel,
  author = {Gretton, Arthur and Borgwardt, Karsten and Rasch, Malte and Sch\"{o}lkopf, Bernhard and Smola, Alex},
  booktitle={Proc. Adv. Neural Inf. Process. Syst.},
  title = {A kernel method for the two-sample problem},
  year={2006},
  volume={19},
  pages={513-520}
}

@INPROCEEDINGS{Mensing2006LM,
  author={Mensing, C. and Plass, S.},
  booktitle={Proc. IEEE Int. Conf. Acoust. Speech Signal Process.},
  title={Positioning algorithms for cellular networks using {TDOA}},
  year={2006},
  volume={4},
  number={},
  pages={4-4},
  doi={10.1109/ICASSP.2006.1661018}
}

@ARTICLE{conti2024dataset,
  author={Conti, Andrea and Torsoli, Gianluca and Gómez-Vega, Carlos A. and Vaccari, Alessandro and Mazzini, Gianluca and Win, Moe Z.},
  journal={{IEEE} Open J. Veh. Technol.},
  title={{3GPP}-compliant datasets for {xG} location-aware networks},
  year={2024},
  volume={5},
  pages={473-484},
  doi={10.1109/OJVT.2023.3340993}
}

@ARTICLE{Zhou2024_6G,
  author={Zhou, Wenqi and Wang, Cheng-Xiang and Huang, Chen and Li, Zheao and Qian, Zhongyu and Lv, Zhen and Chen, Yunfei},
  journal=IEEE_J_IOT,
  title={Channel scenario extensions, identifications, and adaptive modeling for {6G} wireless communications},
  year={2024},
  volume={11},
  number={5},
  pages={7285-7308},
  doi={10.1109/JIOT.2023.3315296}
}

@INPROCEEDINGS{chen1999RWGH,
  author={Chen, Pi-Chun},
  booktitle={Proc. IEEE Wireless Commun. Netw. Conf.},
  title={A non-line-of-sight error mitigation algorithm in location estimation},
  year={1999},
  pages={316-320}
}

@ARTICLE{yin2014EM,
  author={Yin, Feng and Fritsche, Carsten and Gustafsson, Fredrik and Zoubir, Abdelhak M.},
  journal={IEEE Trans. Signal Process.},
  title={{EM}- and {JMAP-ML} based joint estimation algorithms for robust wireless geolocation in mixed {LOS/NLOS} environments},
  year={2014},
  volume={62},
  number={1},
  pages={168-182}
}

@ARTICLE{xiong2021MCC,
  author={Xiong, Wenxin and Schindelhauer, Christian and So, Hing Cheung and Wang, Zhi},
  journal={Circuits Syst. Signal Process.},
  title={Maximum correntropy criterion for robust {TOA}-based localization in {NLOS} environments},
  year={2021},
  volume={40},
  pages={6325-6339},
  doi={10.1007/s00034-021-01800-y}
}

@ARTICLE{rousseeuw1993MAD,
  author={Rousseeuw, Peter J. and Croux, Christophe},
  journal={J. Amer. Statist. Assoc.},
  title={Alternatives to the median absolute deviation},
  year={1993},
  volume={88},
  number={424},
  pages={1273-1283}
}

@ARTICLE{pearson2002hampel,
  author={Pearson, Ronald K.},
  journal={IEEE Trans. Control Syst. Technol.},
  title={Outliers in process modeling and identification},
  year={2002},
  volume={10},
  number={1},
  pages={55-63}
}

@ARTICLE{casas2006LMedS,
  author={Casas, Roberto and Marco, Alvaro and Guerrero, Jose J. and Falc{\'o}, Jorge},
  journal={EURASIP J. Adv. Signal Process.},
  title={Robust estimator for non-line-of-sight error mitigation in indoor localization},
  year={2006},
  volume={2006},
  pages={043429}
}

@ARTICLE{brier1950verification,
  author={Brier, Glenn W.},
  journal={Mon. Weather Rev.},
  title={Verification of forecasts expressed in terms of probability},
  year={1950},
  volume={78},
  number={1},
  pages={1-3}
}

@inproceedings{guo2017calibration,
  author={Guo, Chuan and Pleiss, Geoff and Sun, Yu and Weinberger, Kilian Q.},
  booktitle={Proc. Int. Conf. Mach. Learn.},
  title={On calibration of modern neural networks},
  year={2017},
  pages={1321-1330}
}

@BOOK{rousseeuw1987robust,
  author={Rousseeuw, Peter J. and Leroy, Annick M.},
  title={Robust Regression and Outlier Detection},
  publisher={Wiley},
  address={New York, NY, USA},
  year={1987}
}
\bibliographystyle{IEEEtran}

\clearpage

\appendices
\raggedbottom
\section{Evaluation of the Approximation in P1 and of the Factorization in P2} \label{app:factorization}

The maximum likelihood estimator in \eqref{ML_formulation} requires $\Theta$. \ref{Revised_ML_formulation} replaces $\Theta$ by $U$, the distribution of the PEL set $\mathcal{X}$, and \ref{Decomposed_Problem} factorizes the likelihood over the gNBs so that the LoS/NLoS state of each gNB is decided from $(d_n, \bm r_n, \bm r_n^0)$. This appendix measures the loss of each step in a synthetic setting with known $\Theta$.

\subsection{Synthetic Setting}
The UE location is $\bm p \sim \mathsf{Unif}(\mathcal{A})$, the uniform distribution over the area $\mathcal{A}$ of InF-DH in Sec.~\ref{sec:sim_setup}. The gNB locations $\{\bm z_n\}$ follow the layout of InF-DH: for $N < 18$, a subset of the 18 gNBs is used, and for $N > 18$, gNBs are added between the existing ones. Given $\bm p$, the triplets $(\gamma_n, b_n, w_n)$ in \eqref{eq:distance_estimate} are independent across gNBs \cite{Guvenc2009Survey, Morselli2023SI}. The NLoS indicator $\gamma_n$ of gNB $n$ is Bernoulli with probability $\pi_n(\bm p) = \mathsf{Pr}(\gamma_n = 1 \mid \bm p) = 1 - \exp(-\|\bm z_n - \bm p\|/\kappa)$, which follows the distance-dependent NLoS probability of 3GPP for InF \cite{3GPPTR38.901V19.2.0}. The noise $w_n$ has the PDF $\varphi(w; \mu_0, \sigma_0^2)$ under LoS and $\varphi(w; \mu_1, \sigma_1^2)$ under NLoS, where $\varphi(\cdot\,; \mu, \sigma^2)$ is the Gaussian PDF with mean $\mu$ and variance $\sigma^2$, and the bias $b_n$ has the exponential PDF $\nu e^{-\nu b}$, $b \ge 0$, independent of $w_n$ \cite{chen1999RWGH}. The parameters $(\mu_0, \sigma_0^2, \mu_1, \sigma_1^2, \nu)$ are fitted to the labeled range errors of InF-DH FR1. The parameter $\kappa$ is set so that the NLoS probability averaged over the UE locations and the gNBs equals $\pi$, i.e.,
$
\frac{1}{N |\mathcal{A}|} \sum_{n \in \mathcal{N}} \int_{\mathcal{A}} \pi_n(\bm p)\, \mathrm{d}\bm p = \pi
$,
where $|\mathcal{A}|$ is the area of $\mathcal{A}$. The environmental parameter $\Theta$ is the set $(\kappa, \mu_0, \sigma_0^2, \mu_1, \sigma_1^2, \nu)$, which is known in this setting.

\subsection{Detectors}
Since we compare the detectors by the AUC, each detector outputs the NLoS probability of each gNB instead of the hard decisions of \eqref{ML_formulation}, \ref{Revised_ML_formulation}, and \ref{Decomposed_Problem}. For \eqref{ML_formulation}, we use the exact posterior $\psi_n$ in \eqref{SI_formulation}, which attains the highest AUC of any detector given $\{d_n\}$ and $\Theta$ according to the Neyman--Pearson lemma. For \ref{Revised_ML_formulation} and \ref{Decomposed_Problem}, we use learned detectors.

\subsubsection{Exact Posterior}
Since $\Theta$ is known, $\psi_n$ in \eqref{SI_formulation} can be evaluated exactly. By \eqref{eq:distance_estimate}, the range error $e_n(\bm p) = d_n - \|\bm z_n - \bm p\|$ at the UE location $\bm p$ equals $\gamma_n b_n + w_n$, and its PDFs under LoS and NLoS are
\begin{align}\label{eq:app_pdfs}
f_w(e) &= \varphi(e; \mu_0, \sigma_0^2), \nonumber\\
f_{b+w}(e) &= \int_0^\infty \nu e^{-\nu b}\, \varphi(e - b; \mu_1, \sigma_1^2)\, \mathrm{d}b,
\end{align}
where the second is the convolution of the PDFs of $b_n$ and $w_n$. Let $\Lambda_n(\bm p) = (1 - \pi_n(\bm p)) f_w(e_n(\bm p)) + \pi_n(\bm p) f_{b+w}(e_n(\bm p))$ be the likelihood of $d_n$ at the location $\bm p$, marginalized over the LoS and NLoS states of gNB $n$. Then
\begin{align}\label{eq:app_exact}
\psi_n = \frac{\int_{\mathcal{A}} \pi_n(\bm p) f_{b+w}\big(e_n(\bm p)\big) \prod_{m \neq n} \Lambda_m(\bm p)\, \mathrm{d}\bm p}{\int_{\mathcal{A}} \prod_{m \in \mathcal{N}} \Lambda_m(\bm p)\, \mathrm{d}\bm p},
\end{align}
which is evaluated numerically over $\mathcal{A}$.

\subsubsection{Learned Detector for P1}
We use a \textit{graph neural network} (GNN) with message passing over the complete graph of the gNBs, trained with the cross-entropy loss on LoS/NLoS labels of separate UEs. Each gNB $n$ carries $\bm s_n = (d_n, \bm r_n, \bm r_n^0)$ and receives the message $\mathcal{E}_1(\bm s_n, \bm s_m, \bm z_m - \bm z_n)$ from every other gNB $m$, as shown in Fig.~\ref{fig:app_det_a}, and the GNN outputs the NLoS probability
\begin{align}\label{eq:app_msg}
\begin{split}
\hat\psi_n^{(\mathrm{P1})} &= \mathcal{E}_2(\bm s_n, \bm u_n), \\
\bm u_n &= \frac{1}{N - 1} \sum_{m \in \mathcal{N} \setminus \{n\}} \mathcal{E}_1(\bm s_n, \bm s_m, \bm z_m - \bm z_n),
\end{split}
\end{align}
where $\bm u_n$ is the mean message. $\mathcal{E}_1$ and $\mathcal{E}_2$ are tiny MLPs, shared across all gNB pairs and all gNBs, respectively. Through the messages, the NLoS probability of gNB $n$ depends on the statistics of all gNBs, as in \ref{Revised_ML_formulation}.

\begin{figure}[!t]
    \centering
    \subfloat[]{\begin{tikzpicture}[font=\footnotesize, >=latex,
  mlp/.style={draw=black!70, line width=0.7pt, rounded corners=1.5pt, fill=gray!15, inner sep=0pt, minimum width=0.62cm, minimum height=0.42cm},
  agg/.style={draw=black!70, line width=0.7pt, rounded corners=1.5pt, fill=white, inner sep=0pt, minimum width=0.82cm, minimum height=0.46cm},
  arr/.style={->, draw=black!70, line width=0.55pt},
  lab/.style={inner sep=1pt}]
  \node[lab, anchor=east] (i) at (2.45, 0.15) {$(\bm s_n, \bm s_m, \bm z_m - \bm z_n)$};
  \node[lab, anchor=north east] at (2.45, -0.08) {$m \in \mathcal{N} \setminus \{n\}$};
  \node[mlp] at (3.27, 0.27) {};
  \node[mlp] at (3.19, 0.19) {};
  \node[mlp] (e1) at (3.11, 0.11) {$\mathcal{E}_1$};
  \draw[arr] (2.47, 0.15) -- (2.80, 0.15);
  \node[agg] (m) at (4.55, 0.15) {mean};
  \draw[arr] (3.58, 0.15) -- (m.west);
  \node[mlp, minimum height=1.05cm] (e) at (6.10, -0.10) {$\mathcal{E}_2$};
  \draw[arr] (m.east) -- node[lab, above] {$\bm u_n$} (m.east -| e.west);
  \node[lab] (s) at (4.55, -0.35) {$\bm s_n$};
  \draw[arr] (s.east) -- (s.east -| e.west);
  \node[lab, anchor=west] (o) at (7.00, -0.10) {$\hat\psi_n^{(\mathrm{P1})}$};
  \draw[arr] (e.east) -- (o.west);
\end{tikzpicture}\label{fig:app_det_a}}\\
    \subfloat[]{\begin{tikzpicture}[font=\footnotesize, >=latex,
  mlp/.style={draw=black!70, line width=0.7pt, rounded corners=1.5pt, fill=gray!15, inner sep=0pt, minimum width=0.62cm, minimum height=0.42cm},
  arr/.style={->, draw=black!70, line width=0.55pt},
  lab/.style={inner sep=1pt}]
  \path (0, 0) -- (7.75, 0);
  \node[mlp, minimum height=1.05cm] (e) at (6.10, -0.10) {$\mathcal{E}_2$};
  \node[lab] (z) at (4.55, 0.15) {$\bm 0$};
  \draw[arr] (z.east) -- (z.east -| e.west);
  \node[lab] (s) at (4.55, -0.35) {$\bm s_n$};
  \draw[arr] (s.east) -- (s.east -| e.west);
  \node[lab, anchor=west] (o) at (7.00, -0.10) {$\hat\psi_n^{(\mathrm{P2})}$};
  \draw[arr] (e.east) -- (o.west);
\end{tikzpicture}\label{fig:app_det_b}}
    \caption{Learned detectors for gNB $n$: (a) for \ref{Revised_ML_formulation} and (b) for \ref{Decomposed_Problem}.}
    \label{fig:app_detectors}
\end{figure}

\begin{figure*}[!t]
    \centering
    \subfloat[]{
        \includegraphics[width=0.48\textwidth]{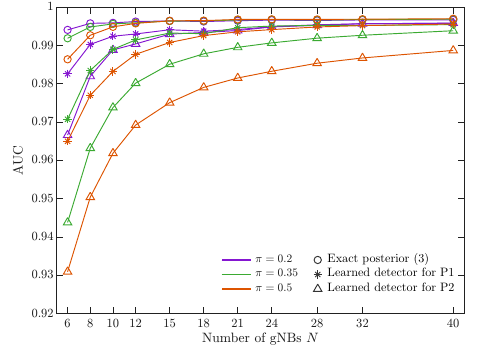}\label{fig:app_fact_auc}
    }
    \hfill
    \subfloat[]{
        \includegraphics[width=0.48\textwidth]{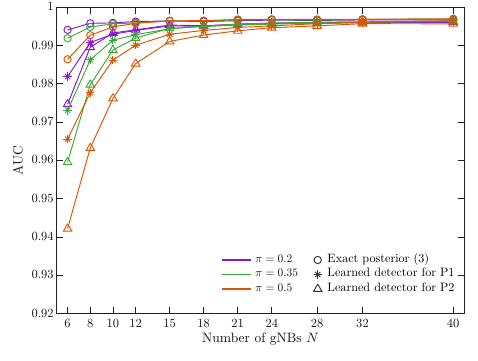}\label{fig:app_fact_auc_ref}
    }
    \caption{AUCs of the exact posterior (circles), the learned detector for \ref{Revised_ML_formulation} (asterisks), and the learned detector for \ref{Decomposed_Problem} (triangles) versus the number of gNBs $N$ for $\pi \in \{0.2, 0.35, 0.5\}$ in synthetic data with the layout and range-error laws of InF-DH FR1, averaged over three replicates. The learned detectors use (a) $(d_n, \bm r_n, \bm r_n^0)$ and (b) $(d_n, \tilde{\bm r}_n, \tilde{\bm r}_n^0)$ after the refinement in Sec.~\ref{sec:iterative_update}.}
    \label{fig:app_fact}
\end{figure*}

\begin{figure*}[!t]
    \centering
    \subfloat[]{
        \includegraphics[width=0.48\textwidth]{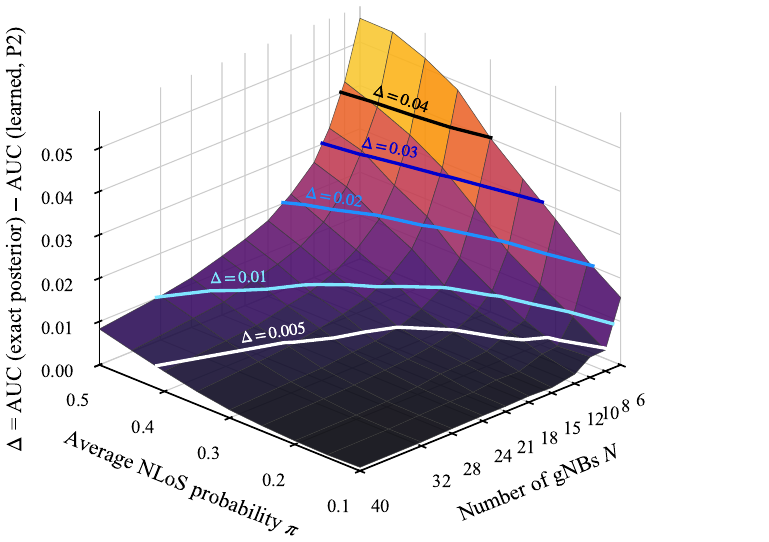}\label{fig:app_fact_ref_a}
    }
    \hfill
    \subfloat[]{
        \includegraphics[width=0.48\textwidth]{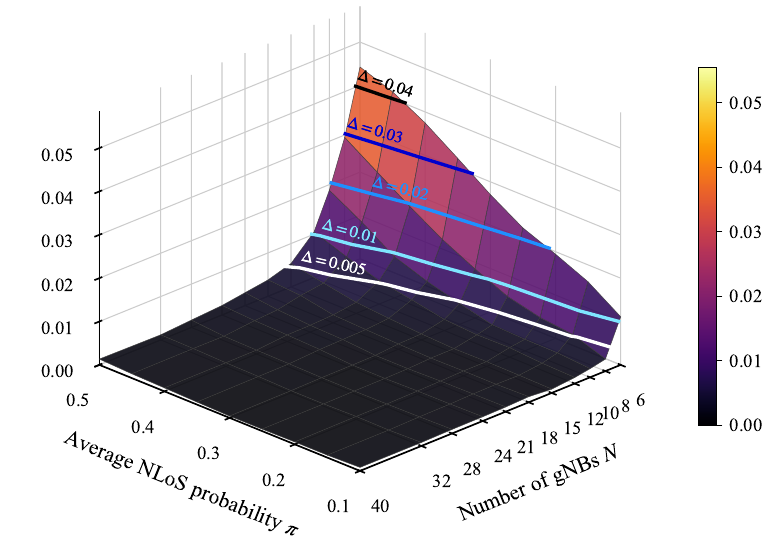}\label{fig:app_fact_ref_b}
    }
    \caption{Difference $\Delta$ between the AUCs of the exact posterior and the learned detector for \ref{Decomposed_Problem} versus the average NLoS probability $\pi$ and the number of gNBs $N$. The detector uses (a) $(d_n, \bm r_n, \bm r_n^0)$ and (b) $(d_n, \tilde{\bm r}_n, \tilde{\bm r}_n^0)$ after the refinement in Sec.~\ref{sec:iterative_update}. The vertical axis and the color scale of (b) are those of (a).}
    \label{fig:app_fact_ref}
\end{figure*}

\subsubsection{Learned Detector for P2}
As shown in Fig.~\ref{fig:app_det_b}, this detector is
\begin{align}\label{eq:app_fact_out}
\hat\psi_n^{(\mathrm{P2})} = \mathcal{E}_2(\bm s_n, \bm 0),
\end{align}
the same GNN trained and tested on the graph without edges, i.e., $\bm u_n = \bm 0$, so that gNB $n$ is disconnected from the other gNBs, as in \ref{Decomposed_Problem}.

\subsection{Results}
All detectors are evaluated by the AUC over all gNBs of the test UEs, with 5,000 training and 1,000 test UEs per $(\pi, N)$ and three replicates, for $\pi \in \{0.1, 0.15, \dots, 0.5\}$ and $N \in \{6, 8, 10, 12, 15, 18, 21, 24, 28, 32, 40\}$. We measure each loss by the AUC difference between two detectors: the loss from replacing $\Theta$ by $U$ between the exact posterior and the learned detector for \ref{Revised_ML_formulation}, the factorization loss between the learned detectors for \ref{Revised_ML_formulation} and \ref{Decomposed_Problem}, and the total loss $\Delta$ between the exact posterior and the learned detector for \ref{Decomposed_Problem}. Note that the learned detectors only approximate the best detectors given their inputs.

As shown in Fig.~\ref{fig:app_fact_auc}, the loss from replacing $\Theta$ by $U$ decreases with $N$ and increases with $\pi$, similar to the slower stabilization of InF-DH in Fig.~\ref{fig:MMD}. For $N \ge 18$, it is at most $0.004$, consistent with the hypothesis that \ref{Revised_ML_formulation} approaches \eqref{ML_formulation} for a large $N$. As anticipated in Remark~\ref{rem:pseudo_ref}, the factorization loss from \ref{Revised_ML_formulation} to \ref{Decomposed_Problem} shows the same trend. As shown in Fig.~\ref{fig:app_fact_ref_a}, the total loss $\Delta$, which includes both losses, also follows this trend.

Figs.~\ref{fig:app_fact_auc_ref} and \ref{fig:app_fact_ref_b} show the results when the inputs $(d_n, \bm r_n, \bm r_n^0)$ of the learned detectors for \ref{Revised_ML_formulation} and \ref{Decomposed_Problem} are replaced by $(d_n, \tilde{\bm r}_n, \tilde{\bm r}_n^0)$ after the refinement in Sec.~\ref{sec:iterative_update}. From Fig.~\ref{fig:app_fact_auc} to Fig.~\ref{fig:app_fact_auc_ref}, the factorization loss for $N \ge 15$ decreases from at most $0.016$ to at most $0.002$. This is because the refinement raises the AUC of the learned detector for \ref{Decomposed_Problem} close to that of the learned detector for \ref{Revised_ML_formulation}. The AUC of the learned detector for \ref{Revised_ML_formulation} changes little, since this detector already uses the statistics of all gNBs. From Fig.~\ref{fig:app_fact_ref_a} to Fig.~\ref{fig:app_fact_ref_b}, $\Delta$ also decreases substantially, especially for larger $\pi$.

\section{Constrained Expectation-Maximization} \label{app:CEM}

To tackle \ref{eq:cem_objective_simplified}, we propose a \textit{constrained expectation-maximization} (CEM) algorithm that explicitly enforces the site-survey prior $\pi$ in \ref{eq:cem_objective_simplified}. 
Let the score-sample set be $\mathcal{Q}\triangleq\{\rho_q\}$, and define the GMM in \eqref{eq:GMM_rep} at iteration $t$ as
\begin{align}
    f_{\mathrm{GMM}}(\rho;\Xi[t]) \triangleq \sum_{k=1}^{K} \alpha_k[t]\,
    \varphi\!\left(\rho;\mu_k[t],(\sigma_k[t])^2\right),
\end{align}
where $\Xi[t]\triangleq\{\alpha_k[t],\mu_k[t],\sigma_k[t]\}_{k=1}^{K}$.

The iterative steps of the proposed CEM are as follows:
\begin{itemize}
    \item \textit{Expectation Step (E-step):} Given the current parameters $\Xi[t]$, we evaluate the responsibility $\zeta_{q,k}[t]$, i.e., the posterior probability that the observation $\rho_q$ originates from the $k$-th Gaussian component:
    \begin{align}
        \zeta_{q,k}[t] 
        = \frac{\alpha_k[t]\varphi\!\left(\rho_q;\mu_k[t],(\sigma_k[t])^2\right)}
        {f_{\mathrm{GMM}}(\rho_q;\Xi[t])}.
    \end{align}
    The aggregate responsibility for each component is then computed as 
    \begin{align}
        S_k[t] = \sum_{q=1}^{|\mathcal{Q}|} \zeta_{q,k}[t],\quad k\in\{1,\dots,K\}.
    \end{align}

    \item \textit{Maximization Step (M-step):} The parameters are updated to $\Xi[t+1]$ to maximize the log-likelihood in \ref{eq:cem_objective_simplified} while enforcing its constraints. 
    First, the means and standard deviations are updated as in the standard EM:
    \begin{subequations}\label{eq:cem_mu_sigma_update_aligned}
    \begin{align}
        \mu_k[t+1] &= \frac{1}{S_k[t]} \sum_{q=1}^{|\mathcal{Q}|} \zeta_{q,k}[t] \rho_q, \\
        \sigma_k[t+1] &= \sqrt{\frac{1}{S_k[t]} \sum_{q=1}^{|\mathcal{Q}|} \zeta_{q,k}[t] \left(\rho_q - \mu_k[t+1]\right)^2 }.
    \end{align}
    \end{subequations}
    
    Next, to satisfy the monotonic constraint in \ref{eq:cem_objective_simplified}, we sort the updated means in ascending order such that
    $\mu_1[t+1] \le \mu_2[t+1] \le \dots \le \mu_K[t+1]$,
    and permute $\{\alpha_k[t],\sigma_k[t+1],S_k[t]\}$ accordingly to keep each component consistent after re-indexing.
    Finally, letting $k^*=\lceil K/2\rceil$ (as used in Sec.~\ref{sec:GMM}), we enforce the prior-mass constraints in \ref{eq:cem_objective_simplified} by renormalizing the mixing weights within each partition:
    \begin{align}
        \alpha_k[t+1] = 
        \begin{cases} 
        (1-\pi)\dfrac{S_k[t]}{\sum_{m=1}^{k^*} S_m[t]} & \text{if } 1 \le k \le k^*, \\[10pt]
        \pi\dfrac{S_k[t]}{\sum_{m=k^*+1}^{K} S_m[t]} & \text{if } k^* < k \le K.
        \end{cases}
    \end{align}

    \item \textit{Convergence Check:} The iterations stop when 
    \begin{align}
        \left| \sum_{q=1}^{|\mathcal{Q}|}\ln f_{\mathrm{GMM}}(\rho_q;\Xi[t+1])
      - \sum_{q=1}^{|\mathcal{Q}|}\ln f_{\mathrm{GMM}}(\rho_q;\Xi[t]) \right|\le \epsilon,
    \end{align}
    where $\epsilon>0$ is a predefined threshold (or when a maximum iteration count is reached).
\end{itemize}

\section{Ablation Study} \label{app:ablation}

To isolate the contribution of each component of CDA-ND, we add the components to HD one at a time:
\begin{itemize}
    \item \textit{HD (Sec.~\ref{sec:nlos-detection}):} HD detects whether each gNB is NLoS by comparing its score $\rho_n$ with the adaptive threshold $\eta$. The UE position is estimated by CDA-ND-RERS (HD), the unweighted median of the retained PELs.
    \item \textit{+ Initial SD function estimation (Sec.~\ref{sec:estimation}):} Inputting the score $\rho_n$ before refinement into the SD function $h^*$ yields the NLoS probability $h^*(\rho_n)$. The UE position is estimated by the weighted median in \eqref{eq:SD_positioning} with $h^*(\rho_n)$ in place of $\psi^*_n$ in \eqref{eq:reliability}.
    \item \textit{+ Score refinement (Sec.~\ref{sec:iterative_update}):} Inputting the refined score $\tilde\rho_n$ into $h^*$ yields the NLoS probability $h^*(\tilde\rho_n)$. The UE position is estimated by the weighted median in \eqref{eq:SD_positioning} with $h^*(\tilde\rho_n)$ in place of $\psi^*_n$ in \eqref{eq:reliability}.
    \item \textit{+ SD function revision (Sec.~\ref{sec:refine_reest}):} Using the revised SD function $\tilde h(\tilde\rho_n)$ in \eqref{eq:reest_SD} for the refined scores yields the final SD value $\psi^*_n$ in \eqref{eq:final_SD_Val}. The UE position is estimated by the weighted median in \eqref{eq:SD_positioning}, i.e., CDA-ND-RERS (SD).
\end{itemize}

\begin{table}[!t]
\centering
\setlength{\tabcolsep}{2.1pt}
\fontsize{6.6pt}{7.3pt}\selectfont
\sisetup{mode=text}%
\caption{Ablation of the components of CDA-ND: detection accuracy (\%), AUC, and Brier score of the NLoS probability, and MAE (m) of positioning.}
\label{tab:ablation}
\begin{tabular}{llcccccccc}
\toprule
\multirow{2}{*}{Scenario} & \multirow{2}{*}{Variant} & \multicolumn{4}{c}{FR1} & \multicolumn{4}{c}{FR2} \\
\cmidrule(lr){3-6}\cmidrule(lr){7-10}
& & Acc. & AUC & Brier & MAE & Acc. & AUC & Brier & MAE \\
\midrule
\multirow{4}{*}{InF-SH} & HD & \num{94.76} & \num{0.967} & -- & \num{0.53} & \num{88.28} & \num{0.918} & -- & \num{0.89} \\
 & + Init.\ SD fn.\ estimation & \num{95.68} & \num{0.966} & \num{0.038} & \num{0.51} & \num{89.12} & \num{0.918} & \num{0.098} & \num{0.71} \\
 & + Score refinement & \num{96.60} & \num{0.985} & \num{0.030} & \num{0.48} & \num{89.84} & \num{0.946} & \num{0.091} & \num{0.67} \\
 & + SD fn.\ revision & \num{96.76} & \num{0.979} & \num{0.030} & \num{0.47} & \num{89.58} & \num{0.939} & \num{0.078} & \num{0.67} \\
\midrule
\multirow{4}{*}{IOO} & HD & \num{84.52} & \num{0.845} & -- & \num{0.90} & \num{78.85} & \num{0.819} & -- & \num{0.96} \\
 & + Init.\ SD fn.\ estimation & \num{85.17} & \num{0.847} & \num{0.118} & \num{0.88} & \num{80.46} & \num{0.820} & \num{0.147} & \num{0.93} \\
 & + Score refinement & \num{90.37} & \num{0.929} & \num{0.080} & \num{0.77} & \num{83.45} & \num{0.882} & \num{0.123} & \num{0.82} \\
 & + SD fn.\ revision & \num{90.76} & \num{0.927} & \num{0.077} & \num{0.79} & \num{85.97} & \num{0.881} & \num{0.116} & \num{0.82} \\
\midrule
\multirow{4}{*}{InF-DH} & HD & \num{78.04} & \num{0.842} & -- & \num{2.12} & \num{78.54} & \num{0.836} & -- & \num{2.19} \\
 & + Init.\ SD fn.\ estimation & \num{78.69} & \num{0.840} & \num{0.158} & \num{1.93} & \num{78.93} & \num{0.835} & \num{0.157} & \num{2.09} \\
 & + Score refinement & \num{91.13} & \num{0.925} & \num{0.127} & \num{1.35} & \num{91.84} & \num{0.936} & \num{0.120} & \num{1.52} \\
 & + SD fn.\ revision & \num{91.50} & \num{0.926} & \num{0.075} & \num{1.30} & \num{92.28} & \num{0.933} & \num{0.069} & \num{1.42} \\
\bottomrule
\end{tabular}
\end{table}

As shown in Table~\ref{tab:ablation}, the initial SD function leaves the AUC nearly unchanged because it preserves the order of the scores within each fold. Its detection gain comes from the decision threshold: a threshold of $0.5$ on the NLoS probability is the maximum a posteriori decision and raises the accuracy slightly in all configurations. Its positioning gain comes from the weighted median: the PELs generated from gNBs with high $h^*(\rho_n)$ receive small weights, which reduces the MAE in all configurations.

Most of the AUC gain comes from the score refinement, which improves the separability of the scores. As noted in Remark~\ref{rem:pseudo_ref}, the score becomes less accurate in harsher NLoS environments, and the gain of the refinement grows accordingly: the AUC increases by up to $\num{0.028}$ in InF-SH and by up to $\num{0.101}$ in InF-DH, with IOO in between. The gains in the accuracy and the MAE follow the same trend: the refinement reduces the MAE by up to \SI{0.04}{m} in InF-SH and by up to \SI{0.58}{m} in InF-DH.

The SD function revision fits the SD function to the refined scores, which improves the reliability of the NLoS probability: the Brier score decreases the most in InF-DH, from $\num{0.127}$ to $\num{0.075}$ in FR1 and from $\num{0.120}$ to $\num{0.069}$ in FR2. The more reliable NLoS probability raises the accuracy, except in InF-SH FR2, while the AUC remains nearly unchanged because $\tilde h$ also preserves the order of the refined scores within each fold. The MAE changes by at most \SI{0.10}{m}.

\section{Effect of the Number of gNBs on Runtime and Performance} \label{app:gnb}
\begin{table*}[!t]
\centering
\setlength{\tabcolsep}{5pt}
\fontsize{7.2pt}{8.5pt}\selectfont
\caption{Average runtime per snapshot (ms) of each step of CDA-ND, with the average number of refinement iterations $t^*$ in parentheses.}
\label{tab:runtime}
\begin{tabular}{lccccccccc}
\toprule
\multirow{3}{*}{Scenario} & \multirow{3}{*}{$\#$ gNBs $N$} & \multirow{3}{*}{$\#$ PELs $L$} & \multicolumn{5}{c}{Step} & \multicolumn{2}{c}{Total} \\
\cmidrule(lr){4-8}\cmidrule(lr){9-10}
 & & & \shortstack{(a) PEL\\generation} & \shortstack{(b) Score gen.\\and HD} & \shortstack{(c) HD\\positioning} & \shortstack{(d) Score\\refinement ($t^*$)} & \shortstack{(e) SD\\positioning} & \shortstack{CDA-ND-\\RERS (HD)} & \shortstack{CDA-ND-\\RERS (SD)} \\
 & & & Sec.~\ref{Sec:Hypothesis} & Secs.~\ref{Sec:Factorization}, \ref{Sec:Threshold-Based Test} & Sec.~\ref{sec:aided_loc} & Sec.~\ref{sec:iterative_update} & Sec.~\ref{sec:SI_positioning} & (a)+(b)+(c) & (a)+(b)+(d)+(e) \\
\midrule
\multirow{6}{*}{\shortstack[l]{InF-SH\\FR1}} & 8 & 52 & \num{0.31} & \num{0.03} & \num{0.01} & \num{0.05} (\num{2.8}) & \num{0.01} & \num{0.35} & \num{0.40} \\
 & 10 & 108 & \num{0.48} & \num{0.04} & \num{0.01} & \num{0.09} (\num{3.5}) & \num{0.01} & \num{0.53} & \num{0.62} \\
 & 12 & 199 & \num{0.66} & \num{0.06} & \num{0.02} & \num{0.14} (\num{3.6}) & \num{0.02} & \num{0.74} & \num{0.88} \\
 & 14 & 328 & \num{0.87} & \num{0.09} & \num{0.02} & \num{0.22} (\num{3.7}) & \num{0.03} & \num{0.98} & \num{1.21} \\
 & 16 & 507 & \num{1.08} & \num{0.13} & \num{0.03} & \num{0.34} (\num{3.7}) & \num{0.04} & \num{1.24} & \num{1.59} \\
 & 18 & 738 & \num{1.33} & \num{0.19} & \num{0.04} & \num{0.51} (\num{3.7}) & \num{0.05} & \num{1.56} & \num{2.08} \\
\midrule
\multirow{6}{*}{\shortstack[l]{InF-DH\\FR1}} & 8 & 52 & \num{0.70} & \num{0.03} & \num{0.01} & \num{0.19} (\num{12.3}) & \num{0.01} & \num{0.74} & \num{0.93} \\
 & 10 & 108 & \num{1.00} & \num{0.04} & \num{0.01} & \num{0.28} (\num{12.9}) & \num{0.01} & \num{1.05} & \num{1.33} \\
 & 12 & 199 & \num{1.28} & \num{0.06} & \num{0.01} & \num{0.41} (\num{12.8}) & \num{0.02} & \num{1.35} & \num{1.77} \\
 & 14 & 328 & \num{1.56} & \num{0.09} & \num{0.01} & \num{0.61} (\num{12.2}) & \num{0.02} & \num{1.66} & \num{2.28} \\
 & 16 & 507 & \num{1.87} & \num{0.13} & \num{0.01} & \num{0.94} (\num{12.3}) & \num{0.02} & \num{2.01} & \num{2.96} \\
 & 18 & 738 & \num{2.21} & \num{0.19} & \num{0.02} & \num{1.34} (\num{11.6}) & \num{0.03} & \num{2.42} & \num{3.77} \\
\bottomrule
\end{tabular}
\end{table*}

\begin{table*}[!t]
\centering
\caption{NLoS detection and positioning performance versus the number of gNBs $N$, averaged over five random subsets of the gNBs for $N<18$.}
\label{tab:gnb}
\begin{tabular}{lccccccccc}
\toprule
\multirow{2}{*}{Scenario} & \multirow{2}{*}{$\#$ gNBs $N$} & \multicolumn{2}{c}{HD} & \multicolumn{2}{c}{SD} & \multicolumn{3}{c}{Mean positioning error (m)} \\
\cmidrule(lr){3-4}\cmidrule(lr){5-6}\cmidrule(lr){7-9}
 & & Accuracy (\%) & AUC & Accuracy (\%) & AUC & CDA-RERS & \textbf{CDA-ND-RERS (HD)} & \textbf{CDA-ND-RERS (SD)} \\
\midrule
\multirow{6}{*}{\shortstack[l]{InF-SH\\FR1}} & 8 & \num{88.54} & \num{0.925} & \num{94.96} & \num{0.969} & \num{2.25} & \num{1.53} & \num{1.04} \\
 & 10 & \num{90.21} & \num{0.937} & \num{95.94} & \num{0.977} & \num{1.30} & \num{1.01} & \num{0.73} \\
 & 12 & \num{92.26} & \num{0.953} & \num{96.67} & \num{0.981} & \num{1.00} & \num{0.80} & \num{0.62} \\
 & 14 & \num{93.06} & \num{0.958} & \num{96.59} & \num{0.980} & \num{0.81} & \num{0.67} & \num{0.55} \\
 & 16 & \num{94.12} & \num{0.964} & \num{96.72} & \num{0.980} & \num{0.68} & \num{0.59} & \num{0.51} \\
 & 18 & \num{94.76} & \num{0.967} & \num{96.76} & \num{0.979} & \num{0.60} & \num{0.53} & \num{0.47} \\
\midrule
\multirow{6}{*}{\shortstack[l]{InF-DH\\FR1}} & 8 & \num{72.86} & \num{0.781} & \num{75.45} & \num{0.822} & \num{12.50} & \num{11.11} & \num{9.23} \\
 & 10 & \num{75.00} & \num{0.801} & \num{83.60} & \num{0.867} & \num{9.10} & \num{8.25} & \num{6.19} \\
 & 12 & \num{75.76} & \num{0.816} & \num{87.09} & \num{0.891} & \num{7.24} & \num{5.85} & \num{4.10} \\
 & 14 & \num{76.86} & \num{0.829} & \num{88.97} & \num{0.906} & \num{5.79} & \num{4.12} & \num{2.69} \\
 & 16 & \num{77.51} & \num{0.834} & \num{90.21} & \num{0.915} & \num{4.72} & \num{2.97} & \num{1.90} \\
 & 18 & \num{78.04} & \num{0.842} & \num{91.50} & \num{0.926} & \num{3.97} & \num{2.12} & \num{1.30} \\
\bottomrule
\end{tabular}
\end{table*}

We evaluate CDA-ND in InF-SH FR1 and InF-DH FR1 with $N\in\{8,10,\dots,18\}$ gNBs, where the results for $N<18$ are averaged over five random subsets of $N$ gNBs selected from the 18 gNBs.

\subsection{Effect of the Number of gNBs on Runtime} \label{app:runtime}
Table~\ref{tab:runtime} lists the average runtime per snapshot of each step of CDA-ND over the 1000 UE realizations. The runtime of each step is measured separately on a single core of an Apple M4 Pro CPU with MATLAB R2025a, where the PELs of all subsets and the (weighted) medians of all gNBs are computed by vectorized operations, and the totals of CDA-ND-RERS for both HD and SD are the sums of the steps involved. The total runtime increases with $N$, e.g., from \SI{0.93}{ms} to \SI{3.77}{ms} for CDA-ND-RERS (SD) in InF-DH FR1. PEL generation takes the largest share of the runtime, and the runtime of score refinement increases with the number of iterations $t^*$, which is larger in InF-DH than in InF-SH.

\subsection{Effect of the Number of gNBs on Performance} \label{app:gnb_perf}
Table~\ref{tab:gnb} lists the NLoS detection and positioning performance. As $N$ decreases, the performance of CDA-ND degrades gradually, while CDA-ND-RERS for both HD and SD achieves lower errors than CDA-RERS for all $N$. SD outperforms HD in all metrics. As $N$ increases from 16 to 18, the accuracy of SD increases by \num{0.04} and \num{1.29} percentage points in InF-SH and InF-DH, respectively, and the mean positioning error of CDA-ND-RERS (SD) decreases by \SI{0.04}{m} and \SI{0.60}{m}, respectively. This is consistent with Fig.~\ref{fig:MMD}, where the PEL distribution stabilizes faster in InF-SH than in InF-DH.

\section{Validation with Real Wi-Fi RTT Measurements} \label{app:wifi}

\begin{table*}[!t]
\centering
\setlength{\tabcolsep}{3.5pt}
\fontsize{7.2pt}{8.5pt}\selectfont
\caption{Walks of the Wi-Fi RTT measurements (MP: measurement position).}
\vspace{-4pt}

\label{tab:wifi_env}
\begin{tabular}{l*{10}{c}}
\toprule
Walk & 1 & 2 & 3 & 4 & 5 & 6 & 7 & 8 & 9 & 10 \\
\midrule
Experiment day & Day 1 & \multicolumn{3}{c}{Day 2} & \multicolumn{6}{c}{Day 3} \\
\cmidrule(lr){2-2}\cmidrule(lr){3-5}\cmidrule(lr){6-11}
Device & \multicolumn{4}{c}{Pixel 2 XL} & \multicolumn{6}{c}{Pixel 3a} \\
\cmidrule(lr){2-5}\cmidrule(lr){6-11}
Path shape & Type 1 & \multicolumn{3}{c}{Type 2} & \multicolumn{3}{c}{Type 3} & \multicolumn{3}{c}{Type 4} \\
\cmidrule(lr){2-2}\cmidrule(lr){3-5}\cmidrule(lr){6-8}\cmidrule(lr){9-11}
Number of MPs & 70 & 34 & 34 & 34 & 83 & 83 & 83 & 95 & 95 & 95 \\
\bottomrule
\end{tabular}
\end{table*}

\begin{table*}[!t]
\centering
\setlength{\tabcolsep}{4.0pt}
\caption{Positioning error on the Wi-Fi RTT measurements: mean error of each walk (columns 1--10, grouped by RTT distribution and path shape) and the mean, standard deviation (Std), and percentiles of the error over all walks. RTT dataset: required (\checkmark) or not (--). Labeling: required labels.}

\vspace{-4pt}

\label{tab:wifi_pos}
\begin{tabular}{lcc*{10}{c}|ccccc}
\toprule
 & \multirow{4}{*}{\shortstack{RTT\\dataset}} & \multirow{4}{*}{Labeling} & \multicolumn{10}{c}{Mean error of each walk (m)} & \multicolumn{5}{c}{All walks (m)} \\
\cmidrule(lr){4-13}\cmidrule(lr){14-18}
RTT distribution & & & Day 1 & \multicolumn{3}{c}{Day 2} & \multicolumn{6}{c}{Day 3} & & & & & \\
\cmidrule(lr){4-4}\cmidrule(lr){5-7}\cmidrule(lr){8-13}
Path shape & & & Type 1 & \multicolumn{3}{c}{Type 2} & \multicolumn{3}{c}{Type 3} & \multicolumn{3}{c}{Type 4} & Mean & Std & 50th & 75th & 95th \\
\cmidrule(lr){4-4}\cmidrule(lr){5-7}\cmidrule(lr){8-10}\cmidrule(lr){11-13}
Method & & & 1 & 2 & 3 & 4 & 5 & 6 & 7 & 8 & 9 & \multicolumn{1}{c}{10} & & & & & \\
\midrule
RWGH \cite{chen1999RWGH} & -- & -- & \num{3.16} & \num{2.48} & \num{2.75} & \num{2.65} & \num{4.61} & \num{4.99} & \num{3.82} & \num{4.11} & \num{5.11} & \num{4.40} & \num{4.10} & \num{4.33} & \num{2.85} & \num{4.87} & \num{11.52} \\
LMedS \cite{casas2006LMedS} & -- & -- & \num{2.29} & \num{2.03} & \num{2.24} & \num{1.59} & \num{3.87} & \num{3.52} & \num{3.07} & \num{3.20} & \num{2.89} & \num{2.92} & \num{2.95} & \num{2.41} & \num{2.24} & \num{3.73} & \num{7.68} \\
EM \cite{yin2014EM} & -- & -- & \num{2.06} & \num{1.99} & \num{1.86} & \num{1.49} & \num{3.29} & \num{2.91} & \num{2.76} & \num{2.81} & \num{2.73} & \num{2.64} & \num{2.62} & \num{2.15} & \num{2.02} & \num{3.27} & \num{7.12} \\
SR-MCC \cite{xiong2021MCC} & -- & -- & \num{4.07} & \num{3.10} & \num{3.40} & \num{3.36} & \num{4.77} & \num{4.74} & \num{4.43} & \num{4.29} & \num{4.03} & \num{4.12} & \num{4.19} & \num{2.96} & \num{3.41} & \num{5.74} & \num{10.18} \\
CDA-RERS \cite{yu2025combinatorial} & -- & -- & \num{2.05} & \num{1.76} & \num{1.80} & \num{1.55} & \num{2.31} & \num{2.24} & \num{2.09} & \num{2.20} & \num{2.20} & \num{2.23} & \num{2.12} & \num{1.47} & \num{1.81} & \num{2.65} & \num{5.54} \\
\midrule
SI \cite{Morselli2023SI} & \checkmark & GT pos. & \num{2.21} & \num{1.31} & \num{1.55} & \num{1.51} & \num{1.05} & \num{1.05} & \num{0.94} & \num{1.10} & \num{1.17} & \num{1.02} & \num{1.23} & \num{1.02} & \num{0.98} & \num{1.69} & \num{3.16} \\
RF & \checkmark & GT pos. & \num{1.87} & \num{1.31} & \num{1.58} & \num{1.09} & \num{1.64} & \num{1.75} & \num{1.49} & \num{1.61} & \num{1.41} & \num{1.56} & \num{1.57} & \num{1.16} & \num{1.29} & \num{1.93} & \num{3.83} \\
MLP & \checkmark & GT pos. & \num{2.04} & \num{1.51} & \num{1.74} & \num{1.17} & \num{1.84} & \num{1.88} & \num{1.61} & \num{1.69} & \num{1.51} & \num{1.64} & \num{1.69} & \num{1.22} & \num{1.38} & \num{2.06} & \num{4.06} \\
\midrule
\textbf{CDA-ND-RERS (HD)} & -- & -- & \textbf{\num{1.89}} & \textbf{\num{1.52}} & \textbf{\num{1.05}} & \textbf{\num{1.26}} & \textbf{\num{1.85}} & \textbf{\num{1.91}} & \textbf{\num{1.73}} & \textbf{\num{1.83}} & \textbf{\num{1.72}} & \textbf{\num{1.72}} & \textbf{\num{1.73}} & \textbf{\num{1.20}} & \textbf{\num{1.44}} & \textbf{\num{2.17}} & \textbf{\num{4.08}} \\
\textbf{CDA-ND-RERS (SD)} & \checkmark & $\pi$ only & \textbf{\num{1.35}} & \textbf{\num{1.28}} & \textbf{\num{0.96}} & \textbf{\num{1.04}} & \textbf{\num{1.83}} & \textbf{\num{1.78}} & \textbf{\num{1.71}} & \textbf{\num{1.63}} & \textbf{\num{1.62}} & \textbf{\num{1.66}} & \textbf{\num{1.58}} & \textbf{\num{1.01}} & \textbf{\num{1.43}} & \textbf{\num{2.04}} & \textbf{\num{3.24}} \\
\bottomrule
\end{tabular}

\vspace{-4pt}

\end{table*}

\begin{table}[!t]
\centering
\setlength{\tabcolsep}{4pt}
\caption{NLoS detection on the Wi-Fi RTT measurements against proxy labels.}
\label{tab:wifi_det}
\begin{tabular}{lcccccc}
\toprule
Method & \shortstack{RTT\\dataset} & \shortstack{LoS/NLoS\\labeling} & Acc. & Rec. & Prec. & AUC \\
\midrule
LMedS \cite{casas2006LMedS} & -- & -- & \num{76.83} & \num{43.10} & \num{83.50} & \num{0.766} \\
EM \cite{yin2014EM} & -- & -- & \num{80.06} & \num{66.57} & \num{74.43} & \num{0.782} \\
\midrule
RF classifier & \checkmark & \checkmark & \num{89.43} & \num{80.77} & \num{88.40} & \num{0.955} \\
MLP classifier & \checkmark & \checkmark & \num{89.42} & \num{81.01} & \num{88.16} & \num{0.958} \\
\midrule
\textbf{Hard Decision} & -- & -- & \num{85.71} & \num{80.29} & \num{79.56} & \num{0.911} \\
\textbf{Soft Decision} & \checkmark & $\pi$ only & \num{90.57} & \num{88.40} & \num{85.47} & \num{0.951} \\
\bottomrule
\end{tabular}
\end{table}

We apply CDA-ND to the Wi-Fi RTT measurements collected at the underground parking lot of Korea Railroad Research Institute, Uiwang, Korea, with $N=10$ access points (Qualcomm IPQ4018) and Google Pixel 2 XL and Pixel 3a smartphones. The dataset consists of 706 range snapshots with ground-truth positions, gathered over ten walks on different days (Table~\ref{tab:wifi_env}), whose detailed path shapes are given in \cite{han2024mobility}. As no LoS/NLoS labels are available, we set the weak prior to $\pi = 0.35$, which corresponds to the fraction of the ranges exceeding the true distances by more than \SI{2}{m}, and use the same criterion as proxy NLoS labels to evaluate the detection. SI, RF, and MLP are trained with nine of the ten walks and tested with the remaining walk, which is repeated for every walk.

Table~\ref{tab:wifi_pos} summarizes the positioning performance. CDA-ND-RERS for both HD and SD reduces the mean error from \SI{2.12}{m} of CDA-RERS \cite{yu2025combinatorial} to \SI{1.73}{m} and \SI{1.58}{m}, respectively, and outperforms the benchmarks without training. Among the benchmarks trained with GT positions, RF and MLP achieve \SI{1.57}{m} and \SI{1.69}{m}. SI achieves the lowest mean error of \SI{1.23}{m}, which is partly attributed to the GT positions overlapping across the path shapes. Against the proxy labels, SD detects the NLoS-biased ranges with an accuracy of \SI{90.57}{\%} and an AUC of 0.951, which are higher than those of LMedS and EM and comparable to those of RF and MLP trained with the proxy labels (Table~\ref{tab:wifi_det}).

\begin{table*}[!t]
\centering
\setlength{\tabcolsep}{4.9pt}
\fontsize{7.2pt}{7.5pt}\selectfont
\caption{Mean, standard deviation, and percentiles of the positioning error (\si{m}) for each scenario--frequency configuration. RTT dataset: required (\checkmark) or not (--). Labeling: required labels.}
\label{tab:si_percentile}
\begin{tabular}{llccccccccccccc}
\toprule
Scenario & Method & \shortstack{RTT\\dataset} & Labeling & Mean & Std & 10th & 30th & 40th & 50th & 60th & 70th & 90th & 95th & 99th \\
\midrule
\multirow{10}{*}{\shortstack[l]{InF-SH \\ FR1}} & RWGH \cite{chen1999RWGH} & -- & -- & \num{1.59} & \num{3.95} & \num{0.32} & \num{0.57} & \num{0.70} & \num{0.82} & \num{1.00} & \num{1.20} & \num{2.28} & \num{4.25} & \num{18.66} \\
 & LMedS \cite{casas2006LMedS} & -- & -- & \num{0.80} & \num{1.01} & \num{0.31} & \num{0.53} & \num{0.62} & \num{0.72} & \num{0.81} & \num{0.91} & \num{1.27} & \num{1.49} & \num{2.16} \\
 & EM \cite{yin2014EM} & -- & -- & \num{0.53} & \num{0.64} & \num{0.16} & \num{0.31} & \num{0.37} & \num{0.44} & \num{0.50} & \num{0.57} & \num{0.87} & \num{1.07} & \num{3.17} \\
 & SR-MCC \cite{xiong2021MCC} & -- & -- & \num{1.26} & \num{3.11} & \num{0.32} & \num{0.57} & \num{0.67} & \num{0.74} & \num{0.86} & \num{0.97} & \num{1.41} & \num{2.25} & \num{16.75} \\
 & CDA-RERS \cite{yu2025combinatorial} & -- & -- & \num{0.60} & \num{0.45} & \num{0.21} & \num{0.39} & \num{0.45} & \num{0.53} & \num{0.61} & \num{0.70} & \num{1.00} & \num{1.21} & \num{1.98} \\
\cmidrule(l){2-15}
 & SI \cite{Morselli2023SI} & \checkmark & GT pos. & \num{0.38} & \num{0.22} & \num{0.13} & \num{0.25} & \num{0.29} & \num{0.34} & \num{0.40} & \num{0.46} & \num{0.67} & \num{0.76} & \num{1.08} \\
 & RF & \checkmark & GT pos. & \num{0.48} & \num{0.35} & \num{0.17} & \num{0.30} & \num{0.36} & \num{0.43} & \num{0.49} & \num{0.56} & \num{0.85} & \num{0.98} & \num{1.48} \\
 & MLP & \checkmark & GT pos. & \num{0.62} & \num{0.50} & \num{0.24} & \num{0.42} & \num{0.49} & \num{0.56} & \num{0.63} & \num{0.71} & \num{0.98} & \num{1.18} & \num{1.76} \\
\cmidrule(l){2-15}
 & \textbf{CDA-ND-RERS (HD)} & -- & -- & \num{0.53} & \num{0.33} & \num{0.18} & \num{0.35} & \num{0.42} & \num{0.49} & \num{0.56} & \num{0.64} & \num{0.91} & \num{1.05} & \num{1.42} \\
 & \textbf{CDA-ND-RERS (SD)} & \checkmark & $\pi$ only & \num{0.47} & \num{0.25} & \num{0.17} & \num{0.31} & \num{0.38} & \num{0.43} & \num{0.50} & \num{0.57} & \num{0.82} & \num{0.96} & \num{1.17} \\
\midrule
\multirow{10}{*}{\shortstack[l]{InF-SH \\ FR2}} & RWGH \cite{chen1999RWGH} & -- & -- & \num{2.61} & \num{6.64} & \num{0.40} & \num{0.72} & \num{0.87} & \num{1.03} & \num{1.27} & \num{1.61} & \num{4.58} & \num{10.44} & \num{29.80} \\
 & LMedS \cite{casas2006LMedS} & -- & -- & \num{0.98} & \num{1.38} & \num{0.31} & \num{0.54} & \num{0.65} & \num{0.75} & \num{0.86} & \num{1.02} & \num{1.50} & \num{1.99} & \num{6.36} \\
 & EM \cite{yin2014EM} & -- & -- & \num{1.57} & \num{4.02} & \num{0.18} & \num{0.37} & \num{0.45} & \num{0.56} & \num{0.66} & \num{0.80} & \num{2.29} & \num{8.74} & \num{19.62} \\
 & SR-MCC \cite{xiong2021MCC} & -- & -- & \num{2.71} & \num{4.99} & \num{0.36} & \num{0.65} & \num{0.80} & \num{0.98} & \num{1.18} & \num{1.55} & \num{6.95} & \num{14.13} & \num{24.15} \\
 & CDA-RERS \cite{yu2025combinatorial} & -- & -- & \num{1.19} & \num{1.83} & \num{0.27} & \num{0.48} & \num{0.59} & \num{0.72} & \num{0.85} & \num{1.04} & \num{2.27} & \num{3.76} & \num{10.04} \\
\cmidrule(l){2-15}
 & SI \cite{Morselli2023SI} & \checkmark & GT pos. & \num{0.58} & \num{0.49} & \num{0.17} & \num{0.33} & \num{0.40} & \num{0.46} & \num{0.55} & \num{0.64} & \num{1.05} & \num{1.39} & \num{2.64} \\
 & RF & \checkmark & GT pos. & \num{0.63} & \num{0.61} & \num{0.20} & \num{0.36} & \num{0.42} & \num{0.50} & \num{0.59} & \num{0.71} & \num{1.10} & \num{1.36} & \num{2.41} \\
 & MLP & \checkmark & GT pos. & \num{0.80} & \num{0.91} & \num{0.27} & \num{0.47} & \num{0.55} & \num{0.64} & \num{0.75} & \num{0.85} & \num{1.31} & \num{1.67} & \num{3.82} \\
\cmidrule(l){2-15}
 & \textbf{CDA-ND-RERS (HD)} & -- & -- & \num{0.89} & \num{1.32} & \num{0.20} & \num{0.40} & \num{0.50} & \num{0.58} & \num{0.71} & \num{0.83} & \num{1.40} & \num{2.38} & \num{7.61} \\
 & \textbf{CDA-ND-RERS (SD)} & \checkmark & $\pi$ only & \num{0.67} & \num{0.45} & \num{0.21} & \num{0.40} & \num{0.49} & \num{0.57} & \num{0.66} & \num{0.79} & \num{1.19} & \num{1.46} & \num{2.44} \\
\midrule
\multirow{10}{*}{\shortstack[l]{IOO \\ FR1}} & RWGH \cite{chen1999RWGH} & -- & -- & \num{2.50} & \num{4.22} & \num{0.47} & \num{0.84} & \num{1.00} & \num{1.18} & \num{1.49} & \num{1.97} & \num{5.28} & \num{8.71} & \num{22.93} \\
 & LMedS \cite{casas2006LMedS} & -- & -- & \num{1.50} & \num{3.78} & \num{0.37} & \num{0.62} & \num{0.76} & \num{0.89} & \num{1.01} & \num{1.18} & \num{2.00} & \num{3.34} & \num{17.63} \\
 & EM \cite{yin2014EM} & -- & -- & \num{1.39} & \num{2.97} & \num{0.26} & \num{0.49} & \num{0.60} & \num{0.72} & \num{0.86} & \num{1.03} & \num{2.40} & \num{4.29} & \num{14.36} \\
 & SR-MCC \cite{xiong2021MCC} & -- & -- & \num{2.90} & \num{4.96} & \num{0.47} & \num{0.83} & \num{0.99} & \num{1.19} & \num{1.47} & \num{2.01} & \num{7.53} & \num{11.87} & \num{25.07} \\
 & CDA-RERS \cite{yu2025combinatorial} & -- & -- & \num{0.95} & \num{1.27} & \num{0.29} & \num{0.52} & \num{0.63} & \num{0.73} & \num{0.82} & \num{0.96} & \num{1.50} & \num{2.32} & \num{5.17} \\
\cmidrule(l){2-15}
 & SI \cite{Morselli2023SI} & \checkmark & GT pos. & \num{0.68} & \num{0.71} & \num{0.21} & \num{0.39} & \num{0.47} & \num{0.55} & \num{0.64} & \num{0.76} & \num{1.14} & \num{1.47} & \num{2.57} \\
 & RF & \checkmark & GT pos. & \num{0.93} & \num{1.45} & \num{0.28} & \num{0.49} & \num{0.58} & \num{0.67} & \num{0.79} & \num{0.91} & \num{1.54} & \num{2.18} & \num{4.60} \\
 & MLP & \checkmark & GT pos. & \num{0.94} & \num{1.15} & \num{0.30} & \num{0.52} & \num{0.61} & \num{0.70} & \num{0.82} & \num{0.97} & \num{1.63} & \num{2.27} & \num{4.65} \\
\cmidrule(l){2-15}
 & \textbf{CDA-ND-RERS (HD)} & -- & -- & \num{0.90} & \num{0.91} & \num{0.29} & \num{0.52} & \num{0.62} & \num{0.71} & \num{0.82} & \num{0.94} & \num{1.51} & \num{2.25} & \num{4.48} \\
 & \textbf{CDA-ND-RERS (SD)} & \checkmark & $\pi$ only & \num{0.79} & \num{0.79} & \num{0.26} & \num{0.48} & \num{0.56} & \num{0.65} & \num{0.76} & \num{0.87} & \num{1.26} & \num{1.66} & \num{3.59} \\
\midrule
\multirow{10}{*}{\shortstack[l]{IOO \\ FR2}} & RWGH \cite{chen1999RWGH} & -- & -- & \num{2.13} & \num{3.39} & \num{0.46} & \num{0.88} & \num{1.02} & \num{1.18} & \num{1.49} & \num{1.89} & \num{4.13} & \num{6.13} & \num{17.36} \\
 & LMedS \cite{casas2006LMedS} & -- & -- & \num{1.31} & \num{1.33} & \num{0.39} & \num{0.72} & \num{0.85} & \num{0.97} & \num{1.11} & \num{1.30} & \num{2.38} & \num{3.70} & \num{7.59} \\
 & EM \cite{yin2014EM} & -- & -- & \num{1.74} & \num{2.71} & \num{0.32} & \num{0.64} & \num{0.84} & \num{1.03} & \num{1.25} & \num{1.58} & \num{3.85} & \num{5.35} & \num{11.77} \\
 & SR-MCC \cite{xiong2021MCC} & -- & -- & \num{4.29} & \num{33.85} & \num{0.47} & \num{0.79} & \num{0.93} & \num{1.09} & \num{1.38} & \num{1.76} & \num{4.18} & \num{5.93} & \num{17.19} \\
 & CDA-RERS \cite{yu2025combinatorial} & -- & -- & \num{1.06} & \num{0.89} & \num{0.32} & \num{0.57} & \num{0.70} & \num{0.82} & \num{0.97} & \num{1.15} & \num{1.99} & \num{2.83} & \num{4.81} \\
\cmidrule(l){2-15}
 & SI \cite{Morselli2023SI} & \checkmark & GT pos. & \num{0.72} & \num{0.54} & \num{0.23} & \num{0.43} & \num{0.50} & \num{0.60} & \num{0.70} & \num{0.83} & \num{1.30} & \num{1.59} & \num{3.01} \\
 & RF & \checkmark & GT pos. & \num{0.98} & \num{1.59} & \num{0.28} & \num{0.53} & \num{0.64} & \num{0.75} & \num{0.86} & \num{1.03} & \num{1.60} & \num{2.26} & \num{4.81} \\
 & MLP & \checkmark & GT pos. & \num{1.16} & \num{1.15} & \num{0.37} & \num{0.64} & \num{0.78} & \num{0.90} & \num{1.03} & \num{1.19} & \num{1.98} & \num{3.06} & \num{6.49} \\
\cmidrule(l){2-15}
 & \textbf{CDA-ND-RERS (HD)} & -- & -- & \num{0.96} & \num{0.71} & \num{0.34} & \num{0.57} & \num{0.68} & \num{0.80} & \num{0.92} & \num{1.08} & \num{1.69} & \num{2.31} & \num{4.03} \\
 & \textbf{CDA-ND-RERS (SD)} & \checkmark & $\pi$ only & \num{0.82} & \num{0.94} & \num{0.27} & \num{0.48} & \num{0.57} & \num{0.68} & \num{0.81} & \num{0.92} & \num{1.35} & \num{1.67} & \num{3.44} \\
\midrule
\multirow{10}{*}{\shortstack[l]{InF-DH \\ FR1}} & RWGH \cite{chen1999RWGH} & -- & -- & \num{12.18} & \num{13.21} & \num{0.99} & \num{3.50} & \num{5.75} & \num{7.92} & \num{11.15} & \num{15.16} & \num{28.03} & \num{36.61} & \num{59.78} \\
 & LMedS \cite{casas2006LMedS} & -- & -- & \num{12.55} & \num{13.36} & \num{0.78} & \num{4.44} & \num{5.78} & \num{7.09} & \num{9.75} & \num{15.20} & \num{30.66} & \num{40.32} & \num{57.76} \\
 & EM \cite{yin2014EM} & -- & -- & \num{7.47} & \num{10.39} & \num{0.30} & \num{0.63} & \num{0.87} & \num{1.86} & \num{4.96} & \num{9.46} & \num{21.18} & \num{26.76} & \num{48.64} \\
 & SR-MCC \cite{xiong2021MCC} & -- & -- & \num{16.89} & \num{12.05} & \num{3.69} & \num{9.70} & \num{12.39} & \num{15.04} & \num{17.69} & \num{20.23} & \num{31.38} & \num{40.12} & \num{60.40} \\
 & CDA-RERS \cite{yu2025combinatorial} & -- & -- & \num{3.97} & \num{5.51} & \num{0.33} & \num{0.70} & \num{0.85} & \num{1.08} & \num{1.57} & \num{3.89} & \num{12.26} & \num{16.08} & \num{24.62} \\
\cmidrule(l){2-15}
 & SI \cite{Morselli2023SI} & \checkmark & GT pos. & \num{1.08} & \num{2.66} & \num{0.21} & \num{0.42} & \num{0.51} & \num{0.61} & \num{0.69} & \num{0.81} & \num{1.32} & \num{2.23} & \num{11.71} \\
 & RF & \checkmark & GT pos. & \num{3.19} & \num{3.75} & \num{0.42} & \num{1.06} & \num{1.46} & \num{1.91} & \num{2.55} & \num{3.37} & \num{7.43} & \num{10.67} & \num{17.68} \\
 & MLP & \checkmark & GT pos. & \num{3.35} & \num{3.94} & \num{0.57} & \num{1.20} & \num{1.58} & \num{2.03} & \num{2.68} & \num{3.47} & \num{7.73} & \num{10.66} & \num{20.11} \\
\cmidrule(l){2-15}
 & \textbf{CDA-ND-RERS (HD)} & -- & -- & \num{2.12} & \num{3.64} & \num{0.38} & \num{0.71} & \num{0.85} & \num{0.95} & \num{1.09} & \num{1.27} & \num{5.62} & \num{9.60} & \num{19.27} \\
 & \textbf{CDA-ND-RERS (SD)} & \checkmark & $\pi$ only & \num{1.30} & \num{2.37} & \num{0.29} & \num{0.52} & \num{0.63} & \num{0.73} & \num{0.83} & \num{0.95} & \num{1.88} & \num{5.15} & \num{12.41} \\
\midrule
\multirow{10}{*}{\shortstack[l]{InF-DH \\ FR2}} & RWGH \cite{chen1999RWGH} & -- & -- & \num{11.71} & \num{13.30} & \num{1.03} & \num{2.94} & \num{4.81} & \num{7.82} & \num{10.70} & \num{14.40} & \num{27.15} & \num{36.37} & \num{60.91} \\
 & LMedS \cite{casas2006LMedS} & -- & -- & \num{11.16} & \num{12.38} & \num{0.81} & \num{3.72} & \num{5.02} & \num{6.71} & \num{8.70} & \num{12.54} & \num{27.27} & \num{34.47} & \num{57.08} \\
 & EM \cite{yin2014EM} & -- & -- & \num{7.11} & \num{10.30} & \num{0.33} & \num{0.67} & \num{0.93} & \num{1.71} & \num{4.63} & \num{8.73} & \num{20.35} & \num{25.55} & \num{42.42} \\
 & SR-MCC \cite{xiong2021MCC} & -- & -- & \num{16.13} & \num{10.60} & \num{3.61} & \num{9.78} & \num{12.46} & \num{14.80} & \num{17.59} & \num{20.50} & \num{29.37} & \num{34.78} & \num{46.96} \\
 & CDA-RERS \cite{yu2025combinatorial} & -- & -- & \num{3.73} & \num{5.42} & \num{0.37} & \num{0.69} & \num{0.84} & \num{1.08} & \num{1.49} & \num{2.95} & \num{12.29} & \num{16.29} & \num{22.73} \\
\cmidrule(l){2-15}
 & SI \cite{Morselli2023SI} & \checkmark & GT pos. & \num{1.33} & \num{3.69} & \num{0.23} & \num{0.42} & \num{0.50} & \num{0.59} & \num{0.69} & \num{0.84} & \num{1.52} & \num{2.92} & \num{20.15} \\
 & RF & \checkmark & GT pos. & \num{2.66} & \num{4.09} & \num{0.41} & \num{0.87} & \num{1.18} & \num{1.49} & \num{1.88} & \num{2.48} & \num{5.81} & \num{8.60} & \num{17.27} \\
 & MLP & \checkmark & GT pos. & \num{2.80} & \num{4.31} & \num{0.46} & \num{0.93} & \num{1.24} & \num{1.61} & \num{2.02} & \num{2.60} & \num{6.03} & \num{9.60} & \num{17.85} \\
\cmidrule(l){2-15}
 & \textbf{CDA-ND-RERS (HD)} & -- & -- & \num{2.19} & \num{3.62} & \num{0.37} & \num{0.67} & \num{0.82} & \num{0.97} & \num{1.12} & \num{1.33} & \num{5.86} & \num{11.45} & \num{17.61} \\
 & \textbf{CDA-ND-RERS (SD)} & \checkmark & $\pi$ only & \num{1.42} & \num{2.62} & \num{0.30} & \num{0.57} & \num{0.66} & \num{0.76} & \num{0.90} & \num{1.05} & \num{2.23} & \num{5.26} & \num{15.48} \\
\bottomrule
\end{tabular}
\end{table*}

\section{Benchmark Settings} \label{app:benchmark}

\subsection{NLoS Detection} \label{app:bench_det}
The RF and MLP classifiers in Sec.~\ref{sec:sim_identification} take as inputs the following quantities of RWGH \cite{chen1999RWGH} and LMedS \cite{casas2006LMedS}:
\begin{itemize}
    \item The range measurement $d_n$.
    \item The residual $e_{\mathrm{LS},n} = d_n - \|\hat{\bm p}_{\mathrm{LS}} - \bm z_n\|$, where $\hat{\bm p}_{\mathrm{LS}}$ is the least-squares position estimate using all $N$ range measurements, as in RWGH \cite{chen1999RWGH}.
    \item The residual $e_{\mathrm{LM},n} = d_n - \|\hat{\bm p}_{\mathrm{LM}} - \bm z_n\|$, where $\hat{\bm p}_{\mathrm{LM}}$ is the position estimate of LMedS \cite{casas2006LMedS}.
    \item The robust scale estimate of LMedS \cite{casas2006LMedS}, $\sigma_{\mathrm{LM}} = 1.48\,(1 + 5/(N-3))\sqrt{r_{\mathrm{med}}}$, where $r_{\mathrm{med}}$ is the minimum median of the squared residuals, which is common to all gNBs of the UE.
    \item The root-mean-square of $\{e_{\mathrm{LS},n}\}_{n=1}^{N}$, i.e., the normalized residual of RWGH \cite{chen1999RWGH}, which is common to all gNBs of the UE.
\end{itemize}
Other features obtainable from RWGH \cite{chen1999RWGH} and LMedS \cite{casas2006LMedS}, and other classifiers are also tested, which yield no noticeable gain in accuracy. Both classifiers are trained and tested with the $10$-fold cross-validation in Sec.~\ref{sec:sim_setup}. The RF consists of $300$ classification trees with a minimum leaf size of $20$. The MLP is a fully connected network with two hidden layers of $64$ and $32$ neurons or three hidden layers of $128$, $64$, and $32$ neurons, selected by validation in each fold, with ReLU activations, trained by minimizing the binary cross-entropy with $\ell_2$ regularization using the L-BFGS algorithm on the standardized inputs.

\subsection{Positioning} \label{app:bench_pos}
The RF and MLP regressors in Sec.~\ref{sec:sim_positioning}, with the same inputs and structures as in Appendix~\ref{app:bench_det}, are trained with the GT positions to predict the range error of each gNB and its magnitude by minimizing the mean squared error.

\section{Detailed Positioning Results} \label{app:positioning}
Table~\ref{tab:si_percentile} lists the mean, standard deviation, and percentiles of the positioning error of the methods in Sec.~\ref{sec:sim_positioning} for each scenario--frequency configuration.

\end{document}